\documentclass[twocolumn,varg]{aa}  
\usepackage{txfonts}

\usepackage{twoopt}
\usepackage{graphicx}
\graphicspath{current_plots/}
\usepackage[switch]{lineno}
\usepackage{verbatim}
\usepackage{pgfplots}
\pgfplotsset{compat=1.7}
\usepackage{txfonts}
\usepackage{graphicx}
\usepackage{xcolor}
\usepackage{subfig}
\usepackage{verbatim}  
\usepackage{textcomp}
\usepackage{graphicx}
\usepackage{upgreek}
\usepackage{amsmath}
\usepackage[pdftex,colorlinks,linkcolor = blue,
            urlcolor  = blue,
            citecolor = blue,
            anchorcolor = blue,
breaklinks=true]{hyperref} 
\bibpunct{(}{)}{;}{a}{}{,}             
\makeatletter
\usepackage{longtable}
\usepackage[export]{adjustbox}
\usepackage{afterpage}
\usepackage{emptypage}
\usepackage[autostyle]{csquotes}
\usepackage{textcomp}
\usepackage{float}
\usepackage{enumitem}
\usepackage{bm}
\usepackage{siunitx}
\usepackage{listings}
\usepackage[breaklinks=true]{hyperref} 
\bibpunct{(}{)}{;}{a}{}{,}             
\makeatletter
\usepackage{subfig}
\usepackage{placeins}
\makeatletter
\setpagewiselinenumbers
\modulolinenumbers[5]
\linenumbers
\nolinenumbers
\begin{document}
   \title{Phenomenological modelling of the Crab Nebula's broadband energy spectrum and its apparent extension}
\titlerunning{Model for SED and extension of Crab Nebula}
   \author{L. Dirson
          \inst{\ref{uhh},\ref{strasb}}
          \and
          D. Horns\inst{\ref{uhh}}
          }

   \institute{\label{uhh}Institute for Experimental Physics, Universität Hamburg,
              Luruper Chaussee 149, 22761 Hamburg, Germany,
              \email{dieter.horns@uni-hamburg.de}
         \and
          \label{strasb} Universit\'e de Strasbourg, CNRS, Observatoire astronomique de Strasbourg, UMR 7550, F-67000 Strasbourg, France. \email{ludmilla.dirson@astro.unistra.fr} 
             }

   \date{Received ; accepted }


  \abstract
   {The Crab Nebula emits exceptionally bright non-thermal radiation across the entire wavelength range 
   from the radio to the most energetic photons. So far, 
   the underlying physical model of a relativistic wind from 
   the pulsar terminating 
   in a hydrodynamic standing shock
    has remained fairly unchanged since the early 1970s when
   it was first introduced. One of the predictions of this model
   is an increase in the toroidal magnetic field downstream from the shock where the flow velocity drops quickly with increasing distance until it reaches its asymptotic value, matching
   the expansion velocity of the nebula.
} %
  {The magnetic field strength in the nebula 
  is poorly known.
    Using the recent measurements of the spatial extension 
  and improved spectroscopy of the gamma-ray
  nebula, it has become --for the first time -- feasible to determine in a robust way both the strength as well as the radial dependence of the 
  magnetic field in the downstream flow.}
   {In this work, we introduce a detailed 
   radiative model which was used
   to calculate the emission from
   non-thermal electrons (synchrotron
   and inverse Compton) as well as from thermal dust  
   present in the Crab Nebula  in a self-consistent way to 
   compare it quantitatively with observational data.
   Special care was given to the radial dependence  of the electron and
   seed field density.}
   { The radiative model was used to estimate the parameters
   related to the electron populations responsible for radio and optical/X-ray synchrotron emission. In this context, the
   mass of cold and warm dust was determined. A combined fit based upon a $\chi^2$ minimisation successfully reproduced
    the complete data set used. For the best-fitting model, the energy density  
   of the magnetic field dominates over the particle energy density up to a distance of $\approx 1.3~r_s$ ($r_s$: distance
   of the termination shock from the pulsar). The very high energy (VHE: $E>100$~GeV) and ultra-high energy (UHE: $E>100$~TeV)
   gamma-ray spectra set the strongest constraints on the radial dependence of the magnetic field, favouring a model 
   where $B(r)=(264\pm9)~\mu\mathrm{G} (r/r_s)^{-0.51\pm0.03}$.
   For a collection of VHE measurements during epochs of higher hard X-ray
   emission, a significantly different solution $B(r)=(167\pm 5)~\mu\mathrm{G} (r/r_s)^{-0.29(+0.03,-0.06)}$ is found.}
  {The high energy (HE: $E>100$~MeV)  and VHE gamma-ray observations of the Crab Nebula lift the degeneracy of
  the synchrotron emission between particle and magnetic field energy density. The reconstructed magnetic field
  and its radial dependence indicates a ratio of Poynting  to kinetic energy flux $\sigma\approx 0.1$
  at the
  termination shock, which is $\approx 30$ times
  larger than estimated up to now.  Consequently, the 
  confinement of the nebula  
  would  require additional mechanisms to slow  the flow 
  down through, for example, excitation of small-scale turbulence with 
  possible dissipation of the magnetic field. 
  }

   \maketitle
%

\section{Introduction}
The Crab Nebula and pulsar are the centrepieces 
of gamma-ray astronomy and high energy (HE) astrophysics in general. The 
discovery of its very high energy (VHE) emission by the pioneering observations with the Whipple air Cherenkov telescope \citep{1989ApJ...342..379W}
marks a breakthrough in the field by establishing the Crab Nebula as the brightest steady gamma-ray source. 

The current understanding of the nebula was already established 
 several decades ago by \citet{RG74}. 
The description of the dynamics of the nebula and
its radiative model dominated by synchrotron emission of relativistic
electrons
\citep{1973ApJ...186..249P} has remained
basically unchanged for the past 50 years. 
In this model, the nebula is powered by a continuous 
cold and ultra-relativistic pulsar wind. 
The wind  terminates
at a distance of about 0.13~pc \citep{2012ApJ...746...41W} to the pulsar in a standing shock. The position
of the termination shock is commonly associated with the bright ring observed at soft X-rays which has
a major axis of $(13.3\pm0.2)^{\prime\prime}$ \citep{2012ApJ...746...41W}\footnote{The pulsar is slightly offset by
$0.9^{\prime\prime}$ from the centre of the ellipse}. 

The  slow expansion of the nebula requires the downstream flow to 
slow down. 
In an ideal magneto-hydrodynamic (MHD) setting as suggested by 
\citet{KennelCoroniti},  the ratio of  magnetic to kinetic energy ($\sigma$ parameter) 
upstream of the shock is estimated to $\sigma=0.003$ to match the boundary conditions of the slow expansion of the outer nebula of $\approx 1500$~km/s. 
In the downstream flow, the toroidal (`wound up') magnetic field  would increase due to magnetic 
flux conservation with increasing distance until
equipartition is reached. 

However, with the three-dimensional treatment of the MHD problem \citep{2013Porth} it has been demonstrated that the magnetic field structure
and magnetic field present in the  plasma  very likely 
deviate from the simplified model of \citet{KennelCoroniti}. Additionally,
the conversion of ordered fields to turbulent fields which 
would be dissipated in non-ideal MHD plasma could change the magnetic field configuration substantially 
\citep{2017MNRAS.470.4066B,2018MNRAS.478.4622T}.

The spatially resolved degree of polarisation and position angle 
in the radio (e.g. \citet{bietenholz1990})  and optical (e.g. 
\citet{1990ApJ...365..224H}) indicate the presence of large-scale
toroidal fields in the region of the torus, while at larger
distances the field orientation becomes predominantly radial as 
traced out in the radio.  The fraction of polarised X-ray emission from the Crab Nebula was
measured to be $(19.2\pm1)~\%$ \citep{1978ApJ...220L.117W, 2020NatAs...4..511F,2021ApJ...922..221L}, which indicates that the magnetic field energy
is roughly equally shared between an ordered and small-scale turbulent field \citep{2017MNRAS.470.4066B}. A more detailed view
of the spatial distribution of the magnetic fields will be possible with the upcoming observations of the Crab Nebula with the
Imaging X-ray Polarimetry Explorer (IXPE) \citep{2016SPIE.9905E..17W}.

With the observation of (inverse Compton) generated gamma-ray emission,
the ideal MHD flow-type solution \citep{KennelCoroniti} has been confirmed to provide a reasonable description of the 
nebula's gamma-ray spectrum \citep{AA1996}. 
Improved spectroscopy at VHEs demonstrates however,
that the simple approach does not provide a reasonable 
description of the data anymore \citep{MagicCrab}. Even though
the ideal MHD flow solution and its siblings 
remain as benchmarks
\citep[see e.g.][]{abdalla_resolving_2020},
the observational data
are of sufficient quality to test the underlying assumptions of the  simple flow model 
 already developed in the 1970s. 

So far, it has not been
possible to spatially resolve the magnetic field strength in the nebula or any other pulsar wind nebulae.
High energy and VHE gamma-ray observations have been used 
to estimate an average magnetic field from the ratio of synchrotron to inverse Compton emission \citep{Aharonianetal2004}, which requires careful
modelling of the seed photon fields in the nebula \citep{meyer}. Even though this method is robust, observational details, 
which are required to match the precision of the observational data with  sufficiently accurate
model calculations,  have been mostly  omitted in past studies.

In this analysis, we take the next step and include the recent measurement of the spatial extension at gamma rays of the nebula 
\citep{paul_paper,abdalla_resolving_2020} in combination with
the most recent spatial and spectral  multi-wavelength data  to 
reconstruct, for the first time, the electron and magnetic field configuration in the nebula in a  consistent phenomenological model
without any underlying assumptions as to the flow properties. 

The underlying parameters were estimated with a $\chi^2$ minimisation method.
This is the first study of this type carried out on any 
gamma-ray emitting source. 
Thanks to the exceptional multi-wavelength observational data, 
the Crab Nebula remains
a corner stone object for the study of non-thermal processes
in relativistic plasma. 

\section{Observations}
\label{section:observations}
In the following, the heliocentric distance of the Crab Nebula is assumed to be $d_\mathrm{Crab}=2~\mathrm{kpc}$ \citep{1973PASP...85..579T}.
The  basis of the compilation  of  multi-wavelength data  used  here  is  summarised  by  \citet{Aharonianetal2004,meyer} and  
references  therein. In the 
past decade, both new observations  (see Table~\ref{tab:observationsSED}) as well as refined 
and relevant corrections on the extinction in the interstellar medium have become available.

\subsection{Radio and infrared observations}
\begin{figure}
\includegraphics[width=\linewidth]{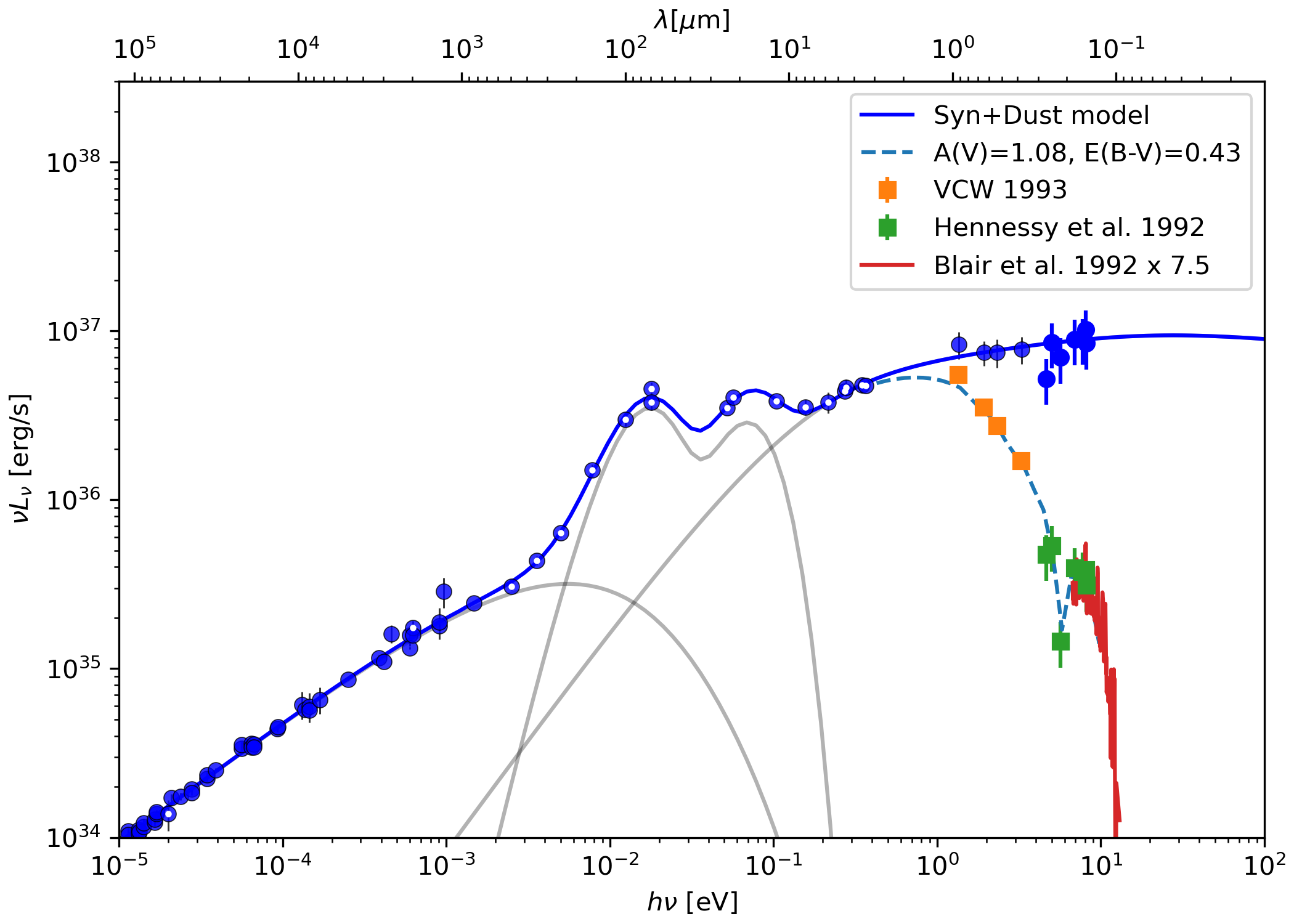}
      \caption{Observed spectral energy distribution (SED) from radio to 
      optical photon energies. The additional data from the recent compilation of 
      photometric data from \citet{2019MNRAS.488..164D} are marked with
      white circles. Optical/UV data are presented 
      with and without extinction correction (see text for more details). The SED has been fit with the superposition of thermal dust emission (see Table~\ref{tab:varIR} for
      a list of parameters and uncertainties) and the underlying synchrotron continuum (see Table~\ref{table:parameters}). 
               \label{dust_data}}
   \end{figure}

The transition between the radio synchrotron component and the thermal dust emission is explored through observations with the Herschel (PACS and SPIRE), Spitzer (MIPS), while
the transition from the dust emission to the optical/wind synchrotron emission is traced with WISE and Spitzer (IRAC) multi-band photometry. The integrated aperture-photometry 
(integrating over an ellipse centred on the pulsar position, with PA=$40^\circ$ and semi-major and minor axes of \SI{245}{\arcsec}$\times$\SI{163}{\arcsec})
is taken
from \citet{2019MNRAS.488..164D}, Table 3. In addition to the continuum component, the emission lines contribute up to 18~\% of the continuum flux in the PACS $\lambda=100~\mu$m 
pass band. The resulting data set is shown in Fig.~\ref{dust_data} in combination with the best-fitting model. 
The new and additional data taken from the compilation of \citet{2019MNRAS.488..164D} are marked with
white circles in Fig.~\ref{dust_data}. 
\subsection{Optical observations}
\label{subsec:optical}
In comparison to previous works \citep{meyer}, the optical data used here have been updated to include a more
recent estimate of the extinction along the line of sight
towards the Crab Nebula. In the original compilation of \citet{2006A&A...457..899A}, the extinction was corrected using  $A(V)=1.5~\mathrm{mag}$
with a colour excess $E(B-V)=0.47~\mathrm{mag}$ 
guided by the discussion of \citet{1993A&A...270..370V}. More recent 3D-reconstruction of the Galactic extinction with Gaia, Pan-STARRS 1, and 2MASS
observations lead to a smaller value of $A(V)=1.2~\mathrm{mag}$  \citep{2019ApJ...887...93G}. We chose here the central value of $A(V)=(1.08\pm0.38)~\mathrm{mag}$
and $E(B-V)=0.43~\mathrm{mag}$ determined by \citet{2019MNRAS.488..164D} from modelling the dust emission along the line of sight towards the Crab Nebula. 
The wavelength dependent extinction
was calculated with the extinction curve from \citet{1994ApJ...422..158O}.  The resulting shape of the visual continuum emission follows a power law, such that $I_\nu\propto \nu^{-s}$
with $s_O\approx 0.8$. This is considerably harder than the X-ray spectrum with $s_X\approx 1.3$. The far-UV spectroscopy of the entire nebula with the Hopkins UV telescope
\citep{1992ApJ...399..611B} confirms the extrapolation of the optical spectrum to the UV. The UV spectrum obtained
in a slit aperture of $17^{\prime\prime}\times116^{\prime\prime}$ displayed in Fig.~\ref{dust_data} was corrected by an ad hoc correction factor of 7.5 
to match the overlapping photometric points of \citet{1992ApJ...395L..13H}. 

\begin{figure}
    \centering
  \includegraphics[width=\linewidth]{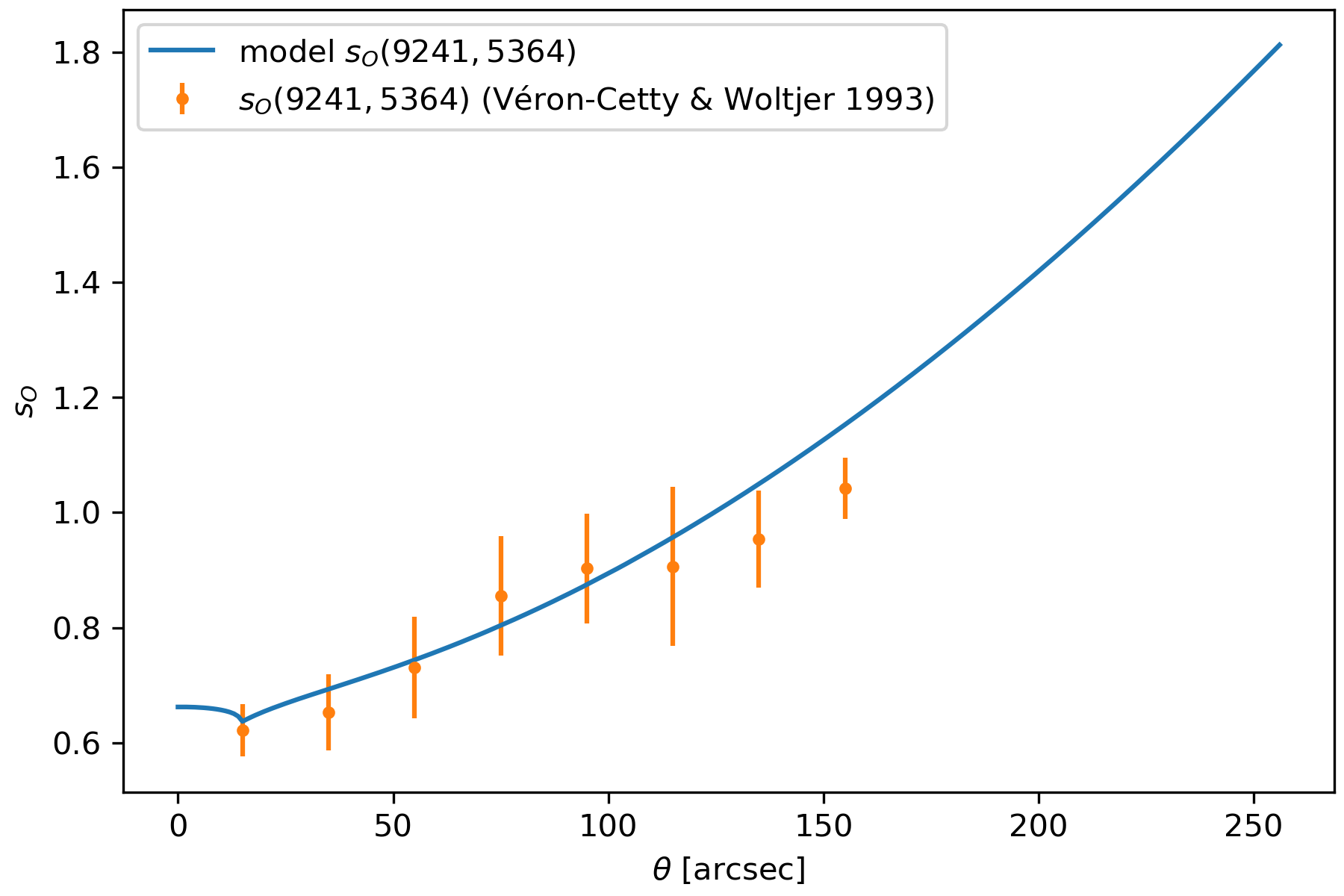}
    \caption{In concentric annuli, the mean and root mean square of the spectral indices $s_O$ ($I_\nu \propto \nu^{-s_O}$) have
    been determined from the data displayed in Fig.~1 of ~\citet{1993A&A...270..370V}. The superimposed curve 
    is calculated from the intensity expected from the synchrotron model as described in Section~\ref{section:fit_sync} with the
    best-fitting parameters listed in Table~\ref{table:parameters}.
    It is important to note that these data have not been used in the fit.}
    \label{fig:optical_index}
\end{figure}

Additionally, we extracted the spectral indices $s(9241-5364)$
obtained between $\lambda=9241\text{\r{A}}$ and
$\lambda=5364\text{\r{A}}$
from Fig.~1 of \citet{1993A&A...270..370V}. The spectral indices were determined within bins of $10^{\prime\prime}\times10^{\prime\prime}$
covering the optical nebula. By averaging over annuli centred on the pulsar
position, we calculated the mean and root mean square of $s_O$. 
The indices change from $\approx 0.6$ in the centre to $\approx 1$ at the periphery of the nebula (see Fig.~\ref{fig:optical_index}). This is well-matched by the 
predicted behaviour of the synchrotron model as indicated by the superimposed blue line in the same Figure.

\subsection{X-ray observations}
\label{sec:xray}
The Crab Nebula has a long history as   a standard candle for X-ray astronomy \citep{1974}. Therefore, for many X-ray telescopes 
the instrumental response functions are tuned to reproduce a particular spectral shape that is deemed to be a canonical X-ray spectrum following a 
power law $dN/dE=N_0 (E/\mathrm{keV})^{-\Gamma}$,
with $\Gamma=2.11$ and $N_0=11$ in units of $\mathrm{cm}^{-2}\mathrm{s}^{-1}\mathrm{keV}^{-1}$, readers
can refer to  \citet{2012ApJ...757..159S}, for example. 

In this spirit, the X-ray spectrum of the Crab Nebula has been
used to cross-calibrate different instruments \citep{2005SPIE.5898...22K,2010ApJ...713..912W}. Given the apparent brightness of the nebula, other weaker non-thermal
X-ray sources are conveniently used for more sensitive instruments which suffer from saturation/pile-up 
for bright sources such as the Crab Nebula. In a recent study of the non-thermal X-ray source G21.5-0.9 \citep{2011A&A...525A..25T}, the findings of earlier comparisons were confirmed that
the energy spectra measured with different instruments in the energy range from 2 keV to 8 keV show relative differences in the normalisation of up to 20~\% and 
differences in the spectral index of $\Delta \Gamma=0.1$. 

Different to most other X-ray instruments in use, the $pn$-camera of the 
XMM-\textit{Newton} X-ray observatory has been  calibrated without relying on the Crab Nebula as a standard candle \citep{2001A&A...365L..18S}. 
The normalisation $N_0= 8.6$ \citep{2005SPIE.5898...22K} found in observations of the nebula with the \textit{pn}-camera in burst mode 
is consistently smaller than the normalisation found with other instruments. 

The challenging task to (cross-)calibrate 
the past, current, and future X-ray missions is carried out by a 
consortium that  formed shortly after the publication of \citet{2005SPIE.5898...22K}. The 
\textit{International Astronomical Consortium for High-Energy Calibration} (IACHEC) recognises these systematic differences \citep{2021arXiv211101613M} but there is still no consensus
which measurement can be considered to be more reliable. 

In the previous works \citep{2006A&A...457..899A,meyer}, we chose to re-scale measurements at higher energies to match the XMM-\textit{Newton} result. 
In view of the recent measurements of
the nebula's X-ray spectrum with the NuSTAR telescope, we updated this approach.
The absolute flux measurement reported by \citet{2017ApJ...841...56M} uses  stray light from the Crab Nebula in the focal plane of the NuSTAR telescope.
Since the X-rays do not pass from the optical mirror assembly, the resulting flux determination does not  suffer 
from the uncertainty of the energy dependent efficiency of the mirror assembly.
Subsequently, the instrumental team of NuSTAR has  changed their canonical Crab spectrum to  $\Gamma = 2.103 \pm 0.001$ and $N_0= 9.69 \pm 0.02$.
This normalisation  is 14~\% larger
than the previously  obtained  value for the $pn$ camera  \citep{2005SPIE.5898...22K}. Moreover, the NuSTAR measurement extends to hard X-rays and is consistent 
with, for example, the results from the coded mask imaging detector and spectrometer SPI on the INTEGRAL observatory \citep{2009ApJ...704...17J}  as well as  
Earth occultation data taken with the BATSE instrument \citep{2003ApJ...598..334L}.  Both measurements indicate a break of the spectrum at an energy of approximately
100~keV. 

Beyond the hard X-ray band, gamma-ray measurements rely on Compton scattering instead of photo-ionisation. The Comptel telescope on-board the Compton-Gamma-Ray-Observatory
(CGRO) observed the Crab Nebula for 9 years \citep{2001A&A...378..918K}. The resulting time-averaged energy spectrum between 0.75 MeV and 30 MeV follows a power law with 
photon index $\Gamma=2.227\pm0.013$ \citep{2001A&A...378..918K}. While the 
power-law index is consistent with the one determined with INTEGRAL-SPI and BATSE data at lower energies,
the flux normalisation of the Comptel measurement is systematically lower than the one obtained with BATSE and SPI data. 
From Fig.~31 of \citet{2001A&A...378..918K}, we estimated possible relative
systematic deviations on the collection area from
two different calibration methods to be $20~\%$. In the data sample used to characterise the synchrotron component of the nebula for the fit, 
we introduced an ad hoc scaling
of the Comptel flux by $1.3$. 

Even though this scaling is larger than the systematic uncertainty of the Comptel flux,
it is well within the range of the combined systematic uncertainty when cross-calibrating
between the very different types of detector.  Clearly, the MeV energy range is most challenging and even though heroic efforts are under way to re-analyse the complete Comptel data
set \citep{2019MmSAI..90..297S}, a new mission is required to improve the situation in this poorly explored energy range (see e.g. the COSI mission \citep{2021arXiv210910403T}).

An additional concern in constructing the X-ray SED 
of the Crab Nebula is the variability of its  X-ray  \citep{1999ApJ...510..305G,2003ApJ...598..334L,2011ApJ...727L..40W} and gamma-ray flux 
\citep{2011Sci...331..736T,2011Sci...331..739A}. The X-ray variability has been observed to occur mainly between 2001 and 2010 with an apparent relative amplitude of 2(4)~\% 
in the energy band $<15$(15-50)~keV on timescales of $\approx 3$ years. In the preceding years of monitoring the Crab Nebula with the RXTE mission, the flux appears to remain
constant \citep{wilson-hodge2012}. The X-ray spectrum softened between
2009 and 2011 with a photon index changing from $2.14$ to $2.17$ as
measured with RXTE-PCA \citep{wilson-hodge2012}. We show
in Fig.~\ref{fig:timeline} the combined hard X-ray light-curve 
between 2000 and beginning of 2022. 

When comparing the recent average flux with the historical measurements of about 50 years ago \citep{1974}, possible long-term flux variations do not exceed the 
variations observed during the 20 years covered by \citet{wilson-hodge2012}. 
The origin of the variability is not clear  but a systematic effect can be firmly rejected as the sole reason given that 
the variability has consistently been observed with independent instruments. 

On a shorter timescale and at larger energies, the Crab Nebula shows rapid variations of the flux between 100 MeV and approximately 1 GeV. The flux can both increase by a factor of 
$\approx 8$ over timescales of days \citep{Buehler2012}
and it can also drop below the average flux within a similar timescale \citep{2020A&A...638A.147Y}. 

For the characterisation of the time-average synchrotron spectral energy distribution we assume that the X-ray variability is averaged out during the long exposure. 
It is to be expected that the
variability measured at soft to hard X-rays should translate in a similar variability in the VHE gamma-ray flux at TeV energies. So far, multi-year observations
of the nebula with ground-based gamma-ray instruments such as HEGRA have only constrained the relative variability to be smaller than 10~\% \citep{2006A&A...457..899A}.  

The 
more volatile situation at the end point of the synchrotron 
spectrum is unlikely to change the inverse Compton emission of the nebula. This has been confirmed by 
simultaneous observations of flares with 
\textit{Fermi}-{LAT} and ground-based gamma-ray telescopes: during these observations no correlated flux changes have been found \citep{2014A&A...562L...4H,2014ApJ...781L..11A}. There may
be however a measurable change of the PeV gamma-ray flux 
during high flux or flaring states (see Sect.~\ref{section:UHE}).

\begin{figure}
   \includegraphics[width=\linewidth]{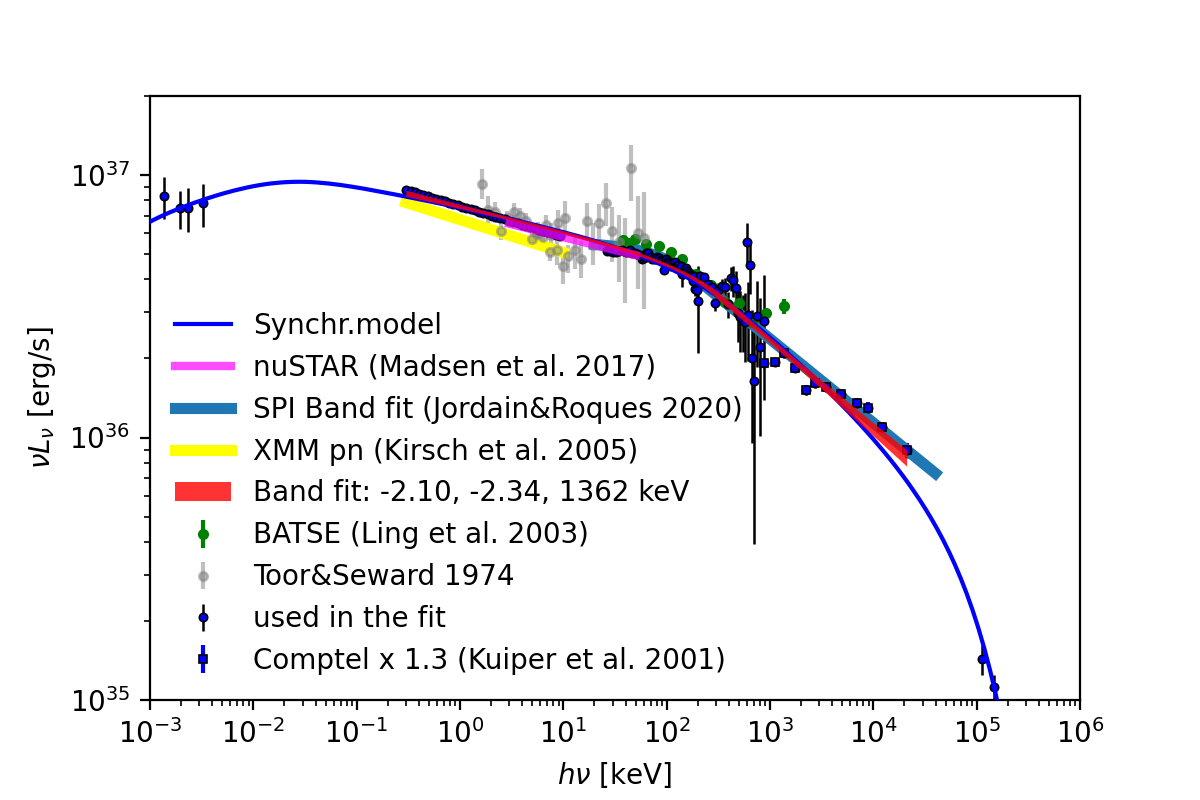}
      \caption{SED of the Crab Nebula centred on the X-ray energy range. 
      The best-fitting  model (blue line, see details on the fitting procedure in Section 4 and 
      Table~\ref{table:parameters})
      is shown along with measurements from different X-ray observatories. 
      The normalisation of the spectrum measured with XMM-\textit{Newton} (yellow band) has been scaled up by 16~\% to
      be consistent with the absolute flux measurement with NuSTAR (magenta band). The COMPTEL gamma-ray flux has been
      scaled up by $30~\%$ to match the extrapolation of the SPI spectrum.}
               \label{xray_data}
   \end{figure}

In Fig.~\ref{xray_data}, the data points used for the fit are shown as blue markers together with  measurements listed above. The blue line indicates the best-fitting synchrotron model 
which includes the \textit{Fermi}-{LAT} data points between 70 MeV and 500 MeV (see below).

With the scaling of the flux measurements to match the absolute flux measurement of NuSTAR applied to the XMM-\textit{Newton} and COMPTEL measurement, we obtain a consistent 
spectral measurement from 0.2~keV to 30 MeV. The spectrum in this energy range can be well characterised with a Band-like model \citep{1993ApJ...413..281B}
\begin{equation}
\frac{dN}{dE} = N_0 
\begin{cases}
\left(
\frac{E}{100~\mathrm{keV}}
\right)^{\Gamma_1}
e^{-E/E_0}
 & \text{if } E\leqslant \: E_0(\Gamma_1-\Gamma_2),\\
 \left[\frac{(\Gamma_1-\Gamma_2) E_0}{100~\mathrm{keV}}
 \right]^{\Gamma_1-\Gamma_2}
e^{\Gamma_2-\Gamma_1}
\left(
\frac{E}{100~\mathrm{keV}}
\right)^{\Gamma_2}
& E > \: E_0(\Gamma_1-\Gamma_2).
\end{cases}
\end{equation}
with parameters
for the photon indices 
$\Gamma_1=-2.095(4)$,  
$\Gamma_2=-2.34(1)$, 
$E_0 = 1.36(17)$~MeV and normalisation at $100$~keV: $N_0=6.40(8)\times 10^{-4}~\mathrm{cm^{-2}\,s^{-1}\,keV^{-1}}$.
The resulting $\chi^2(dof)=97(158)$ indicates a good fit of the Band function. Similar parameters ($\Gamma_1=-1.99(1)$, $\Gamma_2=-2.31(2)$, $E_0=531(30)$, $N_0=7.5(2)\times 10^{-4}~\mathrm{cm^{-2}\,s^{-1}\,keV^{-1}}$) have been obtained from analysing 17 years of INTEGRAL SPI data   
\citep{2020ApJ...899..131J}, see also Fig.~\ref{xray_data}.

\subsection{HE and VHE gamma-ray observations}

Since the work of ~\cite{meyer}, roughly nine more years of \textit{Fermi}-{LAT} observations have been 
presented in recent publications. 
Therefore, the statistical uncertainty on the flux above 100 GeV has been reduced by a factor of three. 
The updated instrumental response function of \textit{Fermi}-{LAT} extends the HE reach from formerly 300 GeV to 500 GeV. 
The systematic uncertainty on the energy scale has 
been  reduced from previously $\Delta E/E =\substack{+5\% \\ -10\%}$ [see e.g. \cite{abdo2009}] 
to $\Delta E/E =\substack{+2\% \\ -5\%}$ ~\citep{ackermann2012}.

 In a recent investigation of the gamma-ray extension of the Crab Nebula,  \citet{paul_paper} analysed the LAT data accumulated over 
 $\approx 9.1$ years with a  refined model for the Crab pulsar's spectrum. 
 They obtain a $>5~$GeV spectrum of FL8Y J0534.5+2201i (the PWN component) 
 which essentially matches the off-pulse spectrum reported by \cite{Buehler2012} but with improved statistical uncertainties and wider energy reach.

 The MAGIC collaboration has published an updated energy spectrum of the Crab nebula overlapping the energy range covered with 
 \textit{Fermi}-{LAT} and 
 reduced uncertainties at energies below 100 GeV \citep{MagicCrab}. 
 The measurement of the flux from the Crab Nebula with the MAGIC instruments spans for the first time from 50 GeV to 30 TeV. In a subsequent analysis
 of data taken with the MAGIC telescope at ultra-large zenith angles, the energy reach has been extended up to 100 TeV \citep{magic_2020}.
 
 The VHE end of the Nebula's spectrum has recently become a focus of
 interest with observational data taken with 
 non-imaging air shower detectors with improved analysis of data taken with the
 HAWC water Cherenkov detector \citep{hawc} or
 with long-term observations with the Tibet AS$\gamma$ air shower array which extends the
 energy spectrum up to 450 TeV \citep{PhysRevLett.123.051101}. The latest
 addition to results on the Crab nebula obtained with the LHAASO KM2A array extends the spectrum beyond 1 PeV \citep{2021Sci...373..425L}. 
 
The  High Energy Stereoscopic System (H.E.S.S.) collaboration has presented
their preliminary results of H.E.S.S. phase II observations of the Crab Nebula
\cite{hess_data} superseding the results obtained with the previous 
setup (phase I) 
which consisted of four imaging Cherenkov telescopes (IACTs) with a
mirror surface area of about 100~m$^2$ each and an energy threshold of approximately 100 GeV. Due to the visibility of the Crab Nebula from the 
southern hemisphere, the energy spectrum obtained with the phase I instruments  
from the Crab Nebula covers the energy range from 440 GeV up to 40 TeV
\citep{2006A&A...457..899A}. The large
telescope of the H.E.S.S. phase II with a mirror surface of 600 m$^2$ 
has lead to a lowered energy threshold of 230 GeV for observations of the Crab Nebula \citep{hess_data}.

Finally, the VERITAS collaboration has reported on observations of the Crab Nebula (55 hours before and 46 hours after the camera update in 2012)  \citep{wells_veritas}.  In a previous analysis, slightly
more data were used \citep{veritas}, but the investigation of \citet{wells_veritas} indicate that the calibration of the 
two different levels of amplification for each channel need to be adapted.
More detailed information on the VHE data sets is provided in Appendix~\ref{appendix:VHE}.
\begin{table}
\caption{Summary of updated observations.}
\begin{center}
 \begin{tabular}{llc}
 \textbf{Energy} & \textbf{Instrument} & \textbf{References}\\
  \hline
 \hline
  radio &  WMAP  &   (1) \\
  &  Planck for the  & (2) \\
    &  HFI instrument & \\
\hline
 Sub-mm/IR  &  Herschel &  (3) \\
&  WISE &  (3) \\
 \hline
 X-ray & Crab   &   (4) \\
   to         &  NuSTAR & (5) \\
  $\gamma-$ ray  & \textit{Fermi}-LAT & (6) \\
  \hline
VHE  &  H.E.S.S. & (7) \\
     &  VERITAS &   (8) \\
     &  MAGIC & (9) \\
     &  Tibet AS$\gamma$&   (10) \\
     &  HAWC &  (11) \\
     & LHAASO WCDA, KM2A & (12) \\

  \hline
\end{tabular}
\end{center}
\tablefoot{All other data shown e.g. in Fig. 7 are taken from
\citet{meyer} and references therein.}
\tablebib{
 (1) \citet{weinland2011}, 
 (2) \citet{PlanckCollab2016,2018A&A...616A..35R}, 
 (3) \citet{GomezDust,2019MNRAS.488..164D}, 
 (4) \cite{1974}, 
 (5) \citet{nustar_2017},
 (6) \citet{Buehler2012},\citet{paul_paper}, 
 (7) \citet{2006A&A...457..899A},
\citet{holler_hess}, \citet{hess_data}, \citet{2014A&A...562L...4H}, (8) \citet{wells_veritas}, Fig. 5.15, \citet{2014ApJ...781L..11A},
(9) \citet{MagicCrab},  \citet{magic_2020}, 
(10) \citet{PhysRevLett.123.051101},
(11) \citet{hawc}, 
(12) \citet{2021Sci...373..425L} }
\label{tab:observationsSED}
\end {table}


\section{Non-thermal emission model}

The specific luminosity $L_\nu$ from the Crab Nebula was 
calculated by integrating the
synchrotron and IC emissivities ($j^{Sy}_\nu$, $j^{IC}_\nu$) over the volume of the nebula (assuming optically thin
plasma and spherical symmetry):
\begin{equation}
L_{\nu}= 4~\pi\int_{V}{\rm d}^3 r \left( j^{Sy}_\nu(r) + j^{IC}_\nu(r) \right),
\end{equation}
where the specific emissivity for inverse Compton emission ($j^{IC}_\nu$) and synchrotron
emission ($j^{Sy}_\nu$) is given by
\begin{equation}
\label{emissivityIC}      
            j_{\nu}^{IC}(r) =\dfrac{1}{4\pi} \int\limits_{\gamma}^{}{\rm d }\gamma n_{el}(r,\gamma) \dfrac{3}{4}\dfrac{\sigma_{T}c}{\gamma^{2}}h^{2}\nu \int\limits_{h\nu/4\gamma^2}^{h\nu}
\dfrac{{\rm d }\varepsilon}{{\varepsilon}}f_{IC}(\varepsilon,\nu,\gamma) n_{seed}(r,\varepsilon),
\end{equation}
\begin{equation}
\label{emissivitySync}
 j_{\nu}^{Sy}(r)  = \dfrac{1}{4\pi} \int\limits_{\gamma} {\rm d}\gamma~ n_{el} (r,\gamma)\dfrac{\sqrt{3}e^{3}B(r)}{mc^2}G\left(\dfrac{\nu}{\nu_{c}}\right),
\end{equation}
with $n_{seed}(r,\varepsilon)$ and $n_{el}(r,\gamma)$
being the differential number densities of seed photons and electrons respectively. 
The resulting specific intensity $I_\nu(\theta)$ was calculated by integrating along a line
of sight subtending an angle $\theta$ with the centre of the nebula. The source region
is optically thin. The external absorption in the interstellar medium
was used to correct the apparent
flux measurements in the optical/UV (de-reddening, see Sect.~\ref{subsec:optical}),
X-ray (photo-ionisation, see Sec.~\ref{sec:xray}), as well as at ultra-high energy ($>100$~TeV)
gamma-rays. The latter effect was re-evaluated here taking into account the anisotropy
of the local radiation field which leads to an increase of the optical depth 
in comparison to previous works (see Appendix~\ref{appendix:tau} for more details).

The function $G(x)$ used for the synchrotron emissivity 
for an isotropic electron distribution in random magnetic fields \citep{1986A&A...164L..16C} has been recently expressed 
in terms of modified Bessel functions as
given in eq. (D5) in \citet{Aharonian_2010}. For the numerical calculations, we  used the convenient approximation  of eq. (D7) in \citet{Aharonian_2010}.

The function $f_{IC}$ is the general function for scattering of electrons in an isotropic photon gas  derived by \citet{PhysRev.167.1159} (Eq. 44).

In order to calculate the emissivity of the electrons  we need to 
determine the electron distribution, seed fields, and magnetic field structure 
to reproduce all available observations. While the observed synchrotron emission 
is mostly degenerate for the choice of magnetic field and particle distribution, the
observed inverse Compton emission breaks the degeneracy to allow us to reconstruct
both the strength and the spatial variation of the magnetic field in the nebula. 

\subsection{Distribution of electrons}
The spectral and spatial distribution of electrons in the nebula is inferred 
by matching a phenomenological model of the electron distribution and its 
synchrotron emission to  
observations. The electron distribution was kept as simple
as possible and as complex as necessary to match the observations.

We assumed two different populations of electrons, 
so-called radio  and 
wind electrons which do not share a common 
origin. While the wind electrons are most likely accelerated
at the wind termination shock, the origin of the radio emitting
electrons is not well understood. This includes both the
generation of sufficient number of electrons/positrons as well 
as their acceleration. Recent suggestions  
include stochastic acceleration \citep{2017ApJ...841...78T},
for example,  in reconnection layers in the turbulent nebula
\citep{2019MNRAS.489.2403L}. 

\subsubsection{Spatial distribution of electrons}
\label{subsec:spatial}
For photon energies below about 70 keV, the  
synchrotron emission is well-resolved by
the available instruments. The observed (projected) structures are characterised by 
small-scale features (e.g. wisps) which evolve dynamically on timescales of weeks
as well as large scale
features including the X-ray bright ring, torus, and jets which remain stable for 
longer time-periods of years.

For the purpose of modelling the energy spectrum and extension of the 
inverse Compton emission of the nebula, we followed an approach similar to 
\citet{JH92, Hillas, meyer} where the electron distribution is found by matching the 
expected synchrotron emission to the observed broadband spectral energy distribution (SED). In a 
second step, the magnetic field was determined by adjusting the  electron distribution 
(mainly the normalisation) to match the observed inverse Compton emission to the data. In order to keep the
model simple, the source was assumed to be spherical. Given the coarse angular resolution of the
instruments used to observe the nebula at the HE end, 
this simplification is acceptable.

An important benefit of the assumption on spherical symmetry is the simple characterisation
of the extension of the
nebula. Here, we used the apparent angular extension $\theta_{68}$ which is defined by the angular
radius which encompasses 68~\% of the total flux. For a helio-centric distance $d_\mathrm{Crab}$ of the Crab Nebula, 
the spatial radius $R_{68}$ relates to the apparent size $ R_{68}\approx  \theta_{68}~d_\mathrm{Crab}$. We assumed a distance $d_\mathrm{Crab}=2~\mathrm{kpc}$.

We collected available observations in order to determine $\theta_{68}$ for different photon energies. The resulting values of $\theta_{68}$ are summarised in Table~\ref{table:theta}.
The Chandra data points \footnote{\url{ivo://ADS/Sa.CXO\#obs/13151} (Chandra ObsId
13151).} indicate the $68\%$ containment angular radius $\theta_{68}$ of the Crab Nebula for
different energy bands.
For each energy band (0.5--1.2; 1.2--2; 2--7; 7--10 keV), the centroid of the background
subtracted image was calculated. 
The value of $\theta_{68}$ for each energy band was determined using 
the method of encircled count fraction (\texttt{ecf\_calc} task of the CIAO 
package v4.13). The uncertainty on $\theta_{68}$ is dominated by systematic uncertainties which were
 estimated by checking against effects of vignetting and background contamination. 
The \textit{NuSTAR} X-ray telescope has been used to image  the 
X-ray nebula for the first time in a broad energy range from 3~keV--78~keV
\citep{Madsen2015}. We estimated $\theta_{68}$ from their Fig.~13 in three energy ranges. In the lowest energy band (3~keV--5~keV), the extension derived
from the \textit{NuSTAR} observations is about 5~\% smaller than the values
obtained with the Chandra image, well within the estimated uncertainties of the two measurements.

\begin{table}[h!]
\caption{Apparent extension of the Crab Nebula.}
\begin{tabular}{crl}
\hline
  $\nu~[$Hz$]$   & $\theta_{68}[^{\prime\prime}]$          &    \\
  \hline
  $5\times 10^{9}$    & $117\pm 1$ & VLA  (1)\\
  $150\times 10^{9}$  & $115\pm 2$ & NIKA (2)\\
  $1.8\times 10^{12}$ & $112\pm 5$ & PACS 160$~\mu$m  (3)\\
  $4.3\times 10^{12}$ & $112\pm 5$ & PACS 60$~\mu$m  (3)\\
  $1.4\times 10^{13}$ & $137\pm5$ & WISE 22$~\mu$m  (4)\\
  $2.5\times 10^{13}$ & $119\pm5$ & WISE 12$~\mu$m  (4)\\
  $6.5\times 10^{13}$ & $140\pm5$ & WISE 4.6$~\mu$m  (4)\\
  $8.9\times 10^{13}$ & $141\pm5$ & WISE 3.3$~\mu$m  (4)\\
  $1.03\times 10^{15}$ & $99\pm5$ & UVOT UVW1 (5) \\
  $2.66\times 10^{17}$ & $46\pm 4$ & Chandra $(0.5-1.5)$ keV (6)\\
  $3.7\times 10^{17}$  & $43 \pm 3$ & Chandra $(1.5-3.0)$ keV (6) \\
  $8\times 10^{17}$    & $39 \pm 3$ &Chandra $(3-6)$ keV   (6)\\
  $1.2\times 10^{18}$  & $37\pm 3 $ & NuSTAR $(5-6)$ keV (7) \\
  $2.4\times 10^{18}$  & $33 \pm 3$ & NuSTAR $(8-12)$ keV (7)\\
  $1.2\times 10^{19}$  & $27\pm 3$  & NuSTAR $(35-80)$ keV (7)\\
  \hline
  $E_\gamma~[$GeV$]$   & $\theta_{68}[^{\prime\prime}]$          &    \\
  \hline
  7                    & $121\pm 47$ & \textit{Fermi}-{LAT} (8)\\
  14                   & $209\pm 22$ & (8)\\
  28                   & $125\pm 26$ & (8)\\
  56                   & $114\pm14$  & (8) \\
  112                  & $95\pm 29$  & (8) \\
  214                  & $83\pm 26$   & (8) \\
  990                  & $121\pm38$   & (8) \\
  1050                 & $79\pm10$    & H.E.S.S. (9) \\
\hline
\end{tabular}
\label{table:theta}
\tablefoot{ The apparent angular size of the nebula was calculated by
integrating the observed intensity over the solid angle up to $\theta_{68}$ which is the percentile 
that encloses $68~\%$ of the total flux. Whenever needed, point sources were excluded and background
subtracted.}
\tablebib{
(1) \citet{Bietenholz2004}, 
(2) Fig.~5 of \citet{2018A&A...616A..35R},
(3) \textit{Herschel} Science Archive (HSA),
(4) Infrared Science Archive (IRSA),
(5) XMM-\textit{Newton} Science Archive (XSA),
(6) \textit{Chandra} Data archive,
(7) \citet{Madsen2015}, 
(8) \citet{paul_paper}, 
(9) \citet{2020NatAs...4..167H}}
\end{table}

The differential electron number density  
$\mathrm{d}n = n_{el}(\gamma,r) \mathrm{d}\gamma \mathrm{d}V$  
was assumed to follow a multiple broken power law 
in Lorentz factor $\gamma$ (see below) with 
a spatial distribution approximated by a radial Gaussian with a 
 the scale length $\rho(\gamma)$ that depends on the electrons'
 energy:
\begin{equation}\label{eq:elec_number}
 n_{el}(\gamma,r)=n(\gamma)e^{-\dfrac{r^{2}}{2\rho(\gamma)^{2}}},
 \end{equation}
  such
that the predicted synchrotron emission matches the observed synchrotron extension.

The observed synchrotron size is mostly 
decreasing with increasing frequency, 
reflecting the cooling of the electrons as they propagate outwards.  
Therefore, we assumed for the wind-electrons a power law for $\rho(\gamma)$:
\begin{equation}
    \rho(\gamma) = \frac{\rho_0}{3600^{\prime\prime}} \cdot \frac{\pi}{180^\circ} \cdot\frac{d_\mathrm{Crab}}{2~\mathrm{kpc}} \left(\frac{B_0}{264~\mathrm{\mu G}}\left(\frac{\gamma}{9\times 10^5}\right)^2\right)^{-\beta},
\end{equation}
with $\rho_0$ (in arc~sec), $B_0$, and $\beta>0$ free parameters in the fit to the data (see below). 

Different to the wind electrons, for the low energy, radio-emitting electrons, 
the radial scale length $\rho_r$  is independent of the electrons'
Lorentz factor. 
The value of $\rho_r$  was left to vary freely in the fit to match the measured synchrotron extension.

\subsubsection{Spectral distribution of electrons}
We considered two distinctly different populations dubbed radio
electrons as they are
mostly emitting in the radio band and so-called wind electrons which are producing
the bulk of the emission from optical to gamma-rays via synchrotron processes:
\begin{equation}
    n_{el}(r,\gamma) = n_{radio}(r,\gamma) + n_{wind}(r,\gamma).
\end{equation}

The differential number density of each population was 
assumed to follow as a power law for the radio electrons 
\begin{equation}
    \label{eqn:radio}
    n_{radio}(r,\gamma) = N_{r,0}\rho_r^{-3} e^{- \frac{r^2}{2\rho_r^2} } \gamma^{-s_r} H(\gamma;\gamma_0, \gamma_1),
\end{equation}
and a broken power law for the wind electrons
\begin{equation}
 \label{eqn:wind}
 \begin{split}
    n_{wind}(r,\gamma) &= N_{w,0}\rho(\gamma)^{-3} e^{- \frac{r^2}{2\rho(\gamma)^2} } \left[
    \left( \frac{\gamma}{\gamma_{w1}} \right)^{-s_1}\right.  
    \left(\frac{\gamma_{w1}}{\gamma_{w2}} \right)^{-s_{2}} H(\gamma;\gamma_{w0},\gamma_{w1}) \\
& + \left.\left(\frac{\gamma}{\gamma_{w2}}\right)^{-s_2} H(\gamma; \gamma_{w1}, \gamma_{w2}) 
       + \left(\frac{\gamma}{\gamma_{w_2}}\right)^{-s_3} H(\gamma; \gamma_{w2}, \gamma_{w3})
       \right]
\end{split}
\end{equation}
respectively.
The auxiliary function $H(\gamma;\gamma_0,\gamma_1) := \Theta(\gamma_1-\gamma)\Theta(\gamma-\gamma_0)$
with the Heaviside function $\Theta(x)$
defines the intervals of the broken power law. 

\begin{figure}
 \includegraphics[width=\linewidth]{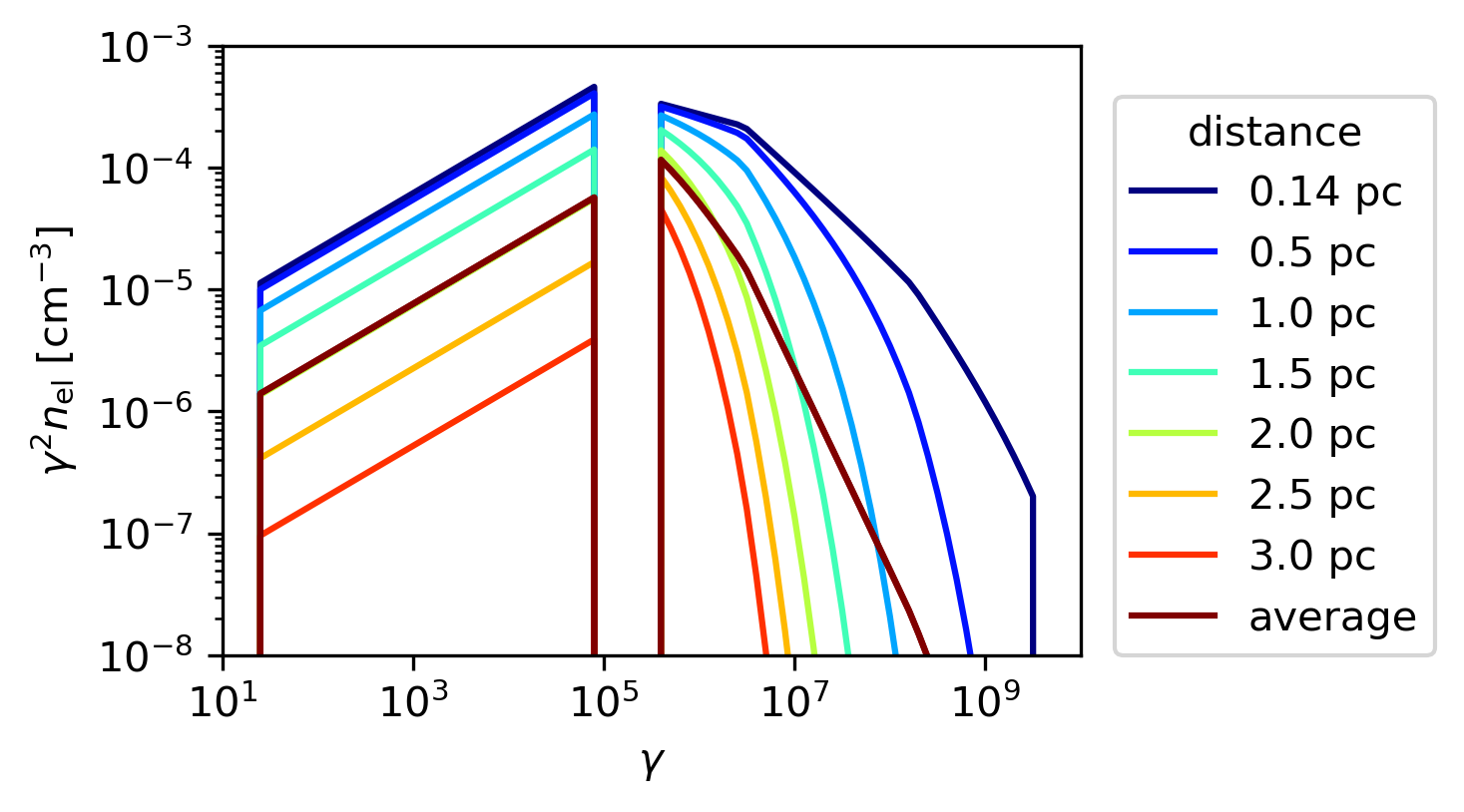}
      \caption{Electron number density $n_{el}$, as a function 
      of $\gamma$ for different values of the distance to the centre 
      of the Crab Nebula $r$ and the volume average. The parameters used here are listed in Table~\ref{table:parameters}.
              }
         \label{Fig:nel_gamma}
   \end{figure}

 A representative electron number density $n_{el}$ 
 at various distances and volume averaged is shown in Fig.~\ref{Fig:nel_gamma}. 
 Note, the distribution of electrons  follows a phenomenological 
 Ansatz
 guided by the notion that electrons lose energy (radiative and adiabatic losses) while expanding in the nebula. 
 However, without further knowledge about the properties 
 and dynamics of the flow
 and the various and competing processes at work, the structure of the nebula can only be 
 generically explained through the idea of a shock-accelerated electron distribution 
 in a  (ideal) MHD  flow of a relativistic plasma as suggested for example by \citet{1974MNRAS.167....1R} and following works \citep[e.g.][]{1984ApJ...283..694K}. 
 
 In the absence of a specific model that matches available observations, we 
 followed the approach of reconstructing the energy distribution of electrons
 and their spatial
 distribution from the observations. In this approach, we do
 not attempt to follow the temporal evolution of the nebula
 through its complex and largely unknown history. 
 
\subsection{Seed photon fields}
The seed photon fields in the nebula were self-consistently calculated and matched
to the observed intensity and brightness of the nebula. Note, in many previous attempts
to model the nebula, the seed photon fields are assumed to be mostly
homogeneous in the nebula and not specifically modelled to match the observations. 

Assuming an isotropic specific volume emissivity $j_\nu$ 
(in $4\pi$ solid angle) in a spherical source, the {spectral number density} 
$\mathrm{d}n = n_\mathrm{seed} (r,\varepsilon) \mathrm{d}V \mathrm{d}\varepsilon$ 
of seed photons at a distance $r$ to the centre of the nebula is:
\begin{equation}
\label{eqn:seed}
n_{seed}(r,\varepsilon)=\dfrac{4\pi}{h\varepsilon}\dfrac{1}{2c} \int_{r_{min}}^{r_{max}}\dfrac{r_{1}}{r} j_\nu(r_{1},\varepsilon) \ln\left( \dfrac{r+r_{1}}{|r-r_{1}|}\right) {\rm d}r_{1}.
\end{equation}

The three main seed photon fields contributing to the IC scattering of the relativistic electrons in the Crab Nebula are: 
(i) the synchrotron radiation; 
(ii) the FIR `excess' radiation which is attributed to the dust emission; 
(iii) the 2.7K cosmic microwave background radiation (CMB).

\subsection{Thermal emission from dusty plasma}
\label{sect:dust}
Observations at sub-millimetre and far-infrared wavelengths 
show  an excess flux in comparison to the extrapolated synchrotron continuum 
alone.   The excess is commonly modelled with 
the thermal emission of spherical, radiatively heated dust 
grains in thermal equilibrium
\citep{Marsden1984,Temim2006,GomezDust,2013ApJ...774....8T,2019MNRAS.488..164D}. A carbon dominated 
composition of the dust has been favoured
in past studies where the temperature
of the dust is either self-consistently modelled 
in combination with a dust grain composition \citep[e.g.][]{2013ApJ...774....8T} or estimated by
fitting a two-temperature population to the data \citep[e.g.][]{2019MNRAS.488..164D}. 
The dust emitting region is concentrated in the 
vicinity of the optical filaments which are forming
a shell with inner radius of $r=0.55$~pc \citep{2004ApJ...609..797C}.

The observed emission in the wavelength band from $\lambda\approx 6~\mu\mathrm{m}$ to $\lambda\approx 1000~\mu\mathrm{m}$ is the superposition of 
continuum synchrotron and thermal dust emission as well
as line-emission from excited atoms in the plasma.
In Fig~\ref{dust_data}, the observed continuum emission 
is shown.
In all previous works,
a quantitative analysis of the dust emission has relied upon
an ad hoc assumption on the underlying continuum emission.

Following the argument of \citet{shell}, we assumed the dust to be predominantly
composed of amorphous carbon grains.
The emissivity of dust grains 
in thermal equilibrium with a mass density
$\rho_d$ in the emitting volume is given by
\begin{equation}
    j_\nu = \rho_d \kappa_\mathrm{abs} B_\nu(T),
\end{equation}
with  $B_\nu(T)$ denoting the black body spectral radiant 
emission 
at temperature $T$ and $\kappa_\mathrm{abs}$ the mass absorption coefficient
specific to the dust grain \citep{2013ApJ...774....8T}. In order to calculate the amount of seed photons 
present in the nebula which is relevant for the inverse Compton emissivity, we
chose to fit a minimum set of free parameters to achieve an acceptable fit to the data. 
The minimal model found here 
is a mixture of dust at two different temperatures and with 
different values of mass ($M_1, T_1, M_2, T_2$). An additional
parameter is the outer radius $r_\mathrm{out}$ of the dust
population. The value of $r_\mathrm{out}$ needs to be adjusted to match the observed
extension which is the result of the superposition of the underlying synchrotron continuum
emission and the thermal emission of the dust. With the emissivity
\begin{equation}
    \label{eqn:dust}
   j_\nu = \frac{3\kappa_\mathrm{abs}}{4\pi(r_\mathrm{out}^3 - r_\mathrm{in}^3)} 
   \left[M_1 B_\nu(T_1) + M_2 B_\nu(T_2)\right] H(r;r_\mathrm{in},r_\mathrm{out})
\end{equation}
of the two temperature dust model for a a fixed set of grain parameters (see next paragraph and  Table~\ref{tab:varIR}) we achieved an acceptable description of the data (see Fig.~\ref{dust_data}).

The mass absorption coefficient $\kappa_\mathrm{abs}$ depends upon the geometry of
the dust grain and the optical constants for the bulk material. Here, we used the measured optical constants 
of the ACAR sample \citep{1996MNRAS.282.1321Z} and calculated
the resulting values of $\kappa_\mathrm{abs}$ assuming a spherical symmetry and following
the Mie-scattering theory in a series expansion. The underlying calculation was carried out 
with the numerical algorithm described in \citet{Bohren1998}, Appendix A. A more thorough
self-consistent modelling of the dust emission by radiatively heated dust grains is left for further work. 
The grains' diameter distribution was  assumed 
to follow a power law  between the minimum 
grain radius $a_\mathrm{min}=10^{-3}~\mu$m and
$a_\mathrm{max}=0.1~\mu$m such that
the probability density of finding a
grain between $a$ and $a+da$ is given by
$p(a)da = p_0 a^{-s}$ with $s=4$ for $a$ in the interval $[a_\mathrm{min},a_\mathrm{max}]$
\citep{2013ApJ...774....8T}. 
In the relevant range of wavelengths (from $\lambda=1~\mu$m to $\lambda=10^3~\mu$m), the resulting opacity is conveniently approximated
with
\begin{equation}
\kappa_\mathrm{abs}= 2.15\times 10^{4} \tfrac{\mathrm{cm}^2}{\mathrm{g}} \left(
\tfrac{\lambda}{\mu\mathrm{m}}\right)^{-1.3},
\end{equation}
with relative deviations between the model
\citep{2013ApJ...774....8T} and the parametrisation of less than 5~\% between
$\lambda=10~\mu$m and $\lambda=200~\mu$m. 

The thermal line emission from the excited plasma
was not considered here. In the visual band, the
 relative contribution of the line emission to the overall flux is approximately 30\% when comparing
 the sum of the equivalent width of the lines listed in Table 2 of  
 \citet{2003MNRAS.346..885S} with the wavelength-band of the measured spectrum. 
\begin{table}[h]
\caption{Parameters of the dust model.}
    \centering
    \begin{tabular}{l|l}
\hline          
\textbf{Dust parameter}                 & \textbf{Value}\\
\hline
    Grain material             & Amorph. Carbon \\
    $\log_{10}(M_1/M_\odot)$   & $-4.4\pm0.1$  \\
    $T_1$ [K]                  & $149\pm 8$  \\
    $\log_{10}(M_2/M_\odot)$   & $-1.2\pm0.1$ \\
    $T_2$ [K]                  & $39\pm 2$    \\
    $r_\mathrm{in}$ [pc]       & $0.55$ \\
    $r_\mathrm{out}$ [pc]      & $1.53\pm0.09$ \\
    $a_\mathrm{min},a_\mathrm{max}$ [$\mu$m]  &$10^{-3},10^{-1}$ \\
    $s$                        & $4$ \\
    \hline  
    \end{tabular}
    \label{tab:varIR}
\end{table}

\subsection{Magnetic field structure}
The widely used  model for the confinement of the Crab Nebula by \citet{1984ApJ...283..694K} is based upon a toroidal 
magnetic field without any turbulence forming. 
Three-dimensional relativistic ideal MHD simulations supports the notion
of disordering of the magnetic field \citep{2013Porth} which has been argued to account for the slowing of the expansion
velocity to match the observations \citep{2018MNRAS.478.4622T} and the comparably small
polarisation observed at X-rays \citep{2017MNRAS.470.4066B,1978ApJ...220L.117W}. 
In non-ideal MHD plasma flows, dissipation of magnetic fields
would need to be considered as well \citep{2018MNRAS.478.4622T, 2020ApJ...896..147L}. In this type of extension of the
pulsar wind nebula scenario, several of the fundamental short comings of the toroidal field model of the Crab Nebula \citep{1984ApJ...283..694K} can be overcome as argued by \citet{2020ApJ...896..147L}.

 In the following, we assumed  a model where the magnetic field is dominated by small-scale turbulence. The 
 root mean square value $B(r)$ follows a power law for its radial dependence:
 \begin{equation}
\label{eqn:B}
  B(r) = B_0 \left(\frac{r}{r_s}\right)^{-\alpha},
 \end{equation}
 with $B_0$ the field strength at the distance of the termination shock $r_s = 0.13$~pc 
 and $\alpha\ge 0$  free parameters of the model. The value $\alpha=0$ corresponds
 to the case of a constant magnetic field as for example
 assumed by \citet{meyer}. The case $\alpha=-1$ corresponds to 
 an MHD flow dominated by
 kinetic energy ($\sigma \ll 1$).

\section{Parameter estimation}
The complete model description of the non-thermal electron distribution (Eqns.~\ref{eqn:radio}, \ref{eqn:wind}), the spatial distribution, temperature, and mass of the dust 
grains (Eqn.~\ref{eqn:dust}) as well as the strength and radial dependence of the magnetic field (Eqn.~\ref{eqn:B}) requires one to choose values for the underlying parameters.
The complete set of parameters  were combined in a tuple  $\mathbf{\Psi}=(\Psi_1,\ldots,\Psi_{21})$ (see Table~\ref{table:parameters} for 
the complete list of parameters and best fitting values found in Sect.~\ref{section:VHE}).

The subset of parameters $\Psi_1,\ldots,\Psi_{14}$ relates
to the electron distribution and is closely related to the choice of the magnetic field with normalisation $\Psi_{20}=B_0$
and power-law index $\Psi_{21}=\alpha$ for the radial distribution (see Eqn.~\ref{eqn:B}). 
The parameters $\Psi_{15},\ldots, \Psi_{19}$ characterise the dust distribution and the
amount and temperature of the two dust populations (see Eqn.~\ref{eqn:dust}).

\subsection{Data sets used}
The method to estimate the  best-fitting values $\mathbf{\Psi}$ 
is based upon the minimisation of  
the sum of $\chi^2$ values
calculated for the two compilations of observational data. 
$\mathcal{D}_\mathrm{sync}$ and $\mathcal{D}_\mathrm{IC}$ (see the following Sect.~\ref{secion:test_statistics} on the definition of the test statistics).

The observational data set 
$\mathcal{D}_\mathrm{sync}$ includes 203 flux measurements $f_\nu$ and
uncertainties $\sigma_f$
\footnote{see the supplementary table available under
\url{https://github.com/dieterhorns/crab_pheno}}
as well as 15 measurements of the angular extent (see Table~\ref{table:theta})
$\theta_{68}^\mathrm{obs}(\nu)$ and corresponding uncertainties 
$\sigma_\theta(\nu)$. The resulting synchrotron 
data set
\begin{equation}
\label{eq:D_sync}
    \mathcal{D}_\mathrm{sync} = \mathcal{D}_{sync,SED} \cup \mathcal{D}_{sync,ext},
\end{equation}
is the union of the spectral measurements 
\begin{equation}
\label{eq:D_syncsed}
  \mathcal{D}_{sync,SED}=\{(f_{\nu,i},\sigma_{f,i})|i=1,\ldots,203\},
\end{equation}
and the measurement of the angular (apparent) extension
\begin{equation}
\label{eq:D_syncext}
  \mathcal{D}_{sync,ext}=\{(\theta_{68}(\nu_j), \sigma_{\theta}(\nu_j))| j=1,\ldots,15\}.
\end{equation}

The spectral data set $\mathcal{D}_{sync,SED}$ comprises a variety of  observations
collected with different instruments. In many instances, the resulting measurements are mutually inconsistent
indicating systematic uncertainties (see also Sect.~\ref{section:observations} for more details). The 
quoted uncertainties on the flux measurements $\sigma_{f,i}$ are limited to the statistical uncertainties. 

While below energies of $1~$GeV, the dominant contribution to the Crab Nebula's luminosity is synchrotron emission of electrons and thermal emission from radiatively
heated dust grains, at energies larger than $1~$GeV, 
the dominant emission process is inverse Compton scattering of the same electron population on various seed photon fields present in the nebula.

For energies above 1~GeV, we compiled a data set 
$\mathcal{D}_\mathrm{IC}$ 
\begin{equation}
\label{eq:D_IC}
\mathcal{D}_\mathrm{IC}= \mathcal{D}_{IC,SED} \cup \mathcal{D}_{IC,ext},
\end{equation}
which includes the  measurements of the energy spectrum using 
\textit{Fermi}-{LAT}
data (25 data points: $\mathcal{D}_{IC,SED}$) and the angular extension ($\mathcal{D}_{IC,ext}$: 8 measurements as listed in Table~\ref{table:theta}) combining \textit{Fermi}-{LAT} \citep{paul_paper} and
H.E.S.S. measurements \citep{2020NatAs...4..167H}.
The spectral measurements obtained with H.E.S.S. and other ground-based instruments were included separately in the fitting procedure (see Sect.~\ref{section:VHE}).

\subsection{Cost functions}
\label{secion:test_statistics}
The approach followed here used combinations of  cost functions  $\chi^2(\mathcal{D}_\mathrm{a},\mathbf{\Psi})$ 
 calculated for the  data sets $\mathcal{D}_\mathrm{a}$,
 where $a$ indicates the underlying observations as given in Eqns.~\ref{eq:D_sync}--\ref{eq:D_IC}; for example, for the synchrotron SED:
 \begin{equation}
 \label{eqn:chisync_sed}
     \chi^2_{sync,SED} = \chi^2(\mathcal{D}_\mathrm{sync,SED},\mathbf{\Psi}) = 
     \sum\limits_{i=1}^{203} 
     \left(\frac{L_{\nu,i}(\mathbf{\Psi}) - 4\pi d_\mathrm{Crab}^2 f_{\nu,i}}
     {4\pi d_\mathrm{Crab}^2 (\sigma_{f,i}^2+(0.05\,f_{\nu,i})^2} \right),
 \end{equation}
 where $L_{\nu,i}$ is the model synchrotron
 luminosity at frequency $\nu_i$. The statistical uncertainty on the flux measurement was adjusted 
 by a systematic uncertainty, added in quadrature. Without further detailed knowledge about the
 contributions to the systematic uncertainty of individual measurements,
 we chose to assume that 
 the flux measurements are subject to relative uncertainty of 5~\%.\footnote{In the previous works of \citet{meyer}, 
 the relative systematic uncertainty was 
 assumed to be $7~\%$.} The best-fitting parameters
 do not change within the accuracy of the minimising routine
 when varying the relative systematic uncertainty in the
 range from $0$~\% to $7$~\%. 
 
 The value of the model luminosity $L_{\nu,i}$ was  calculated for a set of parameters
 $\mathbf{\Psi}$ (see Table~\ref{table:parameters}). 
 Furthermore, we calculated the $\chi^2$-statistics for the angular extension:
 \begin{equation}
 \label{eqn:chisync_ext}
     \chi^2_{sync,ext.} = 
     \chi^2(\mathcal{D}_\mathrm{sync,ext},\mathbf{\Psi}) = 
     \sum\limits_{j=1}^{15}
     \left(\frac{\theta_{68}(\nu_j,\mathbf{\Psi}) - 
                 \theta_{68}(\nu_j)}{\sigma_{\theta}(\nu_j)}\right)^2,
 \end{equation}
 where $\theta_{68}(\nu_j,\mathbf{\Psi})$ was calculated from the intensity $I_{\nu,j}(\theta)$,
 such that 
 \begin{equation}
     0.68 = \frac{\int_{0}^{\theta_{68}(\nu_j)}d\theta\, \sin\theta\,  I_{\nu,j}(\theta)}
                 {\int_{0}^{\theta_\mathrm{max}}d\theta\, \sin\theta\,  I_{\nu,j}(\theta)}.
 \end{equation}
 The maximum value $\theta_\mathrm{max}=250^{\prime\prime}$ was
 chosen such that the calculated angular extension $\theta_{68}$ does
 not change within the numerical accuracy for a choice of a larger value of $\theta_\mathrm{max}$.
 The measured extensions and uncertainties are listed in Table~\ref{table:theta}. The uncertainties on the measured angular extension were
 estimated from the data and do not require additional adjustments as the resulting $\chi^2_{sync,ext}$ (see Section~\ref{section:fit_sync})
 indicates an acceptable goodness of fit.  

The  value of 
\begin{equation}
  \label{eqn:chi_Sy}
    \chi^2_{sync}= \chi^2_{sync,SED}+\chi^2_{sync,ext}
\end{equation}
is the contribution to the cost function related to the synchrotron data set. 

Similarly to the calculation of $\chi^2_{sync}$, the inverse Compton data set $\mathcal{D}_\mathrm{IC}$
was used to calculate separately  $\chi^2_{IC,SED}$ and $\chi^2_{IC,ext}$ and the sum of the two
\begin{equation}
\label{eqn:chi_IC}
    \chi^2_{IC} = \chi^2_{IC,SED}+\chi^2_{IC,ext},
\end{equation}
which is relevant to estimate the parameters related to the strength and radial distribution of magnetic fields in the nebula.

The main systematic uncertainty of the flux measurement with 
\textit{Fermi}-{LAT} (see also Sect.~\ref{section:observations}) is 
related to the energy calibration of the calorimeter.
This known uncertainty
was taken into account by adding
a contribution to $\chi^2_{IC,SED}$ which relates to a relative scaling of the energy (frequency):
\begin{equation*}
    \nu \mapsto \nu^\prime = \nu\,(1+\zeta), 
\end{equation*}
such that
\begin{equation}
\label{eqn:icsed_syst}
\begin{split}
    \chi^2_{IC,SED} &= \sum\limits_{i=1}^{25}  
     \left(\frac{L_{\nu,i}(\mathbf{\Psi}) - 4\pi d_\mathrm{Crab}^2 f_{\nu^\prime,i}}
     {4\pi d_\mathrm{Crab}^2 \sigma_{f,i}}^2 \right) + \\  
     & \Theta(\zeta)\left(\frac{\zeta}{0.02}\right)^2
                                                     +  \Theta(-\zeta)\left(\frac{\zeta}{-0.05}\right)^2.
\end{split}
\end{equation}
The estimated systematic uncertainty 
allows for a relative shift in energy $\zeta$ to
be a random deviate with a asymmetric normal 
distribution given by the quoted uncertainty on $\zeta=0^{+0.02}_{-0.05}$ 
\citep{ackermann2012}. The value of 
$\zeta$ was found
whenever evaluating  $\chi^2_{IC,SED}$ (see Sect.~\ref{section:fitIC}).

Finally, we included the contribution 
of the VHE spectral measurements to the cost function  with 
$\chi^2_{VHE}$  (see Sect.~\ref{section:VHE}). 

In the following sections, we follow three  different minimisation approaches, where the most general fit  
 is presented in Section~\ref{section:VHE}.
Firstly, in Section~\ref{section:fit_sync}, we minimise the cost function $\chi_{sync}^2$ (see Eqn.~\ref{eqn:chi_Sy}) on a grid of 
    values for $B_0$ and $\alpha$, demonstrating, that the synchrotron data set  provides only 
    a weak constraint on the parameter $\alpha$ and
    is degenerate for $B_0$.
    Secondly,  we proceed in Section~\ref{section:fitIC_general} with the inverse 
    Compton component which allows us to lift the degeneracy with respect to $B_0$. In Section~\ref{section:fitIC}, the sum of $\chi^2_{IC}$ (Eqn.~\ref{eqn:chi_IC})
    and $\chi^2_{sync}$ is
     evaluated on the same grid of  $B_0$ and $\alpha$ as in the previous
     approach. 
Finally,  in the most general (time-consuming) fit
    in Section~\ref{section:VHE}, the cost function $\chi^2_{sync}+\chi^2_{IC}$ is used for marginalising over 20 parameters ($\Psi_{1\cdots20}$). 
    The resulting model is used to calculate $\chi^2_{VHE}$ and consequently $\chi^2_{tot}(\alpha)=\chi^2_{sync}+\chi^2_{IC}+\chi^2_{VHE}$
    is used to find the best estimate for $\alpha$. 

\subsection{Fitting the synchrotron spectrum and extension}
\label{section:fit_sync}
Since we are mainly interested in the best-fitting values of 
$\Psi_{20}=B_0$ and $\Psi_{21}=\alpha$, we  marginalised
over $\Psi_{1\dotsc 19}$ such that we used 
the parameters $\check{\Psi}_{1\dotsc 19}$ which minimise $\chi^2_{sync}$  
for a given pair of $\alpha$, $B_0$. The resulting  
\begin{equation}
    \chi^2_{sync}(\alpha,B_0) = \chi^2(\mathcal{D}_\mathrm{sync},\alpha,B_0,\check{\Psi}_{1\dotsc 19}),
\end{equation}
depends on $\alpha$ and $B_0$ only.

\begin{figure*}
    \centering
    \includegraphics[width=\linewidth]{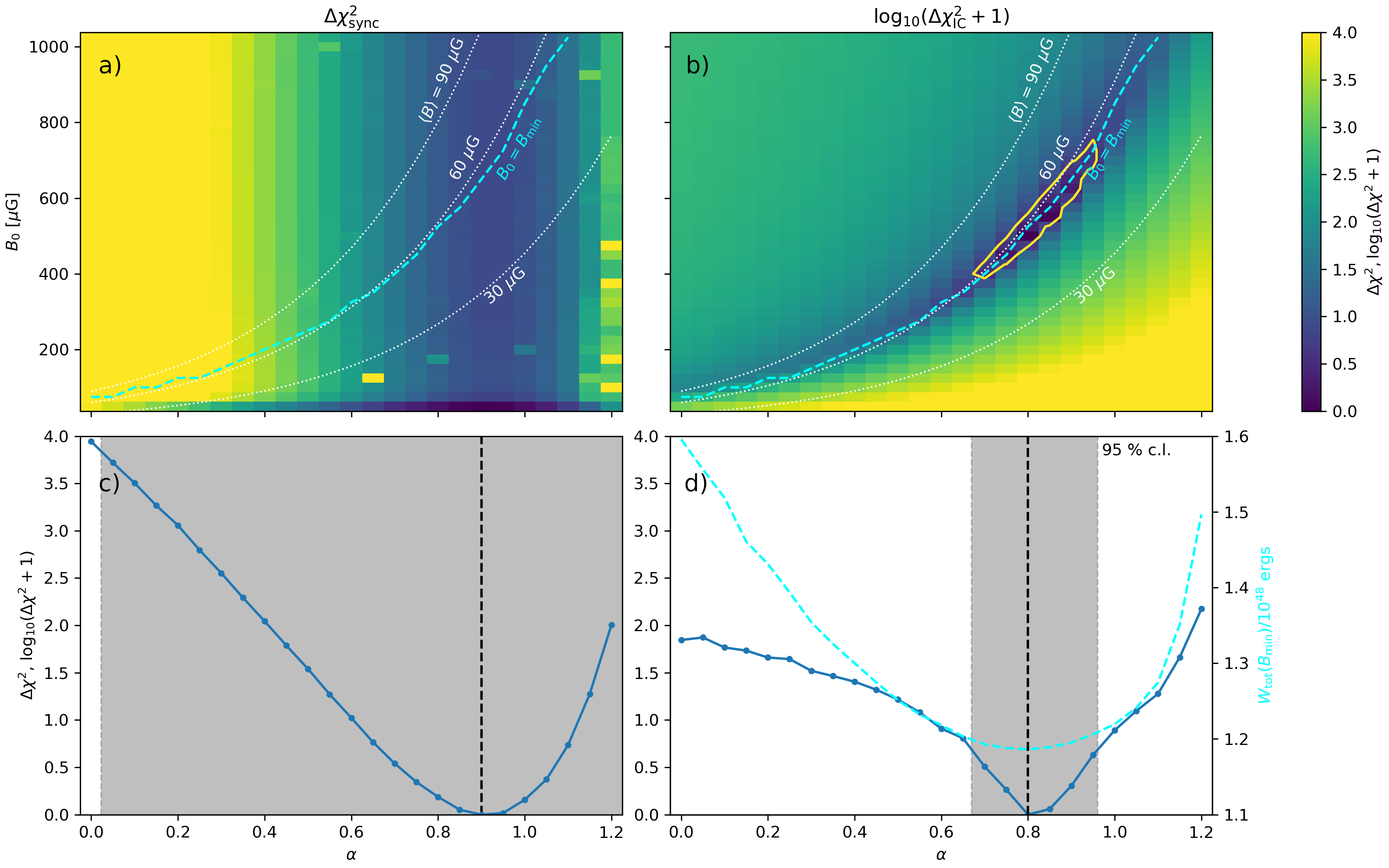}
    \caption{Results of the fitting procedure. Left column: (a) $\Delta \chi^2_{sync}(\alpha,B_0)$ and (c) 
    $\Delta \chi^2_{sync}(\alpha)$. Right column: (b) $\Delta \log_{10}(\chi^2_{IC}(\alpha,B_0)+1)$
    and (d) $\log_{10}(\Delta \chi^2_{IC}(\alpha)+1)$. In the top panels,
    for each combination of $\alpha,B_0$, the
    other parameters of the model are marginalised; in the bottom panels, this includes $B_0$. 
    The dotted white lines in the top panels indicate the 
    contours for constant values of volume averaged $\langle B \rangle=30,60,90~\mu\mathrm{G}$. The cyan dashed line
    indicates the combinations $B_0$ and $\alpha$ with the minimum total energy $W_\mathrm{tot}$ 
    in electrons and magnetic field. The yellow solid contour marks the $95~\%$ confidence level
    error box for $B_0,\alpha$. The grey boxes in the bottom panel indicate the $95~\%$ confidence
    interval for $\alpha$. In the bottom right diagram, the cyan dashed line indicates the minimal total energy $W_\mathrm{tot}$ 
    for a given $\alpha$. For a discussion on the relevance of the co-location of the  global minimum $\chi^2_{IC}$ with the corresponding
    global minimum of $W_\mathrm{tot}$, see the text.
    }
    \label{fig:marginal_x21}
\end{figure*}

As a demonstration of the approach, we show in Fig.~\ref{fig:marginal_x21}a the increment $\Delta \chi^2_{sync}(\alpha,B_0)$ 
relative to the value of the global minimum 
$\chi^2_{sync}(\check{\alpha},\check{B}_{0})$
\begin{equation}
\Delta \chi^2(\alpha,B_0)=\chi^2_{sync}(\alpha,B_0)-
\chi^2_{sync}(\check{\alpha},\check{B}_{0}).
\end{equation}

For a constant  value of $\alpha$, the resulting   $\chi^2_{sync}(\alpha,B_0)$ 
does not change substantially  over the range of
$B_0\in [50~\mu\mathrm{G},1025~\mu\mathrm{G}]$ used in the scan with a step-size of 
$\Delta B_0 = 25~\mu\mathrm{G}$. 

We find a global minimum for  $\chi^2_{sync}$ at $\check{\alpha}=0.9$ with an uncertainty of 
$\sigma_\alpha(95~\% c.l.)\approx 0.8$ (see bottom left panel, Fig.~\ref{fig:marginal_x21}c). 

The parameter $B_0$ is the normalisation of the magnetic field at the  position of 
the shock $r_s$. The resulting  field 
$\langle B \rangle$ averaged over the volume
from $r_s$ to $r_{PWN}=3$~pc was calculated to be
\begin{equation}
    \langle B\rangle = \frac{1}{V} \int~\mathrm{d}V~B(r) = \frac{3~B_0\,r_s^\alpha}{3-\alpha} \frac{r_{PWN}^{3-\alpha}-r_s^{3-\alpha}}{r_{PWN}^3-r_s^{3}}.
\end{equation}
In Fig.~\ref{fig:marginal_x21}a  we plot contours for 
$\langle B\rangle/\mu\mathrm{G}\in\{30,\,60,\,90\}$ for orientation.

\begin{table}[h]
\caption{Best-fitting parameters, and $\chi^2$-values using VHE data (MAGIC, VERITAS, Tibet As$\gamma$, LHAASO).}
    \centering
    \begin{tabular}{|l|c|}
\hline          
\textbf{Parameter}  & \multicolumn{1}{c|}{\textbf{best-fitting values}} \\
                                  & (68~\% c.l.) \\
\hline
\multicolumn{2}{c}{\textit{Radio electrons}} \\
\hline
$\Psi_1 = s_r$          & $1.54\pm0.03$     \\
$\Psi_2 = \ln(N_{r,0})$     & $114.7\pm0.2$  \\
$\Psi_3 = \ln(\gamma_1)$ & $11.4\pm0.1$  \\ 
$\Psi_4 = \rho_r$ [$^{\prime\prime}$] & $89\pm3$    \\
\hline
\multicolumn{2}{c}{\textit{Wind electrons}} \\
\hline
$\Psi_5 = s_1$     & $3.1  \pm 0.2$  \\
$\Psi_6 = s_2$     & $3.45 \pm 0.01$  \\
$\Psi_7 = s_3$     & $3.77 \pm 0.04$  \\
$\Psi_8 = \ln(\gamma_{w0})$ & $12.7\pm0.2$  \\
$\Psi_9 = \ln(\gamma_{w1})$ & $15.6\pm0.8$  \\
$\Psi_{10} = \ln(\gamma_{w2})$ & $19.2\pm0.2$  \\
$\Psi_{11} = \ln(\gamma_{w3})$ & $22.3\pm0.03$  \\
$\Psi_{12} = \ln(N_{w,0})$         & $73.8\pm0.5$  \\
$\Psi_{13} = \beta$        &    $0.15\pm0.01$  \\
$\Psi_{14} = \rho_0$[$^{\prime\prime}$]        & $99\pm4$ \\
\hline 
\multicolumn{2}{c}{\textit{Dust parameters}}  \\
\hline
$\Psi_{15} = r_\mathrm{out}$ [pc] & $1.53\pm0.09$  \\
$\Psi_{16} = \log_{10} (M_1/M_\odot)$ & $-4.4\pm0.1$  \\
$\Psi_{17} = \log_{10} (M_2/M_\odot)$ & $-1.2\pm0.1$  \\
$\Psi_{18} = T_1$ [K] & $149\pm8$  \\
$\Psi_{19} = T_2$ [K] & $39\pm2$  \\
\hline
\multicolumn{2}{c}{\textit{Magnetic field parameters}}\\
\hline
$\Psi_{20} = B_0$ [$\mu$G]    & $264\pm9$   \\
$\Psi_{21} = \alpha$          & $0.51\pm0.03$  \\
\hline
\multicolumn{2}{c}{\textit{Goodness of fit}}\\
\hline
$\chi^2_{sync,SED}(dof)$           & $182$ (184) \\
$\chi^2_{sync,ext}(dof)$           & $16$ (15)   \\
$\chi^2_{IC,SED}(dof)$             & $22$ (23)   \\
$\chi^2_{IC,ext}(dof)$             & $24$ (8)    \\
$\chi^2_{VHE}(dof)$                & $41$ (55)   \\
\hline 
$\chi^2_{tot} (dof)$         & $285$ (285) \\
\hline
\end{tabular}
\tablefoot{See Section 4.4.2 on details of the fit procedure.\label{table:parameters} The value
of $\gamma_0$ for the radio electrons has been kept fixed 
to be $\gamma_0=22$ (higher values are excluded by the
low frequency radio measurements; smaller values are not
constrained by observations). }
\end{table}

For a fixed value of $\alpha$, $\chi^2_\mathrm{sync}$ is degenerate for any choice of the 
    normalisation $B_0$. However, the energy in magnetic field 
\begin{equation}
    W_B = \int \mathrm{d}V \frac{B(r)^2}{8\pi}
\end{equation}
    increases for increasing $B_0$ while
    the normalisation of the electron spectrum decreases and correspondingly, the energy in particles $W_e$ 
    \begin{equation}
    W_e = \int \mathrm{d}V\,\int \mathrm{d}\gamma\, m_e c^2 \gamma n_{el}(r,\gamma) 
\end{equation}
     decreases monotonically 
    with increasing $B_0$.
    
    The 
    minimum energy configuration for the total energy
    \begin{equation}
    W_\mathrm{tot} = W_e + W_B
\end{equation}
 is reached for $B_0=514~\mu\mathrm{G}$, which is slightly lower than the 
 equipartition ($W_e=W_B$) field $B_\mathrm{eq}=558~\mu\mathrm{G}$.

The best-fit result which minimises $\chi^2_{sync}=195.2(197~\mathrm{dof})$ 
is reached for $\check\alpha=0.9$ and large uncertainties
which go well beyond the range of parameters scanned here. 
\begin{figure}
    \centering
    \includegraphics[width=\linewidth]{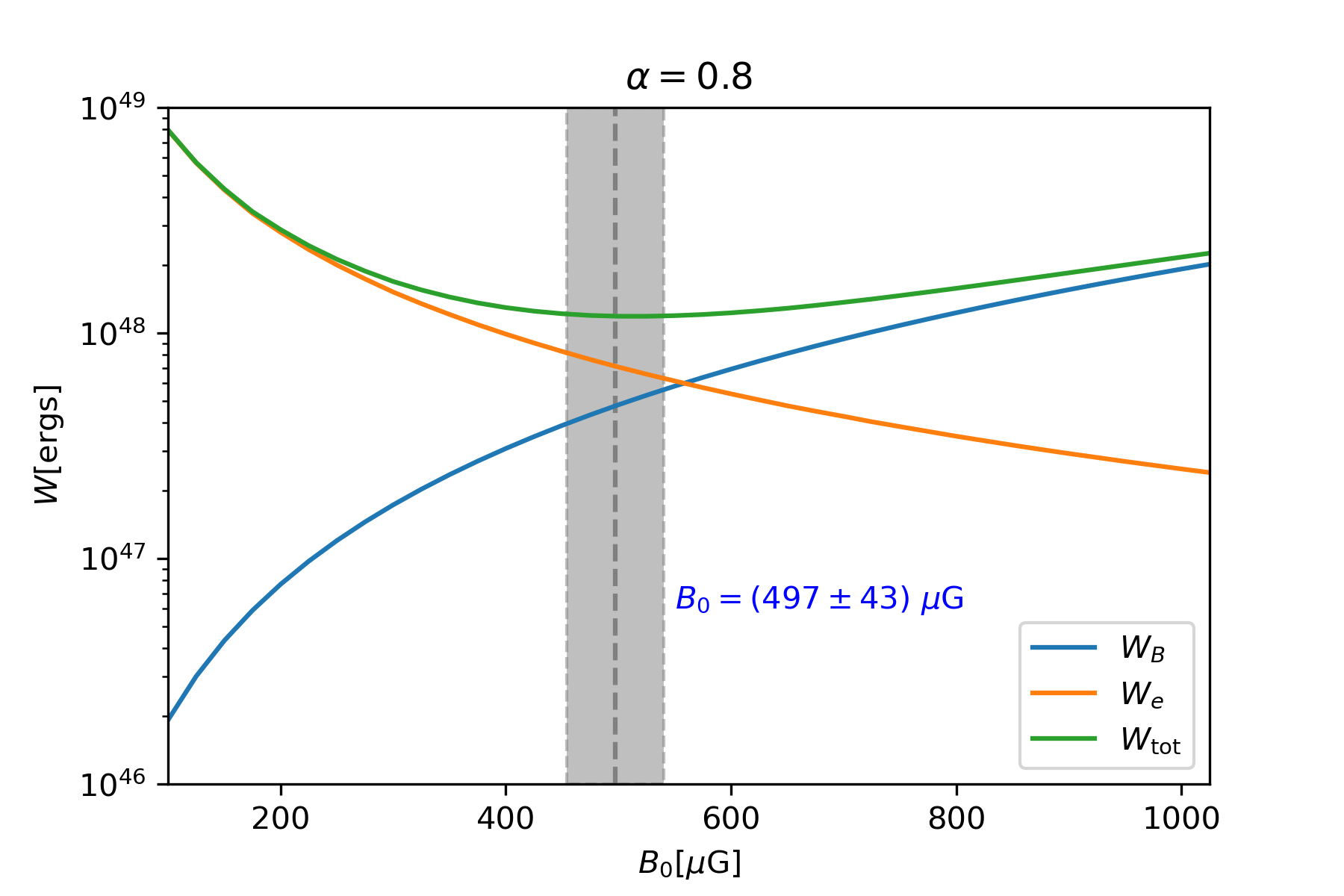}
    \caption{Volume-integrated total energy $W_\mathrm{tot}$ equals the sum of the energy in 
    relativistic electrons $W_e$ and magnetic field $W_B$. For any best-fitting
    set of parameters, the energy partition is a function of the normalisation $B_0$ of
    the magnetic field at the shock distance $r_s$. The best-fitting value is found to be $\check{B_0}=(497\pm 43)~\mu\mathrm{G}$ (90~\% c.l. uncertainty, including systematical uncertainties). This
 value is
    slightly smaller than the  minimal energy configuration at $B_0=514~\mu$G.}
    \label{fig:B_equi}
\end{figure}

The fitting of the model to the synchrotron data $\mathcal{D}_\mathrm{sync}$
provides an overall satisfying goodness of fit. The parameters related to the
electron distribution and dust properties can be estimated accurately.

We identified the combinations of $B_0$ and $\alpha$ which lead to a minimum energy configuration for
each scanned value of $\alpha$. The resulting contour $B_0=B_\mathrm{min}$ is indicated
as a cyan dashed line in Fig.~\ref{fig:marginal_x21}. 

Even though the configuration of the system is expected to be close 
to the minimum or equipartition value of $B_0$,
without additional information the normalisation $B_0$ is  degenerate with the 
number density of the electrons and $\alpha$ is only weakly constrained. 
The latter parameter is mostly degenerate with the
radial distribution of the electron density ($\Psi_{13}=\beta$). 

 Additional information such as the inverse Compton emission
needs to be incorporated in the fit to break these degeneracies and determine
the parameters $B_0$ and $\alpha$.

\subsection{Fitting the inverse Compton spectrum and extension}
\label{section:fitIC_general}
The inverse Compton emissivity of the electrons is proportional to the product of
number density of electrons and the energy density of the seed photon field (in the Thomson limit of inverse Compton scattering). 

The observation of inverse Compton emission and its spatial extension  breaks the
degeneracy of the model parameters. 
Different to a fit to the synchrotron component alone (see previous section), 
no additional constraints as for instance equipartition or minimum energy are required to determine 
$\alpha$ and $B_0$.

 \subsubsection{HE SED and extension}
\label{section:fitIC}

In order to make the fitting numerically tractable and to get
an overview on the best-fitting parameters, we carried out a minimisation 
following these steps\footnote{In the subsequent section, a more general approach is chosen.}:
\begin{enumerate}
    \item Choose $\alpha\in[0,1.2]$, starting at $0$, incremental by $0.05$
    \item For each value of $\alpha$, marginalise over $B_0$:
    $\chi^2(\alpha)=\chi^2(\alpha,\check{B_0})$  
    \item Repeat steps (1) and (2) and locate the global minimum $\chi^2(\check \alpha)$.
\end{enumerate}

The minimisation carried out in step (2) relates to 
$\chi^2=\chi^2_{sync}+\chi^2_{IC}$ with 
$\chi^2_{IC} = \chi^2_{IC,SED}+\chi^2_{IC,ext}$ given by 
\begin{equation}
     \chi^2_{IC,SED} = \chi^2\left(\mathcal{D}_{IC,SED},\alpha, B_0,\check\Psi_{1\dotsc 19}, \check{\zeta}\right),
\end{equation}
and
\begin{equation}
     \chi^2_{IC,ext} = \chi^2\left(\mathcal{D}_{IC,ext},\alpha, B_0,\check\Psi_{1\dotsc 19}\right).
\end{equation}
Note, the values of the parameter estimates related to the
electron and dust population
$\check\Psi_{1\dotsc 19}$ were found 
by minimising $\chi^2_{sync}$ separately for each choice of  parameters 
$\alpha$, $B_0$. 
 This leads to an acceptable
computation time for a parameter scan. 

The resulting $\chi^2_{IC}(\alpha$) shows a 
 minimum for $\check \alpha=0.8$ 
 shown in Fig.~\ref{fig:marginal_x21}d (bottom right panel). 
While $\chi^2_{sync}$ remains essentially constant over the entire
scan range, $\chi^2_{IC}$ rises quickly towards the lower and upper
boundary of the scan range. In order to accommodate the dynamical
range of $\Delta \chi^2_{IC}$, the right hand panels of Fig.~\ref{fig:marginal_x21} show $\log_{10}(\Delta \chi^2_{IC}+1)$.

The best estimates for $\check\alpha=0.8^{+0.16}_{-0.13}$
and $\check B_0=(497\pm43)~\mu\mathrm{G}$ are close to the 
minimum energy condition ($B_\mathrm{0,min}=514~\mu\mathrm{G}$, 
see also cyan lines
in Figures ~\ref{fig:marginal_x21}c,d and \ref{fig:B_equi}). 

For an overview of the best-fitting value $\check B_0$, we
also carried out a scan in $\alpha, B_0$ 
shown in Fig.~\ref{fig:marginal_x21}b. The solid yellow contour 
in the same Figure indicates the
$95~\%$ confidence level parameter region favoured here. The
shape of the favoured parameter region follows the shape of the
cyan line which traces the minimum energy configuration.

The fit to the synchrotron component is 
in comparison to the previous fit to the synchrotron component only 
slightly worse with $\chi^2_{sync}=196.4$ instead of $\chi^2_{sync}=195.2$ found in the previous section.
The inverse Compton component is  fit with $\chi^2_{IC,SED} = 32.0$ and 
$\chi^2_{IC,ext}=17.2$. The total $\chi^2(dof)=245.6(228)$ indicates still an acceptable 
fit with $p(\chi^2>245.6, dof=228)=0.2$. 
However, when considering the fit to the inverse Compton component only, the resulting 
$\chi^2_{IC}(dof)=49.2(29$
indicates that the fit is not acceptable with 
$p(\chi^2>49.2, dof=29)=0.01$.

 Note, the
best fit value for $\check\alpha$ is consistent for the two different estimates from fitting the synchrotron and inverse Compton data.

\subsubsection{VHE SED}
\label{section:VHE}
 The fitting procedure outlined above has been primarily 
 optimised for the synchrotron emission of the
 Crab Nebula (SED, extension).  The inverse Compton component of the SED was
 mainly used to constrain independently 
 from the electron spectral shape the magnetic field $B_0$ and the related parameter $\alpha$.
 
 Both, the synchrotron as well as the inverse Compton SED at energies below the peak 
 in the SED are fairly
 degenerate for different choices of the radial dependence of the magnetic field (see also the following
 Sect.~\ref{section:Varyingalpha}). The resulting uncertainty of the
 power-law index $\alpha$ used for $B(r)\propto r^{-\alpha}$ is comparably large with $\sigma_\alpha\approx 0.15$, 
 even after including the \textit{Fermi}-{LAT} spectral information.  Furthermore, when 
 constraining the electron distribution by matching the synchrotron emission to the SED and 
 determining in a separate step the best-fitting values of $\alpha$ and $B_0$ (see previous section), the resulting goodness of fit to the \textit{Fermi}-{LAT} data is not acceptable.

 The accurate measurement of the IC component at energies covered with ground based VHE gamma-ray
 telescopes bears the potential of determining $\alpha$ with greater precision. This is evident 
 from the wide
 range of flux values and slopes of gamma-ray spectra at energies beyond several 100 GeV predicted 
 for the interval
 of $\alpha$ considered here (see Sect.~\ref{section:Varyingalpha}). 
 Therefore, the VHE measurement of 
 the SED reduces the uncertainty on the value of $\alpha$.
 
 We modified the method used in the previous section {by including 
 the LAT spectral information to estimate the 
 parameters $\Psi_{1\dotsc20}$}.
This numerically expensive approach was necessary
to improve the goodness-of-fit of the 
 inverse Compton method and to include the VHE measurement. 

 The fit is carried out in the following steps:
 \begin{enumerate}
     \item Choose $\alpha \in [0.4,0.8]$, increment by $0.005$
     \item Minimise $\chi^2(\alpha)=\chi^2_{sync}+\chi^2_{IC}$ to estimate 
        $\check{\Psi}_{1\dotsc 20}$
     \item Calculate for each $\alpha$ the predicted VHE spectrum and
        the resulting $\chi^2_{VHE}$.
    \item Repeat until the minimum $\chi^2(\check\alpha)$ is found to estimate
        the parameter $\alpha$.    
 \end{enumerate}
 The third step involves both individual VHE data sets $\mathcal{D}_{VHE,i}$ as 
 well as their combinations. 
 \begin{equation}
     \mathcal{D}_{VHE} = \bigcup_{i} \mathcal{D}_{VHE,i}. 
 \end{equation}
 A complete list of the data sets and individual fit results are presented in Appendix~\ref{appendix:VHE} and Table~\ref{tab:data_sets}.
 In the following, we considered
 a combination of five data sets obtained with MAGIC
 (normal zenith angles as well as very large zenith angles),
 VERITAS, Tibet~
 AS$\gamma$, and LHAASO. With this combination of observations, 
 a consistent model was found. Another consistent solution was found
 for the combination of the data from the HEGRA and HAWC observations
 (see Appendix~\ref{appendix:lowalpha}), while the spectra 
 obtained with the H.E.S.S. I and II telescope arrays cannot be fit well
 by the model presented here (see Table~\ref{tab:data_sets} for the
 individual fit results). 
 
 For each data set $\mathcal{D}_{VHE,i}$, we introduced
 a relative energy scaling 
 $\zeta_i$  as an additional nuisance parameter. 
  This adjustment of the
 relative energy scale compensates for systematic uncertainties related to the
 absolute energy calibration of ground based air Cherenkov telescopes. 
 
 The scaling factors $\zeta_i$ determined for 
 the VHE data  are listed
 in Tab.~\ref{tab:scaling_fac}. The values are close to
 unity with the largest correction applied to the MAGIC data
 set taken at very large zenith angles (MAGIC VLZA: 
 $\zeta=1.14\pm 0.03$). 
 The value found here is well within the error budget
 of the systematic uncertainty related
 to the energy scale which is estimated to be $17~\%$ for this
 data set \citep{magic_2020}. Note, the dramatic reduction of 
 the $\chi^2$ between $\chi^2_{before}=832$ before the scaling
 and $\chi^2_{after}=63.1$ after the scaling (see  Tab.~\ref{tab:scaling_fac} for more details). The introduction of the five additional parameters reduces 
 the number of degrees of freedom from 85 to 80 for the VHE data sets used here.
 
\begin{figure*}[ht!]
\includegraphics[width=\linewidth]{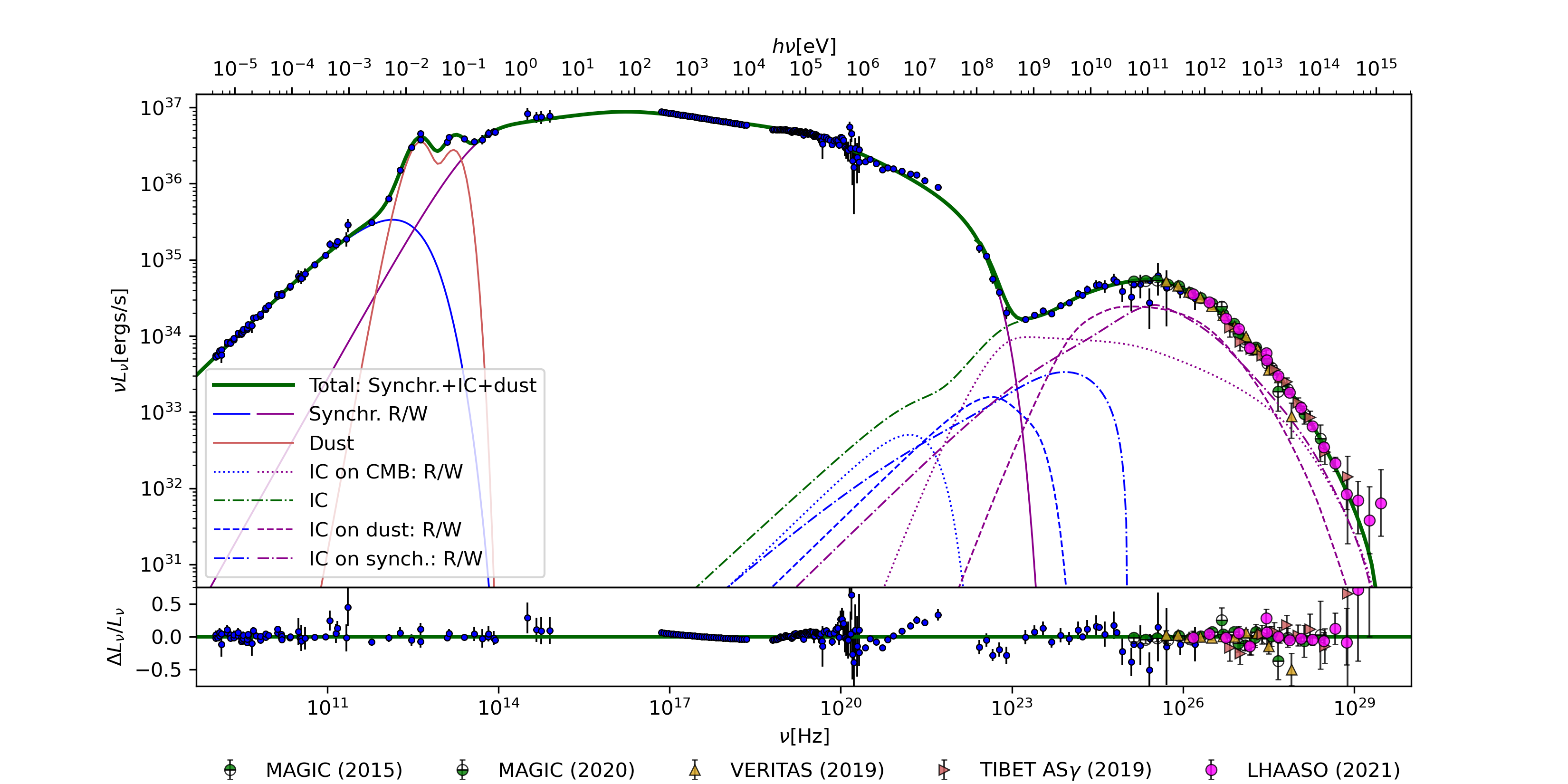}
\caption{The best-fitting model  in comparison to the data.
The upper plot shows the synchrotron and inverse Compton (IC) radiation by radio (R) and wind (W) 
electrons produced in the spatially varying magnetic field and
seed photon field (blue and magenta coloured broken lines indicate the contributions from IC scattering on individual seed photon fields). The blue data points are used in the fit to determine the parameters $\check{\Psi}_{1\ldots 20}$ 
listed in Table~\ref{table:parameters}. 
The VHE data have been scaled using the 
values of $\zeta_i$ determined for each data-set (see text for further details)
to adjust the energy scale for prediction given by the global minimum $\chi^2(\check{\alpha}=0.51)$. In the lower plot, the relative 
residuals are shown 
(statistical uncertainties only).}
\label{fig:SED}
\end{figure*}

 The best-fitting parameters were used to calculate the 
 broadband SED for $\check\alpha=0.51\pm 0.03$ and $\check B_0=(264\pm 9)~\mu\mathrm{G}$ (for a
 complete list of parameters used, see Table~\ref{table:parameters}). 
 The result is shown in Fig.~\ref{fig:SED}.
 The overall fit from radio to PeV energies traces the observational data with a resulting $\chi^2_{SED}=
 \chi^2_{sync,SED}+\chi^2_{IC,SED}+\chi^2_{VHE}=182+24+41=247$
with $184+23+55=262$ degrees of freedom (see Table~\ref{table:parameters}
for the resulting parameter estimates as well as the 
goodness of fit summary).
In the residual shown in the bottom panel of Fig.~\ref{fig:SED}, deviations are visible in 
the MeV spectrum as well as at PeV energies. We  
return to possible explanations in Section~\ref{section:UHE}.

\begin{figure}
\includegraphics[width=\linewidth]{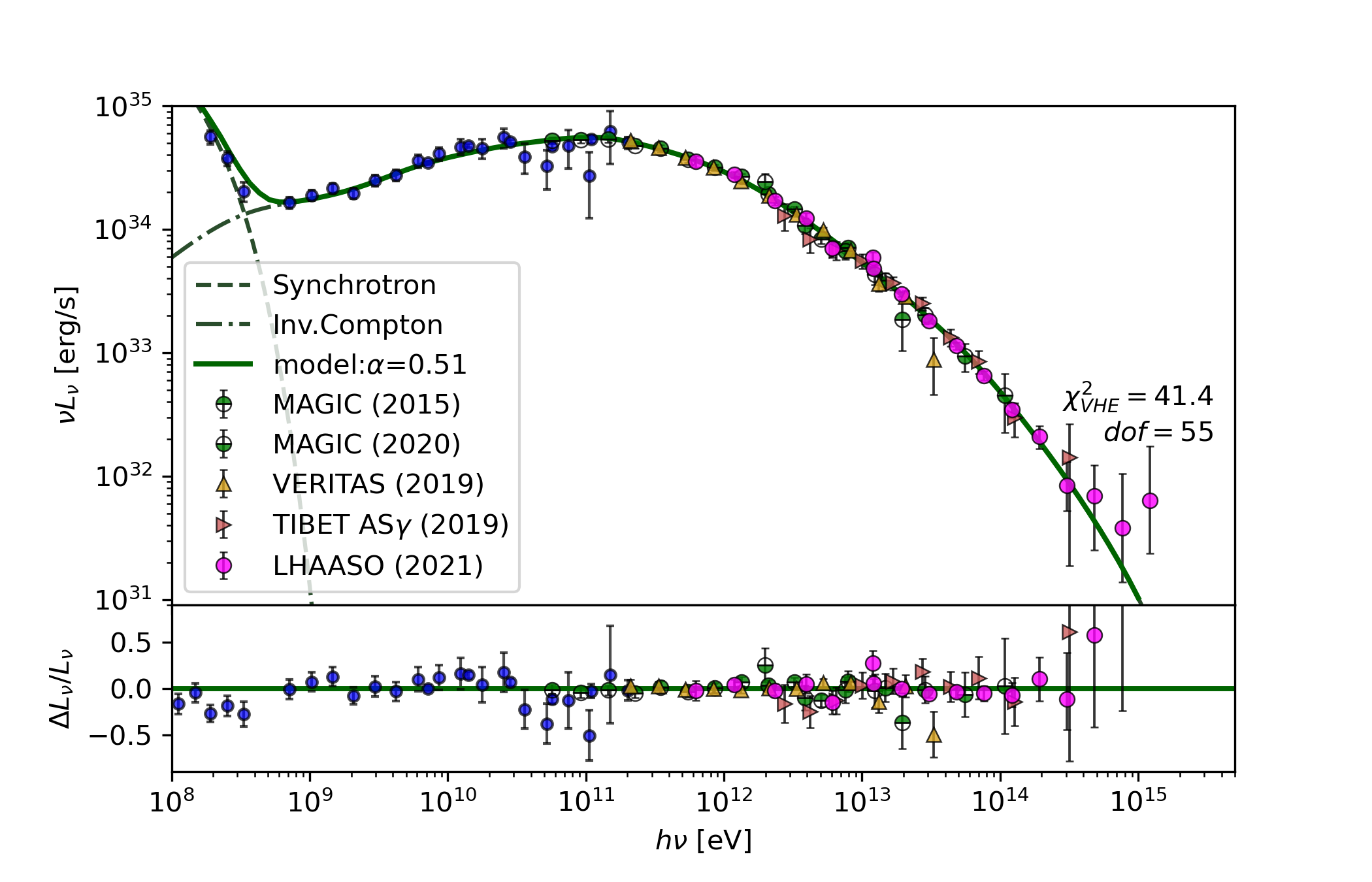}
\caption{Comparison of the best-fitting model with the HE and VHE data. The detailed view confirms the consistency of the 
selected data sets (after scaling) and the good agreement of the model to the data. The selected VHE data sets overlap with
the LAT data and each data set overlaps with at least one other VHE data set.}\label{fig:VHE_hialpha}
\end{figure}

In more detail, the SED at HE and VHE gamma-rays is shown in Fig.~\ref{fig:VHE_hialpha}. The VHE data sets
selected are consistent after the relative energy scales have been 
slightly adjusted to a common energy-scale 
provided by the LAT data. The five different VHE data sets overlap in 
the energy range covered both with the LAT data as well as among each other. The resulting goodness of fit 
is quantified by $\chi^2_{VHE}=41$ for 55 degrees of freedom. Even without additional systematic uncertainties related
to the flux measurement, the statistical uncertainties are not 
sufficiently small to further constrain or even rule out the model
used here.

\begin{figure}
    \centering
    \includegraphics[width=\linewidth]{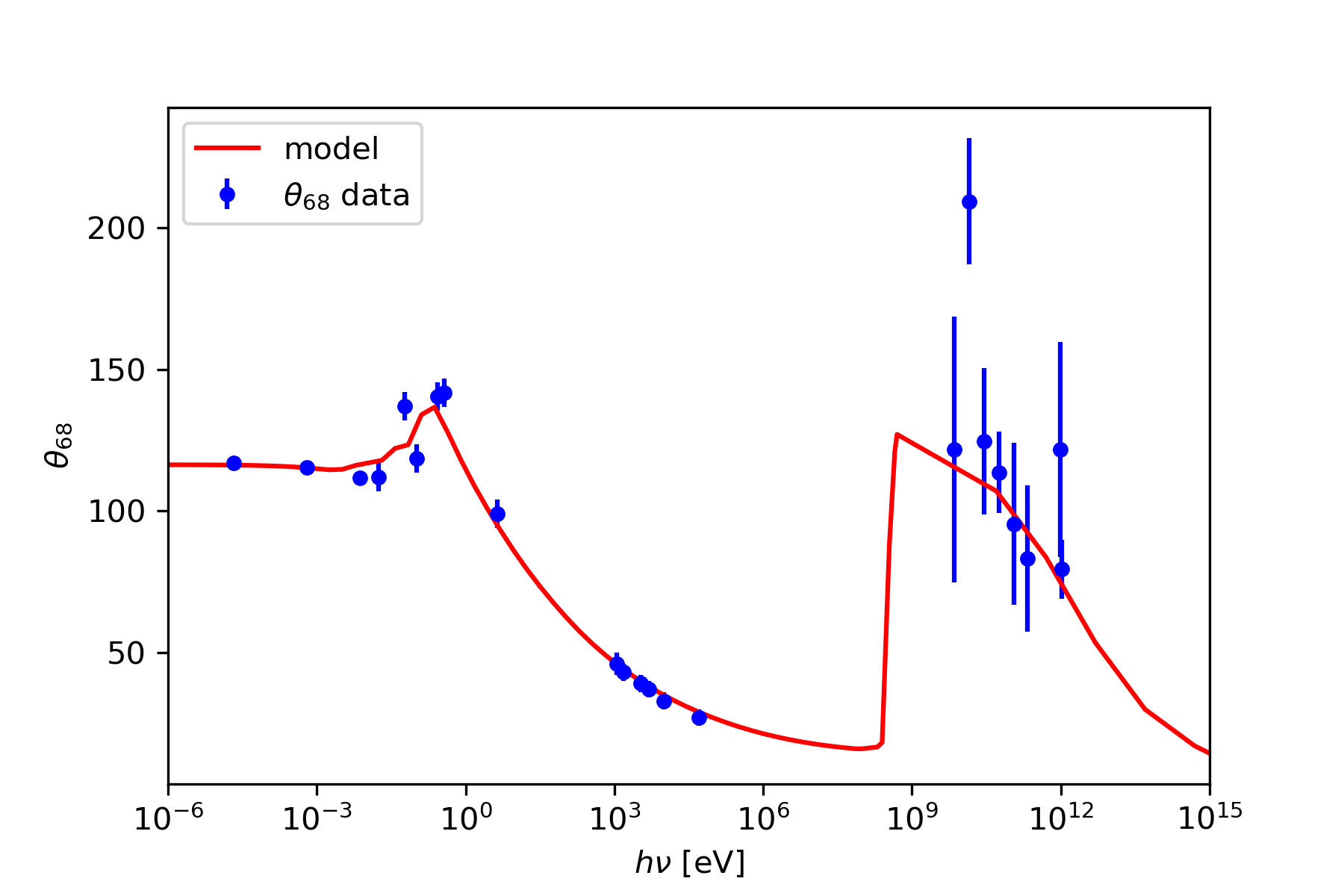}
    \caption{The extension of the model emission is characterised by the radial size $\theta_{68}$. The model curve
    is the calculated extension which matches well the observational data.}
    \label{fig:plot_r68}
\end{figure}

The spatial extension of the model emission is matching the measured extension quite well. 
This is shown in
Fig.~\ref{fig:plot_r68}. The  resulting $\chi^2_{ext}=\chi^2_{sync,ext}+\chi^2_{IC,ext}=16+24=40$ for 15+8=23
data points (the number of degrees of freedom is reduced by the
number of parameters of $\gtrsim 2$) 
indicates at first glance a poor fit: $p(\chi^2>40,dof=21)\approx 1.2\times 10^{-5}$. However, 
the large value of $\chi^2_{IC,ext}$ is mainly driven by one (outlying) measurement at $E=14~\mathrm{GeV}$. When removing that individual point, the
resulting $\chi^2_{ext}(dof)=21(20)$.

After including $\mathcal{D}_{VHE}$, the constraint on 
the magnetic field distribution  improved considerably. The
best-fitting estimate for $\check{\alpha}=0.51\pm0.03$ has a remarkably small statistical uncertainty. The value of $\alpha$ is mainly constrained by the accurate measurement of the 
SED at VHE energies. The measurement of the spatial extension is less constraining (see 
also Section~\ref{section:Varyingalpha}). The difference in the 
best-fitting values for $\alpha$ when comparing the results obtained
in Section~\ref{section:fitIC} and the ones found here
is not significant, 
but will be discussed in the context of an extension of the
model  in Section~\ref{section:magneticfield}.

\subsubsection{Varying $\alpha$: Effect on the IC SED and extension}
\label{section:Varyingalpha}
As discussed in the previous sections, there is a
unavoidable degeneracy of $\chi^2_\mathrm{sync}$ for 
variations of parameters related
to the magnetic field ($\alpha$, $B_0$) and the electron spectrum.

\begin{figure*}
\includegraphics[width=\linewidth]{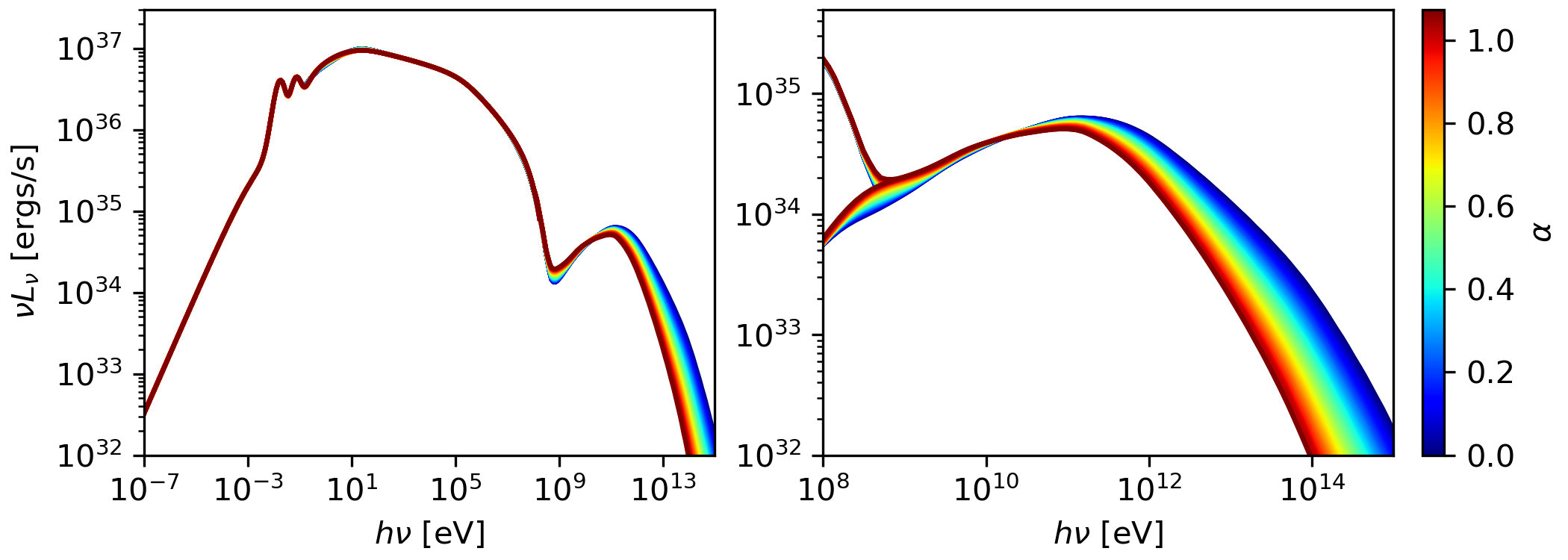}
\caption{\label{fig:sed_varalpha} The overall spectral energy distribution 
for the estimated best-fitting parameters $\check\Psi_{1\cdots 20}$ (after minimising 
$\chi^2_{IC}$ for fixed values of $\alpha$) is shown. The value of $\alpha$
characterises the radial decrease of the magnetic field $B\propto r^{-\alpha}$ 
(see Eqn.~\ref{eqn:B}). While the synchrotron part of the spectral energy distribution
is almost degenerate with respect to the choice of $\alpha$, the inverse Compton dominated
part (right panel) shows noticeable differences up to the peak at approximately $10^{11}$~eV. Note, in the left panel the sum of
synchrotron and inverse Compton emission is shown, while in the right
panel, the SED of the inverse Compton emission is additionally 
superimposed.}
\end{figure*}

As an illustration of this degeneracy with respect to the choice 
of $\alpha$, we
show in Fig.~\ref{fig:sed_varalpha} the
best-fitting SEDs, colour-coded for different values of $\alpha$. 
Each SED is the result of the fitting procedure outlined in Sect.~\ref{section:fitIC} above. 
For clarity, the data points are not included.

The resulting SEDs indicate that the synchrotron
component does not change in a noticeable way when
varying the value of $\alpha$ while the inverse
Compton emission varies in a systematic way: For 
a constant magnetic field ($\alpha=0$), the energy
spectrum is harder than for models where
the magnetic field drops with increasing radius ($\alpha>0$).

The model with the largest value of $\alpha=1.2$ covered in the scan 
has the softest spectrum. The differences between
the models with varying $\alpha$ is most pronounced
at the energies beyond the peak energy of approximately 100~GeV. 

\begin{figure}
    \centering
    \includegraphics[width=\linewidth]{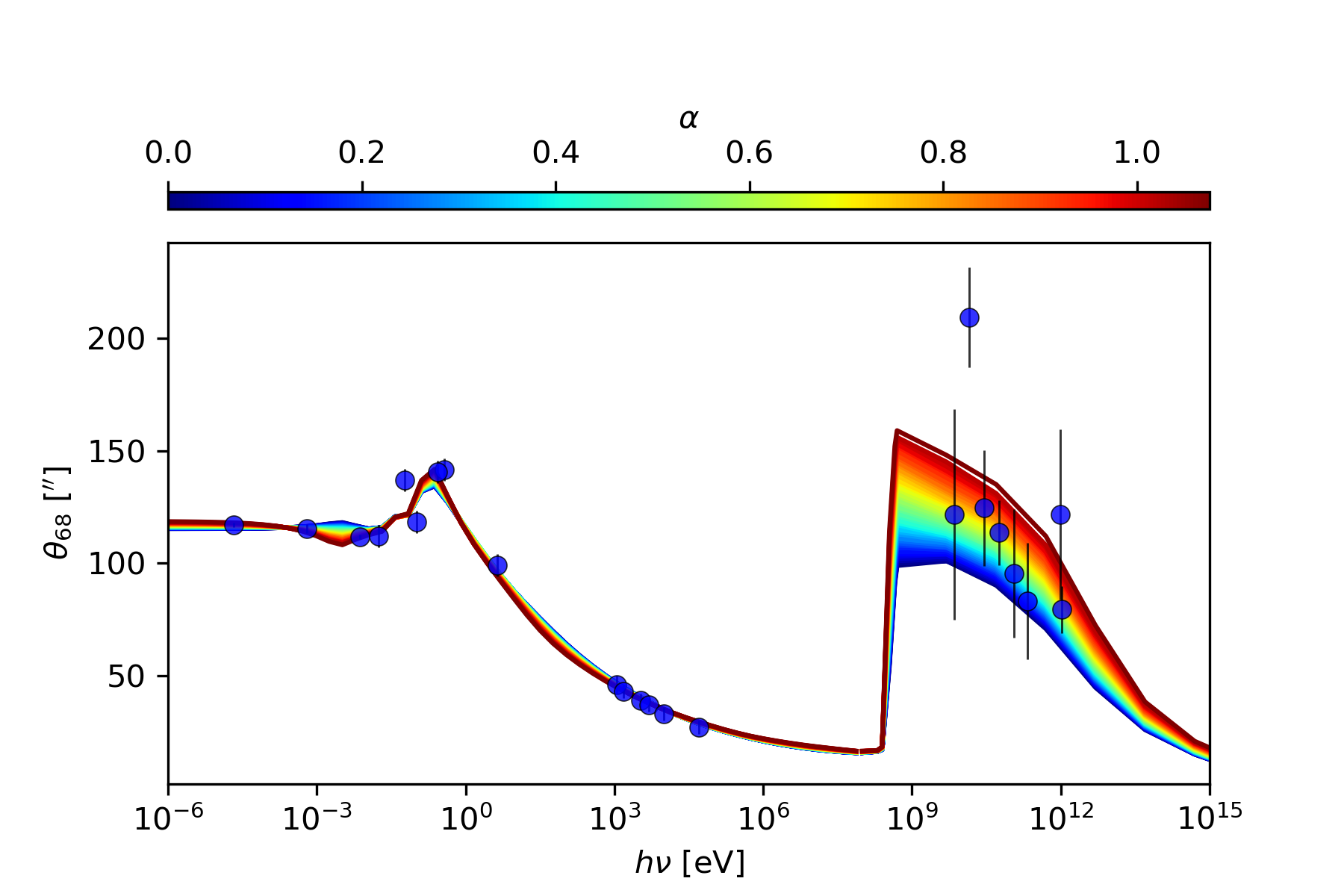}
    \caption{The angular extension from the fitting procedure described in the text. After marginalising
    $\check\Psi_{1\cdots20}$, the predicted angular extension of the synchrotron nebula matches the data well
    for values of $\alpha=0,\ldots,1.1$ with very small variations visible in the transition between 
    thermal and non-thermal emission at near- to far infrared. At higher energies, the predicted size
    differs mostly at the low-energy end of the inverse Compton emission.
    }
    \label{fig:sed_varalpha_size}
\end{figure}

Similar to the SED, the angular extension of the synchrotron nebula
is highly degenerate with respect to variations
of the parameter $\alpha$. This is demonstrated
in Fig.~\ref{fig:sed_varalpha_size}. 

Different to the SED which shows the largest variations of the SED for the VHE part
when changing the value of $\alpha$, the angular extension
varies the strongest for the part of the inverse Compton emission 
covered with HE gamma-ray observations.

At several GeV of gamma-ray energy, for a constant magnetic field, the extension
is the lowest (about 100$^{\prime\prime}$) and
it systematically increases with increasing
$\alpha$, such that for $\alpha=1.2$, 
the extension reaches a value of about $150^{\prime\prime}$.
At the highest energy part, the variation of the extension for different values of $\alpha$ is effectively constant.

The underlying reason for the observed changes in the SED is related to the magnetic field relevant for
electrons of different energies. The most energetic electrons are located close to the shock which features
the largest magnetic field. Consequently, a reduced number density of these electrons is required to match the synchrotron emission 
which in turn leads to a reduced  inverse Compton emission, softening the VHE spectrum with increasing value of $\alpha$. The 
energy range covered with \textit{Fermi}-LAT is less affected by changes of $\alpha$ 
as the underlying electrons populate the entire volume and therefore produce synchrotron emission
in an averaged magnetic field which does not change for the best-fitting model when varying $\alpha$.

\subsubsection{An additional ultra-high energy component}
\label{section:UHE}
The energy spectrum measured at ultra-high energy (UHE) 
gamma-rays above several 100 TeV with the LHAASO array \citep{2021Sci...373..425L} apparently deviates  from the predicted shape 
of the inverse Compton emission at the highest energy bin centred on  $E_\gamma=1.26~\mathrm{PeV}$.\footnote{When
considering the actual photon statistics, the Poissonian probability to observe 2 or more photons in the energy bins around 1 PeV 
is predicted to be 15.6~\% when calculating the expected number of events with the best fitting model and the collection area given by \citet{2021Sci...373..425L}, see Appendix~\ref{appendix:photstat} for
more details.}
It has been speculated that the PeV emission could be due to 
an additional hadronic component, similar to the one possibly present in the Vela-X region \citep{2006A&A...451L..51H}. Previous upper 
limits on the presence of a hadronic wind have constrained the fraction of spin-down power diverted to
the acceleration of relativistic hadrons
to be less than $\approx 20~\%$ up to PeV energies \citep{Aharonianetal2004}. The PeV excess could be produced by as little as 1~\% of the 
spin down power assuming efficient confinement of the protons in the Crab Nebula \citep{2021Sci...373..425L}. 

Here, we invoked a second electron distribution similar to previous attempts of explaining the flaring component of the Crab Nebula 
\citep[see e.g.][]{2011A&A...533A..10L} with the  energetic electrons located in the vicinity of the pulsar or close to the 
shock. The finding of a fast (several days)  {dimming} 
in the Nebula's emission \citep{2020A&A...638A.147Y}  
indicates that most of the highest 
energy synchrotron emission ($>75~\%$) 
is produced in a compact region. 

\begin{figure*}
      \centering
      \includegraphics[width=\linewidth]{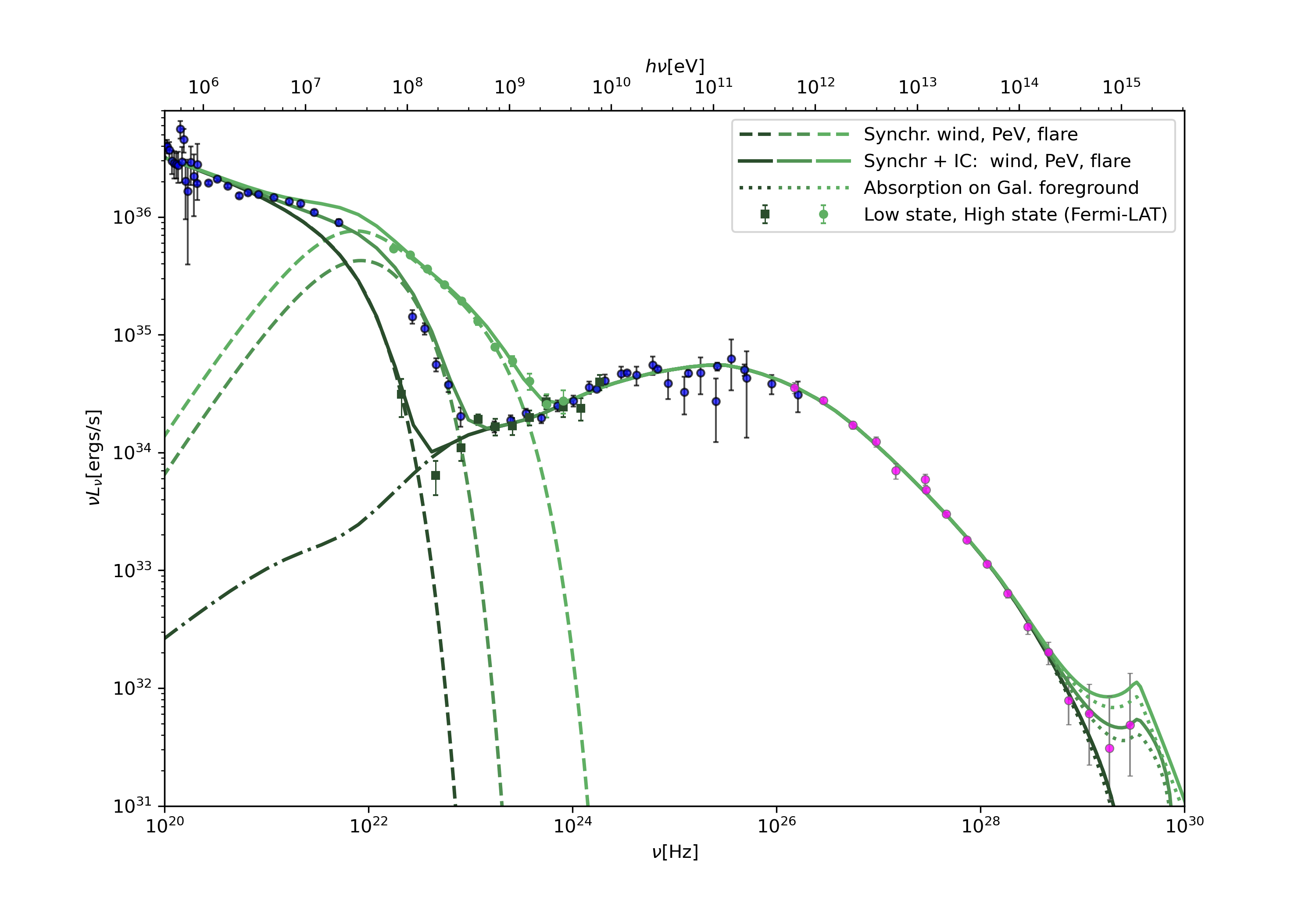}
      \caption{A suggested model for the SED including a secondary electron contribution located at the shock position. 
      The 
     low and high state spectra as measured with \textit{Fermi}-LAT are shown for comparison \citep{2020A&A...638A.147Y}.
      The 
      secondary electrons' spectrum during average (high) state
      is a power-law with $dn/d\gamma \propto \gamma^{-2.3}$ between $\gamma_1 = 3\times10^{9}$ and $\gamma_2=7(18)\times 10^{9}$. The magnetic field
      in the region is chosen to be $85~\mu\mathrm{G}$. The total energy content of the electrons is $2(3)\times 10^{42}$~ergs. The dotted lines indicate the effect of
      absorption on the Galactic foreground emission (see also 
      Appendix~\ref{appendix:tau}).}
      \label{fig:sed_pev_near}
  \end{figure*}

Motivated by these findings and the particular shape of the spectrum measured from the Crab Nebula with the COMPTEL instrument, we 
introduced a secondary PeV electron component 
which (on average) matches the highest energy part of the synchrotron spectrum. In the framework of the model developed here, we reduced 
the maximum energy of the
wind electrons from $\gamma_{w3}=4.8\times 10^{9}$ ($E_{w3}=2.5~\mathrm{PeV}$) to $\gamma_{w3}=2.5\times10^{9}$ 
($E_{w3}=1.3~\mathrm{PeV}$).  This way,  
the presumably steady synchrotron emission cuts off below 10 MeV
 (see Fig.~\ref{fig:sed_pev_near}).

We added a secondary component which is located at a fixed distance where due to radiative cooling, 
the energy is  quickly dissipated.  
In the model considered here, the magnetic field varies as $B(r)=264~\mu\mathrm{G}(r/r_s)^{-0.51}$. In order to fit the synchrotron and 
UHE part, the 
consistent combination of seed photon field and magnetic field strength would favour a  position in
the nebula  at $r\approx 20~r_s$. Since  it seems unlikely that particle acceleration to PeV energies is ongoing in the nebula's 
periphery, 
we explored here the possibility of electron acceleration close to the shock at $r\approx r_s$, possibly in 
a region with weaker magnetic field strength (here, we chose a value of 85~$\mu$G). We considered the seed photon field at the shock self-consistently calculated in our model. We did not attempt a 
quantitative fit here but instead chose representative values in order to match the SED. The average additional electron spectrum was
assumed to follow a power law with $dn/d\gamma\propto \gamma^{-2.3}$ between $\gamma_1=3\times 10^{9}$ and
$\gamma_2=7\times 10^{9}$. 

The resulting spectra for the three flux states: wind-only with
a reduced cut-off energy, average, and flaring PeV component are shown in Fig.~\ref{fig:sed_pev_near} in comparison
to the same data as shown before. 

In addition, we include the low and high-state energy spectra provided by
\citet{2020A&A...638A.147Y} to compare with the respective 
flux state from the model. The total energy required in the
secondary electron distribution is moderate with $W_e=2\times 10^{42}$~ergs. 

The computed SED matches the data quite well. At the MeV energy range, the COMPTEL measurements are described better 
by the superposition of the steady 
wind electron  emission
and the average secondary component. During times of lower activity, the reduced cut-off in the
wind electron spectrum matches the observed low-state spectrum \citep{2020A&A...638A.147Y} (dark green squares in Fig.~\ref{fig:sed_pev_near}).

The inverse Compton emission at  energies beyond 
$\approx 400$ TeV is dominated by the emission of the
secondary component. The emission of the secondary component is produced
by up-scattering  of synchrotron (radio) 
seed photons and to a lesser extent by CMB photons. 

We also included a model for the high flux state (light green curve in Fig.~\ref{fig:sed_pev_near}). The 
minimal modifications
were an increase of maximum energy such that  $\gamma_2=1.8\times 10^{10}$ and an increase
of the energy of the electrons to $3\times 10^{42}$ ergs.
The resulting synchrotron emission matches the observation during the high flux state with \textit{Fermi}-LAT quite well (light green circles). 
The
resulting inverse Compton component during the high flux state is also
shown in Fig.~\ref{fig:sed_pev_near}. In both cases, the 
absorption of gamma-rays due to the anisotropic Galactic radiation field
is included (see Appendix~\ref{appendix:tau} for
details about the calculation carried out here). 

\begin{figure}
   \includegraphics[width=8cm]{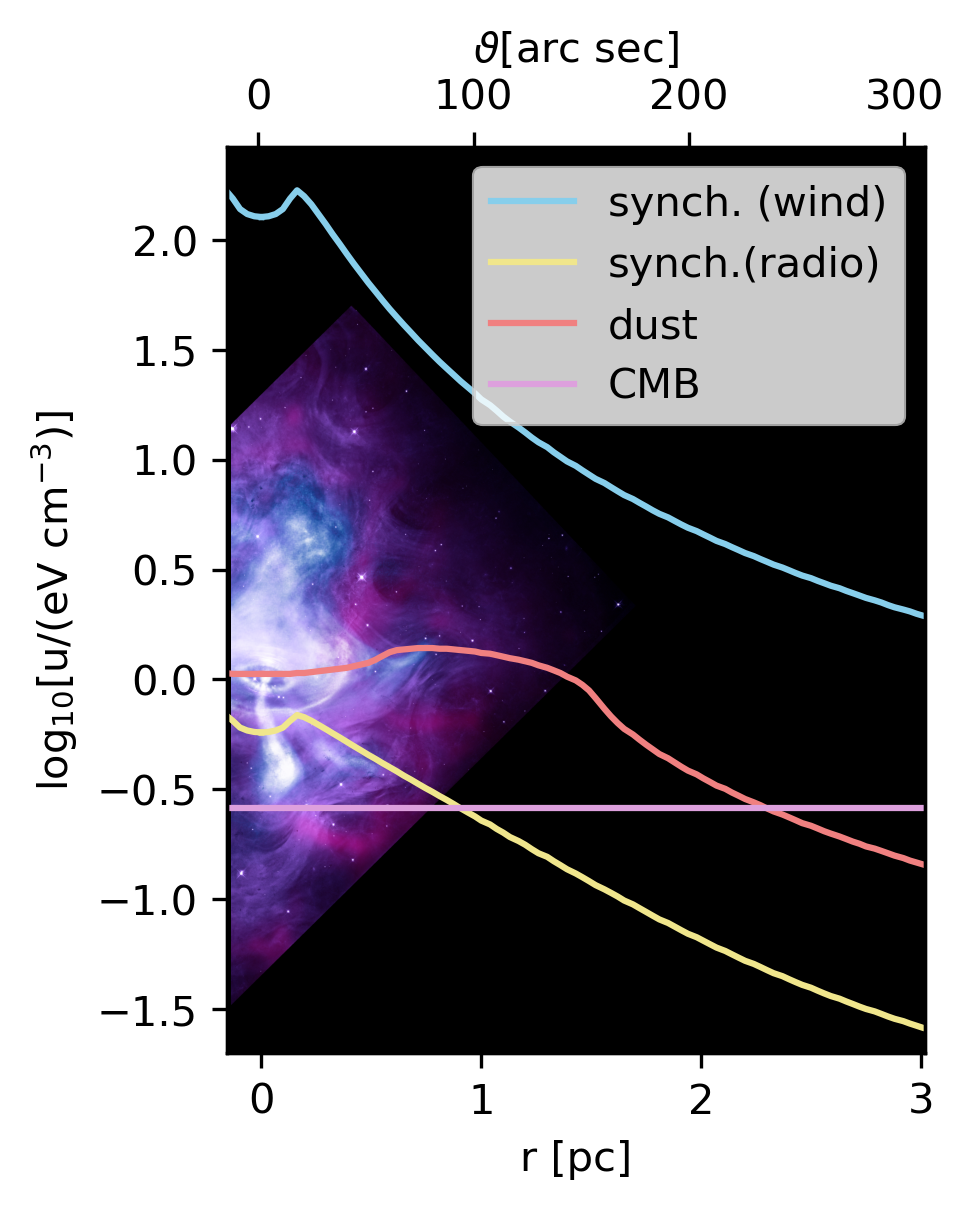}
      \caption{ Energy density of the seed photons as a function of the distance from the pulsar. The dust emissivity was 
      modelled to be confined in a shell surrounding the optical nebula.
      As a comparison, a composite of
       X-ray (Chandra: blue, white)/optical(HST: purple)/ 
       infrared (Spitzer: pink)  composite image  
      from the nebula is shown with approximately 
      matching spatial scale. 
      Background image credit: 
       X-ray: NASA/CXC/SAO; 
      Optical: NASA/STScI; Infrared: NASA-JPL-Caltech)}
         \label{fig:seed_dens}
   \end{figure}

\begin {table*}[t]
\caption{Energy-scaling factors $\zeta_i$ and resulting
$\chi^2$-values before and after changing the energy-scale.}
\begin{center}
\small\tabcolsep=0.15cm
\begin{tabular}{ c|c|c|rr|rr }
 \
 \textbf{Data set} & 
 \multicolumn{1}{p{3cm}|}{\centering \textbf{Scaling factor} $\zeta_i$} &   \multicolumn{1}{p{3cm}|} {\centering \textbf{Statistical error} $\sigma_\zeta$ }  & \multicolumn{2}{p{2.0cm}|}{\centering $\chi^{2}_{before}(dof)$ }  & \multicolumn{2}{p{2.0cm}}{\centering $\chi^{2}_{after}(dof)$ }\\
 ($\check\alpha=0.51$)& & 68~\% c.l. & & \\
 \hline
 \hline
\textit{Fermi}-{LAT}  & 1.00 & fixed  &  21.7&(25) & 21.7&(25)  \\
MAGIC               & 0.891 & 0.006 & 221.0&(14) & 11.7&(13)\\
VERITAS   & 1.050 & 0.003           & 469.9&(12)  & 9.1&(11) \\
MAGIC (VLZA) & 1.141 & 0.025& 55.6&(7)            & 2.6&(6)\\
Tibet AS$\gamma$ & 0.918 & 0.017& 22.8&(10)       & 6.4&(9)\\
LHAASO  & 0.964 & 0.006& 41.0&(17)               & 11.6&(16)\\
\hline
\textbf{Combined} &&       & 832.0&(85) & 63.1&(80) \\
\hline
\end{tabular}
\end{center}
\tablefoot{Energy-scaling factors $\zeta_i$ for the VHE spectra
are derived for the best-fitting model with $\check{\alpha}=0.51$ (the 
corresponding best fit parameters $\mathbf{\Psi}$ are
given in Table~\ref{table:parameters}).} 
\label{tab:scaling_fac}
\end {table*}
\section{Discussion of the parameters}
\subsection{Radio electrons}
The four parameters $\Psi_{1\ldots 4}$ related to the
radio electrons
are well constrained by the observational data.
The total number of radio emitting electrons 
has been found to be $N_r\approx 10^{51}$ up to a maximum 
Lorentz factor of $\gamma_1\approx 90~000$ for $\gamma_0=22$.
The lower bound $\gamma_0=22$ has been
fixed to a value which is not in conflict with the low frequency radio measurements starting at 111 MHz in our sample. A possible smaller value of $\gamma_0$ would increase the 
total number of radio emitting electrons. The value found here for $N_r$ should be
considered a lower limit. 

The resulting total energy in radio electrons is $3.4\times 10^{47}~$ergs (this value
is determined by the value of $\gamma_1$).
The spatial distribution
of radio electrons is found to extend up to $\rho_r=89^{\prime\prime}$
for the best-fitting model of the magnetic field $B(r)\propto r^{-0.51}$. The resulting extension is therefore constant  for
the radio synchrotron emission which is in good agreement with
the radial extension measured to be unchanged between 5~GHz and
150~GHz (see also Table~\ref{table:theta}).

The upper bound
of the radio electron spectrum $\gamma_1$ 
is anti-correlated with the amount of cold dust ($\Psi_{17}$).
The comparably low value for $\gamma_1$ found here is favoured
by the observational data between $\lambda=100~\mu$m and
$1000~\mu$m (see Fig.~\ref{dust_data}) where the dust emission
starts to dominate over the synchrotron emission. 
\subsection{Dust emission}

As a consequence of the comparably low value
of the upper end of the radio electrons' spectrum ($\gamma_1$),
the amount of cold dust required here is larger 
than in previous estimates which were based upon a power-law
extrapolation between the radio and optical. 
Using the same
dust emissivities as  \citet{2019MNRAS.488..164D}, we therefore find
for the cold dust mass a larger value 
 ($\Psi_{17}=\log_{10}(M/M_\odot)=-1.2\pm0.1$) than
previously ($\approx -1.7$). 
In a standard (explosive) nucleosynthesis model \citep{1995ApJS..101..181W}, the progenitor
 star would have to
be sufficiently massive ($M_\mathrm{prog}\ge 12(18)~M_\odot$)
to produce the dust with 75(25)~\% condensation efficiency.
The spatial distribution of the dust is confined to a shell
with fixed inner radius $0.55$~pc (see Sect.~\ref{sect:dust})
and an outer radius estimated to be $\Psi_{15}=1.53\pm0.09$~pc.
This corresponds to an angle of $\approx 150^{\prime\prime}$. This parameter is constrained by the angular
size measured in the infra-red (see Table~\ref{table:theta} and
Fig.~\ref{fig:plot_r68}). The dust emissivity and 
resulting spatial distribution of dust generated
seed photons (see Eqn.~\ref{eqn:dust}) was in turn used for the calculation of the 
IC emissivity (see Eqn.~\ref{eqn:seed}).  The IC emission of
wind electrons up-scattering the dust emission as seed photons
dominates at the peak energy of a few 100 GeV. 
This is distinctly different
from  previous models where  the dust seed photon field
is assumed to maintain a constant energy density of $0.5\ldots 1$
eV/cm$^3$ throughout the nebula.   Since the dust
is formed  in a shell surrounding the pulsar wind nebula, 
the energy density of the dust seed photon field reaches 
a maximum of $\approx 1.5$~eV/cm$^3$ 
at a distance of $\approx 1$~pc (see 
Fig.~\ref{fig:seed_dens}). 

\subsection{Wind electrons}
The spectral and spatial model of the wind electrons includes 
ten parameters ($\Psi_{5\ldots 14}$).  Most of the
electrons' energy is carried by the wind electrons at the lower
boundary $\gamma_{w0}=320~000$ (corresponding to $E_{w0}\approx 200~\mathrm{GeV}$). The total wind electrons' energy in the
nebula is very close to the energy in radio electrons ($3.4\times
10^{47}$ergs). 

There is a considerable gap in energy between 
the two electron components 
($\gamma_1 m_e c^2\approx 50~\mathrm{GeV}$
vs. $\gamma_{w0} m_e c^2\approx 200~\mathrm{GeV}$). 
The de-reddening of the optical nebula emission has revealed the
presence of a further break in the spectrum between the optical and 
X-ray part (spectral index changes from $0.7$ to $1.1$). The
break in the X-ray spectrum at about 120~keV requires a second
break in the electron spectrum, such that we fitted as particle indices
$3.1(2)\rightarrow 3.45(1) \rightarrow 3.77(4)$. The breaks are located
at energies of $\gamma_{w1}m_e c^2\approx 3~\mathrm{TeV}$
and 
$\gamma_{w2} m_e c^2\approx 110~\mathrm{TeV}$. The upper bound
of the energy of wind electrons is fixed by the cut-off in the
average synchrotron spectrum at $\gamma_{w3} m_e c^2 \approx 2.5~\mathrm{PeV}$.
The power-law indices and the positions of the breaks 
are most likely intrinsic to the injected particle spectrum
and the underlying acceleration mechanism. Similar results related on the
position of the lower-energy break have been found by for example \citet{2012MNRAS.427..415M}.
Radiative cooling would
not be sufficient to explain however the appearance of the higher energy break at 110~TeV.

The spatial distribution of the wind electrons was assumed to be
a radial Gaussian with an extension that follows
a power law $\rho(\gamma) = \rho_0 \gamma^{-\beta}$ with index $\beta=0.15$ (see Sect.\ref{subsec:spatial}). The 
pre-factor $\rho_0$ for   the wind electrons' distribution was found to be $(99\pm4)^{\prime\prime}$
which is larger than $\rho_r=(89\pm3)^{\prime\prime}$ characteristic for the radio electrons. This fit
result is a consequence
of the comparably large size ($\approx 145^{\prime\prime}$) found at near infrared in  the WISE images (see Table~\ref{table:theta}). 
The  large value of $\rho_0$ 
is in line with the extension measured at high energies 
with \textit{Fermi}-{LAT} ($\approx 120^{\prime\prime}$).

The observation that the lowest energy wind
electrons with energies of
approximately 200~GeV are distributed within a larger volume than
the radio electrons indicates that the electrons are not
only moving convectively with the wind. 

\subsection{PeV electrons}
The most energetic electrons at PeV energies are traced by their synchrotron emission at
energies of about 100~MeV observed with \textit{Fermi}-{LAT}.  
The light curve of the $80\ldots 800$ MeV synchrotron flux has been found
to include at least three different spectral states (high,
medium, and low state) that
can be identified from \textit{Fermi}-{LAT} observations
\citep{2020A&A...638A.147Y}: 
The high state is evident from the observation of 
flaring activity which occur roughly once per year
\citep{2013ApJ...775L..37M,2020A&A...638A.147Y}. 
In a flaring state, the 
flux increases during several days 
dramatically with a spectral hardening and 
a peak in the SED reaching up to several 100 MeV. The average high flux state
spectrum as defined by \citet{2020A&A...638A.147Y} is shown in Fig.~\ref{fig:sed_pev_near}.
Despite large efforts \citep{2013ApJ...765...56W}, no obvious
counter-part to the flares has been identified, even though
the optical variability (position, size, and 
brightness) of the inner knot does show some correlation with
the changes of the gamma-ray flux \citep{2015ApJ...811...24R}.
This correlation
is consistent with the prediction of \citet{2003MNRAS.344L..93K} that the 
majority of the MeV emission, including the flaring 
activity,
is produced in the inner knot, where supposedly most of the PeV electrons are accelerated. 

The medium state has the highest duty cycle ($>95~\%$) and is therefore
identical to the average flux state shown in the previous sections.
It is characterised by a very soft 
energy spectrum which smoothly connects to the inverse
Compton emission at approximately 1 GeV.

The low state has 
just recently been found to be equally short lived and as rare
as the flaring state. The low state is characterised by a hard spectrum that follows
the extrapolation of the IC spectrum down to an energy of about
200 MeV where the spectrum seems to become  soft again 
\citep{2020A&A...638A.147Y} and see Fig.~\ref{fig:sed_pev_near}. 

Motivated by the apparent excess emission at the UHE 
end of the spectrum, we explore a possible connection of the UHE gamma-ray excess found
with LHAASO and the flux states found in the MeV emission. We therefore introduced  (minimal) modifications to the previous best-fitting model
by (a) reducing the maximum energy of the wind electrons from 2.5 PeV to 1.3~PeV  and  (b) adding
a PeV electron component that  would radiate on  average a synchrotron
spectrum between 1 MeV and about 800 MeV. We assumed the PeV electrons to scatter off the seed photon field
close to the termination shock. With a magnetic field of $85~\mu$G the resulting number of electrons 
radiate sufficient inverse Compton emission to match 
highest energy emission observed. The resulting average state is shown in Fig.~\ref{fig:sed_pev_near}. The 
model matches better the data specifically in the Comptel energy range between 1 and 20 MeV than in the previous model. The
IC emission at PeV energies (after absorption) matches the LHAASO data points well, even though the predicted flux is at the low side. The
IC emission is dominated by the scattering of seed photons produced by the radio electrons and to a lesser degree by CMB emission. This
is a consequence of the distribution of seed photons in the nebula. Close to the termination shock, the radio synchrotron emission dominates\footnote{
Due to the Klein-Nishina effect, the scattering with  optical synchrotron emission is suppressed.}.

In this scenario, the low, medium, and high flux states can be 
successfully described by the dynamics of a single emission region close
to the termination shock. 
The resulting variability at PeV energies when comparing the average and high state is modest (about a factor of 2) and not 
resolvable with the current fluence sensitivity of for example LHAASO. 

At first glance, it seems like a plausible scenario to  include an additional PeV electron population that
 simultaneously explains the hard synchrotron spectrum between 1 and 20 MeV and the PeV excess found with LHAASO
 \footnote{Note, the observed 2 events at around 1 PeV have a probability of $15.6~\%$ to be observed, see
 Appendix~\ref{appendix:photstat} for details.}. 
 Similar scenarios have been  suggested in several studies 
 \citep[see e.g.][]{khangulyan_detection_2020,2021Sci...373..425L,2021ApJ...922..221L,2022ApJ...926....7P}.
 
 However, the common origin of the high, average, and low state emission as suggested by \citet{2020A&A...638A.147Y}
 leads to considerable tension in the magnetic field strength required for confinement and the value estimated above.
The size of the variable emission region is limited to the light crossing 
horizon for the variability time-scale $t_\mathrm{var}\approx
100~\mathrm{ksec}$ (the minimum time-scales for the apparent disappearing of the nebula and for the flares
are similar). This  translates into a minimum magnetic
field to confine the electrons radiating at $E_\mathrm{sync}$ \citep{2011A&A...533A..10L}: 
\begin{equation}
B>500~\mu\mathrm{G} \left( \frac{E_\mathrm{sync}}{100~\mathrm{MeV}}\right)^{-1/3}
\left(\frac{t_\mathrm{var}}{100~\mathrm{ksec}}\right)^{-2/3}.
\end{equation}
 With a magnetic field sufficiently large  to confine the energetic electrons, the expected 
inverse Compton emission would be too small to explain the
PeV emission from the Crab Nebula. This makes it unlikely that
the variable MeV emission is directly linked to the PeV emission.
Other, more complicated scenarios would require additional particle
populations (either leptonic or hadronic, \citep{2021ApJ...922..221L,2022ApJ...926....7P}). However, with the
currently limited statistics of 2  photons detected at PeV energies and an expected number of 0.3 photons in our model, the 
tension between data and model at this point is at the level of one standard deviation. 

\subsection{Magnetic field}
\label{section:magneticfield}
The radial dependence of the magnetic field downstream from the termination shock 
was assumed to follow a power law with a power-law index, such that $B(r)=B_0 (r/r_s)^{-\alpha}$. 
The VHE gamma-ray observations are essential to determine
the value of $\alpha$ as can be seen from Fig.~\ref{fig:sed_varalpha}. A more accurate measurement of the HE extension bears the
potential to measure $\alpha$ independently of the spectral shape (see Fig.~\ref{fig:sed_varalpha_size}). The current precision 
of the gamma-ray extension measurement is however
not sufficient to constrain $\alpha$ strongly. 
Taking the synchrotron and HE gamma-ray measurements alone, a value of $\check\alpha=0.8^{+0.16}_{-0.13}$
and  $\check{B}_0=(497\pm43)~\mu\mathrm{G}$ is found.
This fit is however constrained in the way, that the HE data is only used
after the electron spectra are determined from the  synchrotron SED. 
In the more general fit including the VHE gamma-ray
data, the parameter $\alpha$ is constrained more accurately. The benchmark fit to the combined VHE spectra taken with 
MAGIC, VERITAS, Tibet AS$\gamma$, and LHAASO favours a value of $\check{\alpha}=0.51\pm0.03$ and $\check{B_0}=(264\pm 9)~\mu$G.
The resulting total energy in magnetic field in the nebula amounts to 
$W_B=7.1\times 10^{47}$~ergs which is the same energy as in 
particles present\footnote{the value depends upon the 
volume of integration, here we choose to integrate up to $25r_s$}.

\begin{figure}
    \centering
    \includegraphics[width=\linewidth]{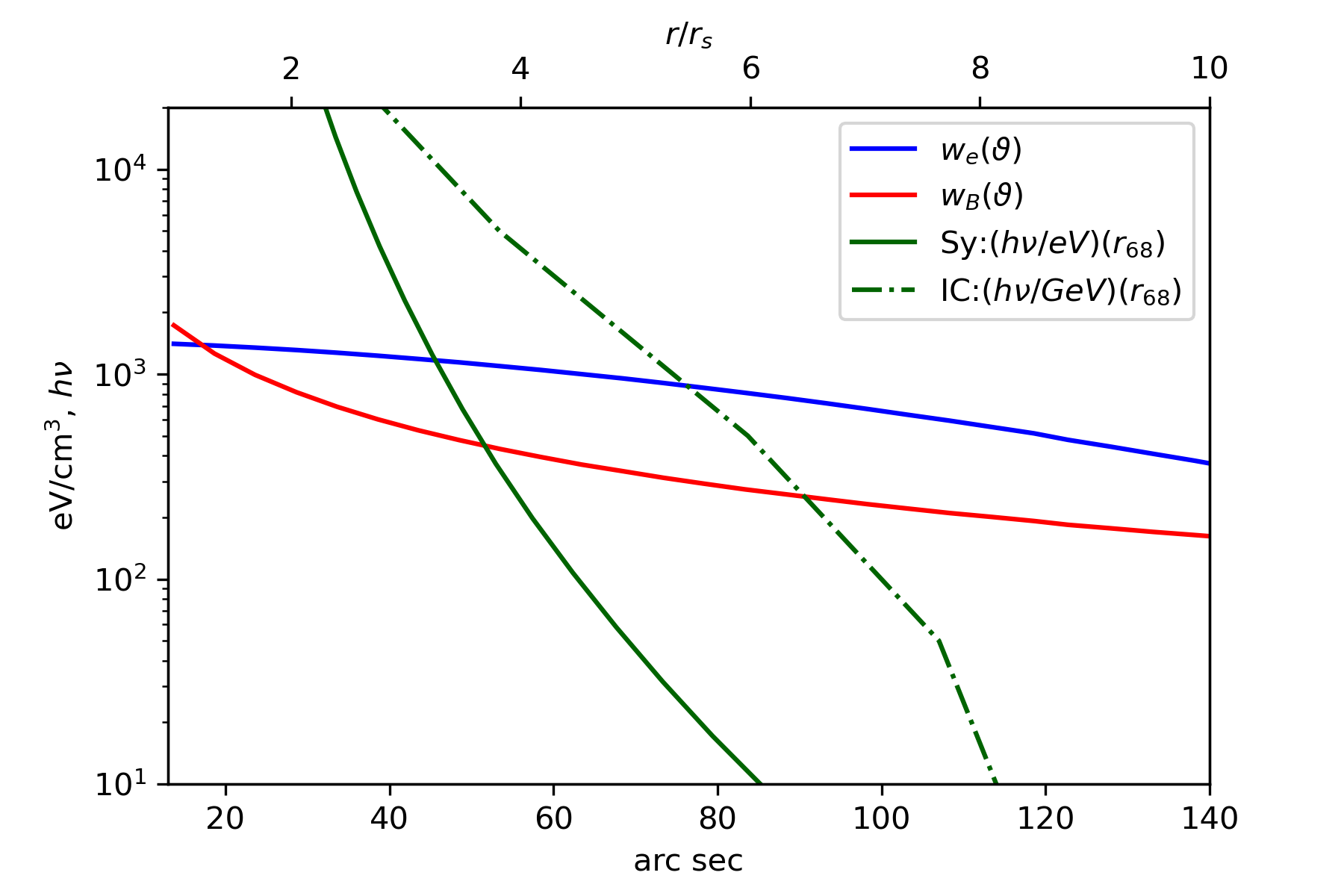}
    \caption{A comparison of the energy density $w_e(\vartheta)$ of the electrons and of the magnetic field $w_B(\vartheta)$.
    The 
    two additional (green) lines indicate the extension $r_{68}$ for the synchrotron and for the inverse Compton component: the
    y-axes indicates the photon energy $h\nu$ in units of eV (synchrotron, solid line) or GeV (IC, dash-dotted line) .}
    \label{fig:sigma}
\end{figure}

The measured value of $\alpha=0.51$
excludes  the assumption of a wound up toroidal magnetic field  (this corresponds to $\alpha=-1$). 
In order to avoid the winding up of the toroidal field in the slowing
wind, the wind has to remain  fast or the magnetic field energy is dissipated (e.g. through reconnection
through tear instabilities) in the plasma. 

The velocity of the downstream flow has been estimated
 using the ratio of X-ray brightness of the advancing and receding part of the torus. The resulting
 estimate of the flow velocity  by
\citet{2003MNRAS.346..841S,2004ApJ...609..186M}  requires a 
faster flow than predicted in the model of \citet{KennelCoroniti}. 
Conversely, the role of dissipation of magnetic fields
in the downstream flow has been discussed in 
the context of confinement of the nebula \citep[see e.g.][]{2018MNRAS.478.4622T} as well as particle acceleration
\citep[see e.g.][]{2017ApJ...841...78T,2019MNRAS.489.2403L, 2020ApJ...896..147L}. 

The fit of the radiative model to the complete data set
provides an estimate of the radial dependence of
the energy density present in particles and in magnetic field.
In Fig.~\ref{fig:sigma}, the energy density $w_B$ of the 
magnetic field dominates over the particle energy density in 
the downstream region until a distance of about $r\approx 1.3~r_s$. The reconstructed magnetic field at the termination
shock of $\check{B}_0=(264\pm9)~\mu\mathrm{G}$ corresponds to a
value of $\sigma\approx 0.1$, substantially larger than
the previous estimates of $\sigma\approx 0.003$. The shock 
is therefore not sufficiently strong (and mainly hydrodynamic)
to slow down the relativistic wind in the downstream region. 

Even though the nominal uncertainty on $\alpha$ is small
($0.51\pm0.0$), the VHE spectral measurements show significant differences
in the shape. 
The estimates of $\alpha$ therefore differ when using different
sets of VHE measurements
(see also Appendix~\ref{appendix:VHE} and Table~\ref{tab:data_sets}). An alternative combination of measurements that
are consistent with each other (see also Appendix~\ref{appendix:lowalpha}) favours a significantly different value of 
$\check \alpha=0.29^{+0.03}_{-0.06}$. Without further data, 
it is impossible to determine
whether these measurements exclude each other or whether the underlying systematic uncertainty on the shape of the energy spectrum is larger than
estimated by the instrumental teams. The contemporaneous data
taken during a flaring episode of the nebula are consistent with each
other (see Table~\ref{tab:data_sets}).
However, when looking at the temporal sequence of the observations
(see Fig.~\ref{fig:timeline}) it appears conceivable that the dynamical changes visible in the relevant part of the pulsar wind
between the termination shock and the torus could 
be caused or accompanied by a change in the dissipation rate of 
magnetic field energy in the flow. 

In addition to the  disparity between the best-fitting
values for $\alpha$ obtained with
different VHE measurements ($0.29\ldots 0.51$), there is also a noticeable (but less
significant) difference to the best-fitting value found for
$\check\alpha=0.8^{+0.16}_{-0.13}$ in Section~\ref{section:fitIC}.
For that fit,  the spectral data from 
\textit{Fermi}-LAT was used only-- 
without including VHE data which covers the 
SED beyond the peak of the inverse-Compton component (see Section~\ref{section:VHE}).
In the context of the discussion related to the potential 
variability of the magnetic field configuration (see above)
and the possible correlation with the hard X-ray flux (see Fig.~\ref{fig:timeline}) it is clear that the value of $0.8$ 
is too far off the range of values found with the VHE data to
be explained by variability. Putting aside the possibility of 
systematic uncertainties, there remains as explanation that the model
is missing an additional ingredient. 
Without attempting an additional fit here, a possible modification would be a more complicated model for the magnetic field strength 
and its variation in the nebula. An obvious modification could be
applied to the radial
dependence. The electrons contributing to the inverse Compton
emission detected with \textit{Fermi}-LAT are spatially more extended
than the electrons radiating at VHE energies. In this sense, the
value of $\alpha$ could smoothly change, steepening with increasing
distance from a value of $0.5$ to $1$. In this way, a more accurate
spectral measurement in the energy range covered with \textit{Fermi}-LAT could establish a more detailed picture of the magnetic field
distribution in the Crab Nebula. 
\section{Summary and conclusion}
The observed broadband emission and spatial extension of the Crab Nebula have been 
used to reconstruct the contemporaneous properties of the relativistic electrons, magnetic field, and dust content in the pulsar wind nebula. 
The main findings can be summarised in the following points:
\begin{enumerate}
    \item The  assembled observational 
    data for the Crab Nebula 
    is unique as it covers an extremely broad energy range
    with an accuracy that is generally not appreciated when
    attempting to model the data.
    \item The recent measurement of the spatial extension of the
    inverse Compton nebula provides insights into the
    magnetic field strength in the downstream region of the 
    termination shock. 
    \item A detailed radiative model has been developed to 
    estimate a minimal set of parameters related to the
    spatial and spectral distribution of electrons, dust,
    and the magnetic field under the assumption of radial symmetry.
    \item The best-fitting model provides a good fit
    to the data. The
    $\chi^2$ minimisation approach provides  
    an accurate estimate
    of all the underlying parameters without degeneracies. 
    \item The  VHE gamma-ray data provide the
    strongest constraint on  the magnetic field strength
    beyond the termination shock at $r_s$:
    $B(r)=B_0 (r/r_s)^{-\alpha}$ with $\check \alpha=0.51\pm0.03$
    and $\check B_0=(264\pm 9)~\mu\mathrm{G}$.
    \item The long-standing assumption introduced about 50 years ago that the downstream magnetic field increases in a kinetically dominated pulsar wind ($\sigma=0.003$)  is
    ruled out by the data. 
    \item The downstream flow is dominated by magnetic energy
    density up to $r\approx 1.3~r_s$. For larger distances,
    the particle energy density dominates until equipartition
    is recovered in the periphery of the PWN.
    \item VHE data taken during an increased hard X-ray flux state
    favour a value of $\check\alpha=0.29^{+0.03}_{-0.05}$, with an overall slightly worse goodness of fit. 
    \item The simultaneous fit of the synchrotron and
    dust emission  provides an estimate of the total dust mass, temperature,
    and spatial distribution in the nebula. The resulting dust mass is approximately five times larger than previous estimates 
    \citep{2019MNRAS.488..164D}. The larger dust mass  increases the required efficiency for dust formation.
    \item The production process of the two PeV photons detected
    from the Crab Nebula remains of uncertain origin. The simplest
    model would link the PeV photon production to inverse Compton emission
    of PeV electrons which radiate 80-800 MeV synchrotron emission. 
    Since the majority ($>75~\%$)of this energetic synchrotron emission has been found to be produced in a compact
    region \citep{2020A&A...638A.147Y} with fast variability,  the
    resulting constraint on the minimum (confining) magnetic field ($>500~\mu$G)
    is in conflict
    with the comparably low value of magnetic field ($\approx 85~\mu$G) required to fit  the inverse Compton and synchrotron
    emission simultaneously. 
\end{enumerate}

The results obtained here mainly rely  on the assumptions of a 
spherical and static system. 
In principle, the model can be extended
to axial symmetry and to include the effects of the flow in 
the radiative treatment. 
On the observational side, multi-instrument 
simultaneous VHE observations of the Crab Nebula
are crucial  to verify the effect of spectral variations. The upcoming spatially resolved observations of the X-ray polarisation 
of the Crab Nebula with the IXPE
satellite \citep{2016SPIE.9905E..17W} will provide a measurement of the degree of small-scale turbulence in the flow \citep{2017MNRAS.470.4066B}
which lead to a  reduced fraction  of polarised synchrotron 
emission.

\begin{acknowledgements}
      Part of this work was supported by the German
      \emph{Deut\-sche For\-schungs\-ge\-mein\-schaft, DFG\/} project
      number 284334853: HO 3305/2 and
      by  ANR-20-CE31-0010. This research has made use of NASA's Astrophysics Data System Bibliographic Services. The authors thank the developers of  matplotlib \citep{Hunter:2007} for making their code available on a free and open-source basis. DH  acknowledges support by the DFG under Germany’s Excellence Strategy – EXC 2121 „Quantum Universe“ – 390833306. We thank the referee for the very useful reports and for encouraging us to investigate the differences found between the \textit{Fermi}-LAT and VHE fit results. 
\end{acknowledgements}

  \bibliographystyle{aa} 
   \bibliography{biblio.bib} 

\begin{thebibliography}{105}
\expandafter\ifx\csname natexlab\endcsname\relax\def\natexlab#1{#1}\fi

\bibitem[{Abdalla {et~al.}(2020)Abdalla, Aharonian, Ait~Benkhali, Angüner,
  Arakawa, Arcaro, Armand, Backes, Barnard, Becherini, Becker~Tjus, Berge,
  Bernlöhr, Blackwell, Böttcher, Boisson, Bolmont, Bonnefoy, Bordas, Bregeon,
  Brun, Brun, Bryan, Büchele, Bulik, Bylund, Capasso, Caroff, Carosi,
  Casanova, Cerruti, Chakraborty, Chand, Chandra, Chaves, Chen, Colafrancesco,
  Condon, Davids, Deil, Devin, deWilt, Dirson, Djannati-Ataï, Dmytriiev,
  Donath, Doroshenko, Drury, Dyks, Egberts, Emery, Ernenwein, Eschbach, Fegan,
  Fiasson, Fontaine, Funk, Füßling, Gabici, Gallant, Gaté, Giavitto,
  Glawion, Glicenstein, Gottschall, Grondin, Hahn, Haupt, Heinzelmann, Henri,
  Hermann, Hinton, Hofmann, Hoischen, Holch, Holler, Horns, Huber, Iwasaki,
  Jacholkowska, Jamrozy, Jankowsky, Jankowsky, Jouvin, Jung-Richardt,
  Kastendieck, Katarzyński, Katsuragawa, Katz, Khangulyan, Khélifi, King,
  Klepser, Kluźniak, Komin, Kosack, Kraus, Lamanna, Lau, Lefaucheur, Lemière,
  Lemoine-Goumard, Lenain, Leser, Lohse, López-Coto, Lypova, Malyshev,
  Marandon, Marcowith, Mariaud, Martí-Devesa, Marx, Maurin, Meintjes,
  Mitchell, Moderski, Mohamed, Mohrmann, Moore, Moulin, Murach, Nakashima,
  de~Naurois, Ndiyavala, Niederwanger, Niemiec, Oakes, O’Brien, Odaka, Ohm,
  Ostrowski, Oya, Panter, Parsons, Perennes, Petrucci, Peyaud, Piel, Pita,
  Poireau, Priyana~Noel, Prokhorov, Prokoph, Pühlhofer, Punch, Quirrenbach,
  Raab, Rauth, Reimer, Reimer, Renaud, Rieger, Rinchiuso, Romoli, Rowell,
  Rudak, Ruiz-Velasco, Sahakian, Saito, Sanchez, Santangelo, Sasaki,
  Schlickeiser, Schüssler, Schulz, Schutte, Schwanke, Schwemmer,
  Seglar-Arroyo, Senniappan, Seyffert, Shafi, Shilon, Shiningayamwe, Simoni,
  Sinha, Sol, Specovius, Spir-Jacob, Stawarz, Steenkamp, Stegmann, Steppa,
  Takahashi, Tavernet, Tavernier, Taylor, Terrier, Tiziani, Tluczykont,
  Trichard, Tsirou, Tsuji, Tuffs, Uchiyama, van~der Walt, van Eldik, van
  Rensburg, van Soelen, Vasileiadis, Veh, Venter, Vincent, Vink, Voisin, Völk,
  Vuillaume, Wadiasingh, Wagner, Wagner, White, Wierzcholska, Yang, Yoneda,
  Zaborov, Zacharias, Zanin, Zdziarski, Zech, Ziegler, Zorn, Żywucka, \&
  {H.E.S.S. Collaboration}}]{abdalla_resolving_2020}
Abdalla, H., Aharonian, F., Ait~Benkhali, F., {et~al.} 2020, Nature Astronomy,
  4, 167

\bibitem[{{Abdo} {et~al.}(2011){Abdo}, {Ackermann}, {Ajello}, {Allafort},
  {Baldini}, {Ballet}, {Barbiellini}, {Bastieri}, {Bechtol}, {Bellazzini},
  {Berenji}, {Blandford}, {Bloom}, {Bonamente}, {Borgland}, {Bouvier},
  {Brandt}, {Bregeon}, {Brez}, {Brigida}, {Bruel}, {Buehler}, {Buson},
  {Caliandro}, {Cameron}, {Cannon}, {Caraveo}, {Casandjian}, {{\c{C}}elik},
  {Charles}, {Chekhtman}, {Cheung}, {Chiang}, {Ciprini}, {Claus},
  {Cohen-Tanugi}, {Costamante}, {Cutini}, {D'Ammando}, {Dermer}, {de Angelis},
  {de Luca}, {de Palma}, {Digel}, {do Couto e Silva}, {Drell}, {Drlica-Wagner},
  {Dubois}, {Dumora}, {Favuzzi}, {Fegan}, {Ferrara}, {Focke}, {Fortin},
  {Frailis}, {Fukazawa}, {Funk}, {Fusco}, {Gargano}, {Gasparrini}, {Gehrels},
  {Germani}, {Giglietto}, {Giordano}, {Giroletti}, {Glanzman}, {Godfrey},
  {Grenier}, {Grondin}, {Grove}, {Guiriec}, {Hadasch}, {Hanabata}, {Harding},
  {Hayashi}, {Hayashida}, {Hays}, {Horan}, {Itoh}, {J{\'o}hannesson},
  {Johnson}, {Johnson}, {Khangulyan}, {Kamae}, {Katagiri}, {Kataoka}, {Kerr},
  {Kn{\"o}dlseder}, {Kuss}, {Lande}, {Latronico}, {Lee}, {Lemoine-Goumard},
  {Longo}, {Loparco}, {Lubrano}, {Madejski}, {Makeev}, {Marelli}, {Mazziotta},
  {McEnery}, {Michelson}, {Mitthumsiri}, {Mizuno}, {Moiseev}, {Monte},
  {Monzani}, {Morselli}, {Moskalenko}, {Murgia}, {Nakamori}, {Naumann-Godo},
  {Nolan}, {Norris}, {Nuss}, {Ohsugi}, {Okumura}, {Omodei}, {Ormes}, {Ozaki},
  {Paneque}, {Parent}, {Pelassa}, {Pepe}, {Pesce-Rollins}, {Pierbattista},
  {Piron}, {Porter}, {Rain{\`o}}, {Rando}, {Ray}, {Razzano}, {Reimer},
  {Reimer}, {Reposeur}, {Ritz}, {Romani}, {Sadrozinski}, {Sanchez},
  {Parkinson}, {Scargle}, {Schalk}, {Sgr{\`o}}, {Siskind}, {Smith}, {Spandre},
  {Spinelli}, {Strickman}, {Suson}, {Takahashi}, {Takahashi}, {Tanaka},
  {Thayer}, {Thompson}, {Tibaldo}, {Torres}, {Tosti}, {Tramacere}, {Troja},
  {Uchiyama}, {Vandenbroucke}, {Vasileiou}, {Vianello}, {Vitale}, {Wang},
  {Wood}, {Yang}, \& {Ziegler}}]{2011Sci...331..739A}
{Abdo}, A.~A., {Ackermann}, M., {Ajello}, M., {et~al.} 2011, Science, 331, 739

\bibitem[{{Abdo} {et~al.}(2009){Abdo}, {Ackermann}, {Ajello}, {Atwood},
  {Axelsson}, {Baldini}, {Ballet}, {Barbiellini}, {Bastieri}, \&
  {Battelino}}]{abdo2009}
{Abdo}, A.~A., {Ackermann}, M., {Ajello}, M., {et~al.} 2009, \prl, 102, 181101

\bibitem[{{Abeysekara} {et~al.}(2019){Abeysekara}, {Albert}, {Alfaro},
  {Alvarez}, {Álvarez}, {Camacho}, {Arceo}, {Arteaga-Velázquez}, {Arunbabu},
  {Avila Rojas}, {Ayala Solares}, {Baghmanyan}, {Belmont-Moreno}, {BenZvi},
  {Brisbois}, {Caballero-Mora}, {Capistrán}, {Carramiñana}, {Casanova},
  {Cotti}, {Cotzomi}, {Coutiño de León}, {De la Fuente}, {de León},
  {Dichiara}, {Dingus}, {DuVernois}, {Díaz-Vélez}, {Ellsworth}, {Engel},
  {Espinoza}, {Fick}, {Fleischhack}, {Fraija}, {Galván-Gámez},
  {García-González}, {Garfias}, {González}, {Goodman}, {Harding},
  {Hernandez}, {Hinton}, {Hona}, {Hueyotl-Zahuantitla}, {Hui}, {Hüntemeyer},
  {Iriarte}, {Jardin-Blicq}, {Joshi}, {Kaufmann}, {Kieda}, {Lara}, {Lee},
  {León Vargas}, {Linnemann}, {Longinotti}, {Luis-Raya}, {Lundeen}, {Malone},
  {Marinelli}, {Martinez}, {Martinez-Castellanos}, {Martínez-Castro},
  {Martínez-Huerta}, {Matthews}, {Miranda-Romagnoli}, {Morales-Soto},
  {Moreno}, {Mostafá}, {Nayerhoda}, {Nellen}, {Newbold}, {Nisa},
  {Noriega-Papaqui}, {Peisker}, {Pérez-Pérez}, {Pretz}, {Ren}, {Rho},
  {Rivière}, {Rosa-González}, {Rosenberg}, {Ruiz-Velasco}, {Salazar}, {Salesa
  Greus}, {Sandoval}, {Schneider}, {Schoorlemmer}, {Seglar Arroyo}, {Sinnis},
  {Smith}, {Springer}, {Surajbali}, {Tabachnick}, {Tanner}, {Tibolla},
  {Tollefson}, {Torres}, {Weisgarber}, {Westerhoff}, {Wood}, {Yapici},
  {Zepeda}, {Zhou}, \& {HAWC Collaboration}}]{hawc}
{Abeysekara}, A.~U., {Albert}, A., {Alfaro}, R., {et~al.} 2019, \apj, 881, 134

\bibitem[{{Ackermann} {et~al.}(2012){Ackermann}, {Ajello}, {Allafort},
  {Atwood}, {Axelsson}, {Baldini}, {Barbiellini}, {Bastieri}, {Bechtol},
  {Bellazzini}, {Berenji}, {Bloom}, {Bonamente}, {Borgland }, {Bouvier},
  {Bregeon}, {Brez}, {Brigida}, {Bruel}, {Buehler}, {Buson}, {Caliandro},
  {Cameron}, {Caraveo}, {Casand jian}, {Cecchi}, {Charles}, {Chekhtman},
  {Chiang}, {Ciprini}, {Claus}, {Cohen-Tanugi}, {Cutini}, {de Palma}, {Dermer},
  {Digel}, {Do Couto E Silva}, {Drell}, {Drlica-Wagner}, {Dubois}, {Enoto},
  {Falletti}, {Favuzzi}, {Fegan}, {Focke}, {Fortin}, {Fukazawa}, {Funk},
  {Fusco}, {Gargano}, {Gehrels}, {Germani}, {Giglietto}, {Giordano},
  {Giroletti}, {Glanzman}, {Godfrey}, {Grenier}, {Grove}, {Guiriec}, {Hadasch},
  {Hayashida}, {Hays}, {Hughes}, {Jóhannesson}, {Johnson}, {Johnson}, {Kamae},
  {Katagiri}, {Kataoka}, {Knödlseder}, {Kuss}, {Lande}, {Latronico}, {Lee},
  {Longo}, {Loparco}, {Lovellette}, {Lubrano}, {Madejski}, {Mazziotta},
  {McEnery}, {Michelson}, {Mizuno}, {Moiseev}, {Monte}, {Monzani}, {Morselli},
  {Moskalenko}, {Murgia}, {Nakamori}, {Naumann-Godo}, {Nolan}, {Norris},
  {Nuss}, {Ohsugi}, {Okumura}, {Omodei}, {Orlando}, {Ormes}, {Ozaki},
  {Paneque}, {Panetta}, {Parent}, {Pesce-Rollins}, {Pierbattista}, {Piron},
  {Rainò}, {Rando}, {Razzano}, {Reimer}, {Reimer}, {Reposeur}, {Ritz},
  {Rochester}, {Sgrò}, {Siskind}, {Smith}, {Spand re}, {Spinelli}, {Suson},
  {Takahashi}, {Tanaka}, {Thayer}, {Thayer}, {Thompson}, {Tibaldo}, {Tosti},
  {Troja}, {Usher}, {Vandenbroucke}, {Vasileiou}, {Vianello}, {Vilchez},
  {Vitale}, {Waite}, {Wang}, {Winer}, {Wood}, {Yang}, \&
  {Zimmer}}]{ackermann2012}
{Ackermann}, M., {Ajello}, M., {Allafort}, A., {et~al.} 2012, Astroparticle
  Physics, 35, 346–353

\bibitem[{{Aharonian} {et~al.}(2004){Aharonian}, {Akhperjanian}, {Beilicke},
  {Bernlöhr}, {Börst}, {Bojahr}, {Bolz}, {Coarasa}, {Contreras}, {Cortina},
  {Denninghoff}, {Fonseca}, {Girma}, {Götting}, {Heinzelmann}, {Hermann},
  {Heusler}, {Hofmann}, {Horns}, {Jung}, {Kankanyan}, {Kestel}, {Kohnle},
  {Konopelko}, {Kranich}, {Lampeitl}, {Lopez}, {Lorenz}, {Lucarelli}, {Mang},
  {Mazin}, {Meyer}, {Mirzoyan}, {Moralejo}, {Oña-Wilhelmi}, {Panter},
  {Plyasheshnikov}, {Pühlhofer}, {de los Reyes}, {Rhode}, {Ripken}, {Rowell},
  {Sahakian}, {Samorski}, {Schilling}, {Siems}, {Sobzynska}, {Stamm},
  {Tluczykont}, {Vitale}, {Völk}, {Wiedner}, \& {Wittek}}]{Aharonianetal2004}
{Aharonian}, F., {Akhperjanian}, A., {Beilicke}, M., {et~al.} 2004, \apj, 614,
  897–913

\bibitem[{{Aharonian} {et~al.}(2006){Aharonian}, {Akhperjanian}, {Bazer-Bachi},
  {Beilicke}, {Benbow}, {Berge}, {Bernl{\"o}hr}, {Boisson}, {Bolz}, {Borrel},
  {Braun}, {Breitling}, {Brown}, {B{\"u}hler}, {B{\"u}sching}, {Carrigan},
  {Chadwick}, {Chounet}, {Cornils}, {Costamante}, {Degrange}, {Dickinson},
  {Djannati-Ata{\"\i}}, {O'C. Drury}, {Dubus}, {Egberts}, {Emmanoulopoulos},
  {Espigat}, {Feinstein}, {Ferrero}, {Fiasson}, {Fontaine}, {Funk}, {Funk},
  {Gallant}, {Giebels}, {Glicenstein}, {Goret}, {Hadjichristidis}, {Hauser},
  {Hauser}, {Heinzelmann}, {Henri}, {Hermann}, {Hinton}, {Hofmann}, {Holleran},
  {Horns}, {Jacholkowska}, {de Jager}, {Kh{\'e}lifi}, {Komin}, {Konopelko},
  {Kosack}, {Latham}, {Le Gallou}, {Lemi{\`e}re}, {Lemoine-Goumard}, {Lohse},
  {Martin}, {Martineau-Huynh}, {Marcowith}, {Masterson}, {McComb}, {de
  Naurois}, {Nedbal}, {Nolan}, {Noutsos}, {Orford}, {Osborne}, {Ouchrif},
  {Panter}, {Pelletier}, {Pita}, {P{\"u}hlhofer}, {Punch}, {Raubenheimer},
  {Raue}, {Rayner}, {Reimer}, {Reimer}, {Ripken}, {Rob}, {Rolland}, {Rowell},
  {Sahakian}, {Saug{\'e}}, {Schlenker}, {Schlickeiser}, {Schwanke}, {Sol},
  {Spangler}, {Spanier}, {Steenkamp}, {Stegmann}, {Superina}, {Tavernet},
  {Terrier}, {Th{\'e}oret}, {Tluczykont}, {van Eldik}, {Vasileiadis}, {Venter},
  {Vincent}, {V{\"o}lk}, {Wagner}, \& {Ward}}]{2006A&A...457..899A}
{Aharonian}, F., {Akhperjanian}, A.~G., {Bazer-Bachi}, A.~R., {et~al.} 2006,
  \aap, 457, 899

\bibitem[{Aharonian {et~al.}(2010)Aharonian, Kelner, \&
  Prosekin}]{Aharonian_2010}
Aharonian, F.~A., Kelner, S.~R., \& Prosekin, A.~Y. 2010, Physical Review D, 82

\bibitem[{{Aleksić} {et~al.}(2015){Aleksić}, {Ansoldi}, {Antonelli},
  {Antoranz}, {Babic}, {Bangale}, {Barrio}, {Becerra González}, {Bednarek},
  {Bernardini}, {Biasuzzi}, {Biland}, {Blanch}, {Bonnefoy}, {Bonnoli},
  {Borracci}, {Bretz}, {Carmona}, {Carosi}, {Colin}, {Colombo}, {Contreras},
  {Cortina}, {Covino}, {Da Vela}, {Dazzi}, {De Angelis}, {De Caneva}, {De
  Lotto}, {de Oña Wilhelmi}, {Delgado Mendez}, {Doert}, {Dominis Prester},
  {Dorner}, {Doro}, {Einecke}, {Eisenacher}, {Elsaesser}, {Fonseca}, {Font},
  {Frantzen}, {Fruck}, {Galindo}, {García López}, {Garczarczyk}, {Garrido
  Terrats}, {Gaug}, {Godinović}, {González Muñoz}, {Gozzini}, {Hadasch},
  {Hanabata}, {Hayashida}, {Herrera}, {Hildebrand}, {Hose}, {Hrupec}, {Idec},
  {Kadenius}, {Kellermann}, {Kodani}, {Konno}, {Krause}, {Kubo}, {Kushida}, {La
  Barbera}, {Lelas}, {Lewandowska}, {Lindfors}, {Lombardi}, {López},
  {López-Coto}, {López-Oramas}, {Lorenz}, {Lozano}, {Makariev}, {Mallot},
  {Maneva}, {Mankuzhiyil}, {Mannheim}, {Maraschi}, {Marcote}, {Mariotti},
  {Martínez}, {Mazin}, {Menzel}, {Miranda}, {Mirzoyan}, {Moralejo},
  {Munar-Adrover}, {Nakajima}, {Niedzwiecki}, {Nilsson}, {Nishijima}, {Noda},
  {Nowak}, {Orito}, {Overkemping}, {Paiano}, {Palatiello}, {Paneque},
  {Paoletti}, {Paredes}, {Paredes-Fortuny}, {Persic}, {Prada Moroni},
  {Prandini}, {Preziuso}, {Puljak}, {Reinthal}, {Rhode}, {Ribó}, {Rico},
  {Rodriguez Garcia}, {Rügamer}, {Saggion}, {Saito}, {Saito}, {Satalecka},
  {Scalzotto}, {Scapin}, {Schultz}, {Schweizer}, {Shore}, {Sillanpää},
  {Sitarek}, {Snidaric}, {Sobczynska}, {Spanier}, {Stamatescu}, {Stamerra},
  {Steinbring}, {Storz}, {Strzys}, {Takalo}, {Takami}, {Tavecchio}, {Temnikov},
  {Terzić}, {Tescaro}, {Teshima}, {Thaele}, {Tibolla}, {Torres}, {Toyama},
  {Treves}, {Uellenbeck}, {Vogler}, {Wagner}, {Zanin}, {Horns}, {Martín}, \&
  {Meyer}}]{MagicCrab}
{Aleksić}, J., {Ansoldi}, S., {Antonelli}, L.~A., {et~al.} 2015, Journal of
  High Energy Astrophysics, 5, 30–38

\bibitem[{{Aliu} {et~al.}(2014){Aliu}, {Archambault}, {Aune}, {Benbow},
  {Berger}, {Bird}, {Bouvier}, {Buckley}, {Bugaev}, {Byrum}, {Cerruti}, {Chen},
  {Ciupik}, {Connolly}, {Cui}, {Dumm}, {Errando}, {Falcone}, {Federici},
  {Feng}, {Finley}, {Fortin}, {Fortson}, {Furniss}, {Galante}, {Gillanders},
  {Griffin}, {Griffiths}, {Grube}, {Gyuk}, {Hanna}, {Holder}, {Hughes},
  {Humensky}, {Kaaret}, {Kertzman}, {Khassen}, {Kieda}, {Krennrich}, {Kumar},
  {Lang}, {Lyutikov}, {Maier}, {McArthur}, {McCann}, {Meagher}, {Millis},
  {Moriarty}, {Mukherjee}, {O'Faol{\'a}in de Bhr{\'o}ithe}, {Ong}, {Otte},
  {Park}, {Perkins}, {Pohl}, {Popkow}, {Quinn}, {Ragan}, {Rajotte}, {Reyes},
  {Reynolds}, {Richards}, {Roache}, {Sembroski}, {Sheidaei}, {Smith},
  {Staszak}, {Telezhinsky}, {Theiling}, {Tucci}, {Tyler}, {Varlotta}, {Wakely},
  {Weekes}, {Weinstein}, {Welsing}, {Williams}, {Zajczyk}, \&
  {Zitzer}}]{2014ApJ...781L..11A}
{Aliu}, E., {Archambault}, S., {Aune}, T., {et~al.} 2014, \apjl, 781, L11

\bibitem[{Amenomori {et~al.}(2019)Amenomori, Bao, Bi, Chen, Chen, Chen, Chen,
  Chen, Cirennima, Cui, Danzengluobu, Ding, Fang, Fang, Feng, Feng, Feng, Gao,
  Gou, Guo, He, He, Hibino, Hotta, Hu, Hu, Huang, Jia, Jiang, Jin, Kajino,
  Kasahara, Katayose, Kato, Kato, Kawata, Kozai, Labaciren, Le, Li, Li, Li,
  Lin, Liu, Liu, Liu, Liu, Lou, Lu, Meng, Mitsui, Munakata, Nakamura, Nanjo,
  Nishizawa, Ohnishi, Ohta, Ozawa, Qian, Qu, Saito, Sakata, Sako, Sengoku,
  Shao, Shibata, Shiomi, Sugimoto, Takita, Tan, Tateyama, Torii, Tsuchiya, Udo,
  Wang, Wu, Xue, Yagisawa, Yamamoto, Yang, Yuan, Zhai, Zhang, Zhang, Zhang,
  Zhang, Zhang, Zhang, Zhang, Zhaxisangzhu, \& Zhou}]{PhysRevLett.123.051101}
Amenomori, M., Bao, Y.~W., Bi, X.~J., {et~al.} 2019, Phys. Rev. Lett., 123,
  051101

\bibitem[{{Atoyan} \& {Aharonian}(1996)}]{AA1996}
{Atoyan}, A.~M. \& {Aharonian}, F.~A. 1996, \mnras, 278, 525–541

\bibitem[{{Band} {et~al.}(1993){Band}, {Matteson}, {Ford}, {Schaefer},
  {Palmer}, {Teegarden}, {Cline}, {Briggs}, {Paciesas}, {Pendleton}, {Fishman},
  {Kouveliotou}, {Meegan}, {Wilson}, \& {Lestrade}}]{1993ApJ...413..281B}
{Band}, D., {Matteson}, J., {Ford}, L., {et~al.} 1993, \apj, 413, 281

\bibitem[{{Bietenholz} {et~al.}(2004){Bietenholz}, {Hester}, {Frail}, \&
  {Bartel}}]{Bietenholz2004}
{Bietenholz}, M.~F., {Hester}, J.~J., {Frail}, D.~A., \& {Bartel}, N. 2004,
  \apj, 615, 794–804

\bibitem[{{Bietenholz} \& {Kronberg}(1990)}]{bietenholz1990}
{Bietenholz}, M.~F. \& {Kronberg}, P.~P. 1990, \apjl, 357, L13

\bibitem[{{Blair} {et~al.}(1992){Blair}, {Long}, {Vancura}, {Bowers}, {Conger},
  {Davidsen}, {Kriss}, \& {Henry}}]{1992ApJ...399..611B}
{Blair}, W.~P., {Long}, K.~S., {Vancura}, O., {et~al.} 1992, \apj, 399, 611

\bibitem[{Bohren \& Huffman(1998)}]{Bohren1998}
Bohren, C. \& Huffman, D.~R. 1998, Absorption and Scattering of Light by Small
  Particles, ed. C.~Bohren \& D.~R. Huffman (Wiley Science Paperback Series)

\bibitem[{{Bucciantini} {et~al.}(2017){Bucciantini}, {Bandiera}, {Olmi}, \&
  {Del Zanna}}]{2017MNRAS.470.4066B}
{Bucciantini}, N., {Bandiera}, R., {Olmi}, B., \& {Del Zanna}, L. 2017, \mnras,
  470, 4066

\bibitem[{{Buehler} {et~al.}(2012){Buehler}, {Scargle}, {Blandford}, {Baldini},
  {Baring}, {Belfiore}, {Charles}, {Chiang}, {D'Ammando}, \&
  {Dermer}}]{Buehler2012}
{Buehler}, R., {Scargle}, J.~D., {Blandford}, R.~D., {et~al.} 2012, \apj, 749,
  26

\bibitem[{{Crusius} \& {Schlickeiser}(1986)}]{1986A&A...164L..16C}
{Crusius}, A. \& {Schlickeiser}, R. 1986, \aap, 164, L16

\bibitem[{{de Jager} \& {Harding}(1992)}]{JH92}
{de Jager}, O.~C. \& {Harding}, A.~K. 1992, \apj, 396, 161–172

\bibitem[{{De Looze} {et~al.}(2019){De Looze}, {Barlow}, {Bandiera}, {Bevan},
  {Bietenholz}, {Chawner}, {Gomez}, {Matsuura}, {Priestley}, \&
  {Wesson}}]{2019MNRAS.488..164D}
{De Looze}, I., {Barlow}, M.~J., {Bandiera}, R., {et~al.} 2019, \mnras, 488,
  164

\bibitem[{{Feng} {et~al.}(2020){Feng}, {Li}, {Long}, {Bellazzini}, {Costa},
  {Wu}, {Huang}, {Jiang}, {Minuti}, {Wang}, {Xu}, {Yang}, {Baldini}, {Citraro},
  {Nasimi}, {Soffitta}, {Muleri}, {Jung}, {Yu}, {Jin}, {Zeng}, {An}, {Brez},
  {Latronico}, {Sgro}, {Spandre}, \& {Pinchera}}]{2020NatAs...4..511F}
{Feng}, H., {Li}, H., {Long}, X., {et~al.} 2020, Nature Astronomy, 4, 511

\bibitem[{{Gomez} {et~al.}(2012){Gomez}, {Krause}, {Barlow}, {Swinyard},
  {Owen}, {Clark}, {Matsuura}, {Gomez}, {Rho}, \& {Besel}}]{GomezDust}
{Gomez}, H.~L., {Krause}, O., {Barlow}, M.~J., {et~al.} 2012, \apj, 760, 96

\bibitem[{{Gould} \& {Schr{\'e}der}(1967)}]{1967PhRv..155.1404G}
{Gould}, R.~J. \& {Schr{\'e}der}, G.~P. 1967, Physical Review, 155, 1404

\bibitem[{{Green} {et~al.}(2019){Green}, {Schlafly}, {Zucker}, {Speagle}, \&
  {Finkbeiner}}]{2019ApJ...887...93G}
{Green}, G.~M., {Schlafly}, E., {Zucker}, C., {Speagle}, J.~S., \&
  {Finkbeiner}, D. 2019, \apj, 887, 93

\bibitem[{{Greiveldinger} \& {Aschenbach}(1999)}]{1999ApJ...510..305G}
{Greiveldinger}, C. \& {Aschenbach}, B. 1999, \apj, 510, 305

\bibitem[{{H.~E.~S.~S. Collaboration}(2020)}]{2020NatAs...4..167H}
{H.~E.~S.~S. Collaboration}. 2020, Nature Astronomy, 4, 167

\bibitem[{{H.~E.~S.~S. Collaboration} {et~al.}(2014){H.~E.~S.~S.
  Collaboration}, {Abramowski}, {Aharonian}, {Ait Benkhali}, {Akhperjanian},
  {Ang{\"u}ner}, {Anton}, {Balenderan}, {Balzer}, {Barnacka}, {Becherini},
  {Becker Tjus}, {Bernl{\"o}hr}, {Birsin}, {Bissaldi}, {Biteau},
  {B{\"o}ttcher}, {Boisson}, {Bolmont}, {Bordas}, {Brucker}, {Brun}, {Brun},
  {Bulik}, {Carrigan}, {Casanova}, {Cerruti}, {Chadwick}, {Chalme-Calvet},
  {Chaves}, {Cheesebrough}, {Chr{\'e}tien}, {Colafrancesco}, {Cologna},
  {Conrad}, {Couturier}, {Cui}, {Dalton}, {Daniel}, {Davids}, {Degrange},
  {Deil}, {deWilt}, {Dickinson}, {Djannati-Ata{\"\i}}, {Domainko}, {Drury},
  {Dubus}, {Dutson}, {Dyks}, {Dyrda}, {Edwards}, {Egberts}, {Eger}, {Espigat},
  {Farnier}, {Fegan}, {Feinstein}, {Fernandes}, {Fernandez}, {Fiasson},
  {Fontaine}, {F{\"o}rster}, {F{\"u}{\ss}ling}, {Gajdus}, {Gallant},
  {Garrigoux}, {Giavitto}, {Giebels}, {Glicenstein}, {Grondin},
  {Grudzi{\'n}ska}, {H{\"a}ffner}, {Hahn}, {Harris}, {Heinzelmann}, {Henri},
  {Hermann}, {Hervet}, {Hillert}, {Hinton}, {Hofmann}, {Hofverberg}, {Holler},
  {Horns}, {Jacholkowska}, {Jahn}, {Jamrozy}, {Janiak}, {Jankowsky}, {Jung},
  {Kastendieck}, {Katarzy{\'n}ski}, {Katz}, {Kaufmann}, {Kh{\'e}lifi},
  {Kieffer}, {Klepser}, {Klochkov}, {Klu{\'z}niak}, {Kneiske}, {Kolitzus},
  {Komin}, {Kosack}, {Krakau}, {Krayzel}, {Kr{\"u}ger}, {Laffon}, {Lamanna},
  {Lefaucheur}, {Lemi{\`e}re}, {Lemoine-Goumard}, {Lenain}, {Lennarz}, {Lohse},
  {Lopatin}, {Lu}, {Marandon}, {Marcowith}, {Marx}, {Maurin}, {Maxted},
  {Mayer}, {McComb}, {M{\'e}hault}, {Meintjes}, {Menzler}, {Meyer}, {Moderski},
  {Mohamed}, {Moulin}, {Murach}, {Naumann}, {de Naurois}, {Niemiec}, {Nolan},
  {Oakes}, {Ohm}, {de O{\~n}a Wilhelmi}, {Opitz}, {Ostrowski}, {Oya}, {Panter},
  {Parsons}, {Paz Arribas}, {Pekeur}, {Pelletier}, {Perez}, {Petrucci},
  {Peyaud}, {Pita}, {Poon}, {P{\"u}hlhofer}, {Punch}, {Quirrenbach}, {Raab},
  {Raue}, {Reimer}, {Reimer}, {Renaud}, {de los Reyes}, {Rieger}, {Rob},
  {Romoli}, {Rosier-Lees}, {Rowell}, {Rudak}, {Rulten}, {Sahakian}, {Sanchez},
  {Santangelo}, {Schlickeiser}, {Sch{\"u}ssler}, {Schulz}, {Schwanke},
  {Schwarzburg}, {Schwemmer}, {Sol}, {Spengler}, {Spies}, {Stawarz},
  {Steenkamp}, {Stegmann}, {Stinzing}, {Stycz}, {Sushch}, {Szostek},
  {Tavernet}, {Tavernier}, {Taylor}, {Terrier}, {Tluczykont}, {Trichard},
  {Valerius}, {van Eldik}, {van Soelen}, {Vasileiadis}, {Venter}, {Viana},
  {Vincent}, {V{\"o}lk}, {Volpe}, {Vorster}, {Vuillaume}, {Wagner}, {Wagner},
  {Ward}, {Weidinger}, {Weitzel}, {White}, {Wierzcholska}, {Willmann},
  {W{\"o}rnlein}, {Wouters}, {Zabalza}, {Zacharias}, {Zajczyk}, {Zdziarski},
  {Zech}, \& {Zechlin}}]{2014A&A...562L...4H}
{H.~E.~S.~S. Collaboration}, {Abramowski}, A., {Aharonian}, F., {et~al.} 2014,
  \aap, 562, L4

\bibitem[{{Hennessy} {et~al.}(1992){Hennessy}, {O'Connell}, {Cheng}, {Bohlin},
  {Collins}, {Gull}, {Hintzen}, {Isensee}, {Landsman}, {Roberts}, {Smith},
  {Smith}, \& {Stecher}}]{1992ApJ...395L..13H}
{Hennessy}, G.~S., {O'Connell}, R.~W., {Cheng}, K.~P., {et~al.} 1992, \apjl,
  395, L13

\bibitem[{{Hickson} \& {van den Bergh}(1990)}]{1990ApJ...365..224H}
{Hickson}, P. \& {van den Bergh}, S. 1990, \apj, 365, 224

\bibitem[{{Hillas} {et~al.}(1998){Hillas}, {Akerlof}, {Biller}, {Buckley},
  {Carter-Lewis}, {Catanese}, {Cawley}, {Fegan}, {Finley}, {Gaidos},
  {Krennrich}, {Lamb}, {Lang}, {Mohanty}, {Punch}, {Reynolds}, {Rodgers},
  {Rose}, {Rovero}, {Schubnell}, {Sembroski}, {Vacanti}, {Weekes}, {West}, \&
  {Zweerink}}]{Hillas}
{Hillas}, A.~M., {Akerlof}, C.~W., {Biller}, S.~D., {et~al.} 1998, \apj, 503,
  744–759

\bibitem[{Hodge(2012)}]{wilson-hodge2012}
Hodge, C. A.~W. 2012, {When a Standard Candle Flickers: Crab Nebula Variations
  in Hard X-rays}, Tech. rep., Nasa Marshall Space Flight Center Huntsville

\bibitem[{Holler(2014)}]{holler_hess}
Holler, M. 2014, PhD thesis, Universität Potsdam

\bibitem[{{Holler} {et~al.}(2015){Holler}, {Berge}, {van Eldik}, {Lenain},
  {Marandon}, {Murach}, {de Naurois}, {Parsons}, {Prokoph}, \&
  {Zaborov}}]{hess_data}
{Holler}, M., {Berge}, D., {van Eldik}, C., {et~al.} 2015, arXiv e-prints,
  arXiv:1509.02902

\bibitem[{{Horns} {et~al.}(2006){Horns}, {Aharonian}, {Santangelo}, {Hoffmann},
  \& {Masterson}}]{2006A&A...451L..51H}
{Horns}, D., {Aharonian}, F., {Santangelo}, A., {Hoffmann}, A.~I.~D., \&
  {Masterson}, C. 2006, \aap, 451, L51

\bibitem[{Hunter(2007)}]{Hunter:2007}
Hunter, J.~D. 2007, Computing In Science \& Engineering, 9, 90

\bibitem[{Jones(1968)}]{PhysRev.167.1159}
Jones, F.~C. 1968, Phys. Rev., 167, 1159

\bibitem[{{Jourdain} \& {Roques}(2009)}]{2009ApJ...704...17J}
{Jourdain}, E. \& {Roques}, J.~P. 2009, \apj, 704, 17

\bibitem[{{Jourdain} \& {Roques}(2020)}]{2020ApJ...899..131J}
{Jourdain}, E. \& {Roques}, J.~P. 2020, \apj, 899, 131

\bibitem[{{Kennel} \& {Coroniti}(1984{\natexlab{a}})}]{KennelCoroniti}
{Kennel}, C.~F. \& {Coroniti}, F.~V. 1984{\natexlab{a}}, \apj, 283, 694–709

\bibitem[{{Kennel} \& {Coroniti}(1984{\natexlab{b}})}]{1984ApJ...283..694K}
{Kennel}, C.~F. \& {Coroniti}, F.~V. 1984{\natexlab{b}}, \apj, 283, 694

\bibitem[{{Khangulyan} {et~al.}(2020){Khangulyan}, {Arakawa}, \&
  {Aharonian}}]{khangulyan_detection_2020}
{Khangulyan}, D., {Arakawa}, M., \& {Aharonian}, F. 2020, \mnras, 491, 3217

\bibitem[{{Kirsch} {et~al.}(2005){Kirsch}, {Briel}, {Burrows}, {Campana},
  {Cusumano}, {Ebisawa}, {Freyberg}, {Guainazzi}, {Haberl}, {Jahoda},
  {Kaastra}, {Kretschmar}, {Larsson}, {Lubi{\'n}ski}, {Mori}, {Plucinsky},
  {Pollock}, {Rothschild}, {Sembay}, {Wilms}, \&
  {Yamamoto}}]{2005SPIE.5898...22K}
{Kirsch}, M.~G., {Briel}, U.~G., {Burrows}, D., {et~al.} 2005, in Society of
  Photo-Optical Instrumentation Engineers (SPIE) Conference Series, Vol. 5898,
  UV, X-Ray, and Gamma-Ray Space Instrumentation for Astronomy XIV, ed.
  O.~H.~W. {Siegmund}, 22--33

\bibitem[{{Komissarov} \& {Lyubarsky}(2003)}]{2003MNRAS.344L..93K}
{Komissarov}, S.~S. \& {Lyubarsky}, Y.~E. 2003, \mnras, 344, L93

\bibitem[{{Krimm} {et~al.}(2013){Krimm}, {Holland}, {Corbet}, {Pearlman},
  {Romano}, {Kennea}, {Bloom}, {Barthelmy}, {Baumgartner}, {Cummings},
  {Gehrels}, {Lien}, {Markwardt}, {Palmer}, {Sakamoto}, {Stamatikos}, \&
  {Ukwatta}}]{2013ApJS..209...14K}
{Krimm}, H.~A., {Holland}, S.~T., {Corbet}, R.~H.~D., {et~al.} 2013, \apjs,
  209, 14

\bibitem[{{Kuiper} {et~al.}(2001){Kuiper}, {Hermsen}, {Cusumano}, {Diehl},
  {Sch{\"o}nfelder}, {Strong}, {Bennett}, \& {McConnell}}]{2001A&A...378..918K}
{Kuiper}, L., {Hermsen}, W., {Cusumano}, G., {et~al.} 2001, \aap, 378, 918

\bibitem[{{LHAASO Collaboration} {et~al.}(2021){LHAASO Collaboration}, {Cao},
  {Aharonian}, {An}, {Axikegu}, {Bai}, {Bai}, {Bao}, {Bastieri}, {Bi}, {Bi},
  {Cai}, {Cai}, {Cao}, {Chang}, {Chang}, {Chen}, {Chen}, {Chen}, {Chen},
  {Chen}, {Chen}, {Chen}, {Chen}, {Chen}, {Chen}, {Chen}, {Chen}, {Chen},
  {Chen}, {Cheng}, {Cheng}, {Cui}, {Cui}, {Cui}, {D'Ettorre Piazzoli}, {Dai},
  {Dai}, {Dai}, {Danzengluobu}, {Della Volpe}, {Dong}, {Duan}, {Fan}, {Fan},
  {Fan}, {Fang}, {Fang}, {Feng}, {Feng}, {Feng}, {Feng}, {Gao}, {Gao}, {Gao},
  {Gao}, {Gao}, {Ge}, {Geng}, {Gong}, {Gou}, {Gu}, {Guo}, {Guo}, {Guo}, {Guo},
  {Guo}, {Han}, {He}, {He}, {He}, {He}, {He}, {He}, {Heller}, {Hor}, {Hou},
  {Hou}, {Hu}, {Hu}, {Hu}, {Hu}, {Huang}, {Huang}, {Huang}, {Huang}, {Huang},
  {Huang}, {Ji}, {Ji}, {Jia}, {Jiang}, {Jiang}, {Jin}, {Ke}, {Kuleshov},
  {Levochkin}, {Li}, {Li}, {Li}, {Li}, {Li}, {Li}, {Li}, {Li}, {Li}, {Li},
  {Li}, {Li}, {Li}, {Li}, {Li}, {Li}, {Li}, {Li}, {Liang}, {Liang}, {Lin},
  {Liu}, {Liu}, {Liu}, {Liu}, {Liu}, {Liu}, {Liu}, {Liu}, {Liu}, {Liu}, {Liu},
  {Liu}, {Liu}, {Liu}, {Liu}, {Liu}, {Long}, {Lu}, {Lv}, {Ma}, {Ma}, {Ma},
  {Mao}, {Masood}, {Min}, {Mitthumsiri}, {Montaruli}, {Nan}, {Pang},
  {Pattarakijwanich}, {Pei}, {Qi}, {Qi}, {Qiao}, {Qin}, {Ruffolo}, {Rulev},
  {Saiz}, {Shao}, {Shchegolev}, {Sheng}, {Shi}, {Song}, {Stenkin}, {Stepanov},
  {Su}, {Sun}, {Sun}, {Sun}, {Tam}, {Tang}, {Tian}, {Wang}, {Wang}, {Wang},
  {Wang}, {Wang}, {Wang}, {Wang}, {Wang}, {Wang}, {Wang}, {Wang}, {Wang},
  {Wang}, {Wang}, {Wang}, {Wang}, {Wang}, {Wang}, {Wang}, {Wang}, {Wang},
  {Wang}, {Wei}, {Wei}, {Wei}, {Wen}, {Wu}, {Wu}, {Wu}, {Wu}, {Wu}, {Xi},
  {Xia}, {Xia}, {Xiang}, {Xiao}, {Xiao}, {Xiao}, {Xin}, {Xin}, {Xing}, {Xu},
  {Xu}, {Xue}, {Yan}, {Yan}, {Yang}, {Yang}, {Yang}, {Yang}, {Yang}, {Yang},
  {Yang}, {Yao}, {Yao}, {Ye}, {Yin}, {Yin}, {You}, {You}, {Yu}, {Yuan}, {Zeng},
  {Zeng}, {Zeng}, {Zeng}, {Zha}, {Zhai}, {Zhang}, {Zhang}, {Zhang}, {Zhang},
  {Zhang}, {Zhang}, {Zhang}, {Zhang}, {Zhang}, {Zhang}, {Zhang}, {Zhang},
  {Zhang}, {Zhang}, {Zhang}, {Zhang}, {Zhang}, {Zhang}, {Zhang}, {Zhao},
  {Zhao}, {Zhao}, {Zhao}, {Zhao}, {Zheng}, {Zheng}, {Zhou}, {Zhou}, {Zhou},
  {Zhou}, {Zhou}, {Zhou}, {Zhu}, {Zhu}, {Zhu}, {Zhu}, \&
  {Zuo}}]{2021Sci...373..425L}
{LHAASO Collaboration}, {Cao}, Z., {Aharonian}, F., {et~al.} 2021, Science,
  373, 425

\bibitem[{{Ling} \& {Wheaton}(2003)}]{2003ApJ...598..334L}
{Ling}, J.~C. \& {Wheaton}, W.~A. 2003, \apj, 598, 334

\bibitem[{{Liu} \& {Wang}(2021)}]{2021ApJ...922..221L}
{Liu}, R.-Y. \& {Wang}, X.-Y. 2021, \apj, 922, 221

\bibitem[{{Lobanov} {et~al.}(2011){Lobanov}, {Horns}, \&
  {Muxlow}}]{2011A&A...533A..10L}
{Lobanov}, A.~P., {Horns}, D., \& {Muxlow}, T.~W.~B. 2011, \aap, 533, A10

\bibitem[{{Luo} {et~al.}(2020){Luo}, {Lyutikov}, {Temim}, \&
  {Comisso}}]{2020ApJ...896..147L}
{Luo}, Y., {Lyutikov}, M., {Temim}, T., \& {Comisso}, L. 2020, \apj, 896, 147

\bibitem[{{Lyutikov} {et~al.}(2019){Lyutikov}, {Temim}, {Komissarov}, {Slane},
  {Sironi}, \& {Comisso}}]{2019MNRAS.489.2403L}
{Lyutikov}, M., {Temim}, T., {Komissarov}, S., {et~al.} 2019, \mnras, 489, 2403

\bibitem[{{Madsen} {et~al.}(2021){Madsen}, {Burwitz}, {Forster}, {Grant},
  {Guainazzi}, {Kashyap}, {Marshall}, {Miller}, {Natalucci}, {Plucinsky}, \&
  {Terada}}]{2021arXiv211101613M}
{Madsen}, K.~K., {Burwitz}, V., {Forster}, K., {et~al.} 2021, arXiv e-prints,
  arXiv:2111.01613

\bibitem[{{Madsen} {et~al.}(2017{\natexlab{a}}){Madsen}, {Forster},
  {Grefenstette}, {Harrison}, \& {Stern}}]{2017ApJ...841...56M}
{Madsen}, K.~K., {Forster}, K., {Grefenstette}, B.~W., {Harrison}, F.~A., \&
  {Stern}, D. 2017{\natexlab{a}}, \apj, 841, 56

\bibitem[{{Madsen} {et~al.}(2017{\natexlab{b}}){Madsen}, {Forster},
  {Grefenstette}, {Harrison}, \& {Stern}}]{nustar_2017}
{Madsen}, K.~K., {Forster}, K., {Grefenstette}, B.~W., {Harrison}, F.~A., \&
  {Stern}, D. 2017{\natexlab{b}}, \apj, 841, 56

\bibitem[{{Madsen} {et~al.}(2015){Madsen}, {Reynolds}, {Harrison}, {An},
  {Boggs}, {Christensen}, {Craig}, {Fryer}, {Grefenstette}, {Hailey},
  {Markwardt}, {Nynka}, {Stern}, {Zoglauer}, \& {Zhang}}]{Madsen2015}
{Madsen}, K.~K., {Reynolds}, S., {Harrison}, F., {et~al.} 2015, \apj, 801, 66

\bibitem[{{MAGIC Collaboration} {et~al.}(2020){MAGIC Collaboration}, {Acciari},
  {Ansoldi}, {Antonelli}, {Arbet Engels}, {Baack}, {Babi{\'c}}, {Banerjee},
  {Barres de Almeida}, {Barrio}, {Becerra Gonz{\'a}lez}, {Bednarek},
  {Bellizzi}, {Bernardini}, {Berti}, {Besenrieder}, {Bhattacharyya},
  {Bigongiari}, {Biland}, {Blanch}, {Bonnoli}, {Bo{\v{s}}njak}, {Busetto},
  {Carosi}, {Ceribella}, {Chai}, {Chilingaryan}, {Cikota}, {Colak}, {Colin},
  {Colombo}, {Contreras}, {Cortina}, {Covino}, {D'Elia}, {da Vela}, {Dazzi},
  {de Angelis}, {de Lotto}, {Delfino}, {Delgado}, {Depaoli}, {di Pierro}, {di
  Venere}, {Do Souto Espi{\~n}eira}, {Dominis Prester}, {Donini}, {Dorner},
  {Doro}, {Elsaesser}, {Fallah Ramazani}, {Fattorini}, {Ferrara}, {Fidalgo},
  {Foffano}, {Fonseca}, {Font}, {Fruck}, {Fukami}, {Garc{\'\i}a L{\'o}pez},
  {Garczarczyk}, {Gasparyan}, {Gaug}, {Giglietto}, {Giordano}, {Godinovi{\'c}},
  {Green}, {Guberman}, {Hadasch}, {Hahn}, {Herrera}, {Hoang}, {Hrupec},
  {H{\"u}tten}, {Inada}, {Inoue}, {Ishio}, {Iwamura}, {Jouvin}, {Kerszberg},
  {Kubo}, {Kushida}, {Lamastra}, {Lelas}, {Leone}, {Lindfors}, {Lombardi},
  {Longo}, {L{\'o}pez}, {L{\'o}pez-Coto}, {L{\'o}pez-Oramas}, {Loporchio},
  {Machado de Oliveira Fraga}, {Maggio}, {Majumdar}, {Makariev}, {Mallamaci},
  {Maneva}, {Manganaro}, {Mannheim}, {Maraschi}, {Mariotti}, {Mart{\'\i}nez},
  {Mazin}, {Mi{\'c}anovi{\'c}}, {Miceli}, {Minev}, {Miranda}, {Mirzoyan},
  {Molina}, {Moralejo}, {Morcuende}, {Moreno}, {Moretti}, {Munar-Adrover},
  {Neustroev}, {Nigro}, {Nilsson}, {Ninci}, {Nishijima}, {Noda}, {Nogu{\'e}s},
  {Nozaki}, {Paiano}, {Palacio}, {Palatiello}, {Paneque}, {Paoletti},
  {Paredes}, {Pe{\~n}il}, {Peresano}, {Persic}, {Prada Moroni}, {Prandini},
  {Puljak}, {Rhode}, {Rib{\'o}}, {Rico}, {Righi}, {Rugliancich}, {Saha},
  {Sahakyan}, {Saito}, {Sakurai}, {Satalecka}, {Schmidt}, {Schweizer},
  {Sitarek}, {{\v{S}}nidari{\'c}}, {Sobczynska}, {Somero}, {Stamerra}, {Strom},
  {Strzys}, {Suda}, {Suri{\'c}}, {Takahashi}, {Tavecchio}, {Temnikov},
  {Terzi{\'c}}, {Teshima}, {Torres-Alb{\`a}}, {Tosti}, {Vagelli}, {van
  Scherpenberg}, {Vanzo}, {Vazquez Acosta}, {Vigorito}, {Vitale}, {Vovk},
  {Will}, \& {Zari{\'c}}}]{magic_2020}
{MAGIC Collaboration}, {Acciari}, V.~A., {Ansoldi}, S., {et~al.} 2020, \aap,
  635, A158

\bibitem[{{Marsden} {et~al.}(1984){Marsden}, {Gillett}, {Jennings}, {Emerson},
  {de Jong}, \& {Olnon}}]{Marsden1984}
{Marsden}, P.~L., {Gillett}, F.~C., {Jennings}, R.~E., {et~al.} 1984, \apj,
  278, L29–L32

\bibitem[{{Mart{\'\i}n} {et~al.}(2012){Mart{\'\i}n}, {Torres}, \&
  {Rea}}]{2012MNRAS.427..415M}
{Mart{\'\i}n}, J., {Torres}, D.~F., \& {Rea}, N. 2012, \mnras, 427, 415

\bibitem[{{Mayer} {et~al.}(2013){Mayer}, {Buehler}, {Hays}, {Cheung}, {Dutka},
  {Grove}, {Kerr}, \& {Ojha}}]{2013ApJ...775L..37M}
{Mayer}, M., {Buehler}, R., {Hays}, E., {et~al.} 2013, \apjl, 775, L37

\bibitem[{Meagher(2015)}]{veritas}
Meagher, K. 2015, {Six years of VERITAS observations of the Crab Nebula}

\bibitem[{{Meyer} {et~al.}(2010){Meyer}, {Horns}, \& {Zechlin}}]{meyer}
{Meyer}, M., {Horns}, D., \& {Zechlin}, H.-S. 2010, \aap, 523, A2

\bibitem[{{Mori} {et~al.}(2004){Mori}, {Burrows}, {Hester}, {Pavlov},
  {Shibata}, \& {Tsunemi}}]{2004ApJ...609..186M}
{Mori}, K., {Burrows}, D.~N., {Hester}, J.~J., {et~al.} 2004, \apj, 609, 186

\bibitem[{{O'Donnell}(1994)}]{1994ApJ...422..158O}
{O'Donnell}, J.~E. 1994, \apj, 422, 158

\bibitem[{{Owen} \& {Barlow}(2015)}]{shell}
{Owen}, P.~J. \& {Barlow}, M.~J. 2015, \apj, 801, 141

\bibitem[{{Pacini} \& {Salvati}(1973)}]{1973ApJ...186..249P}
{Pacini}, F. \& {Salvati}, M. 1973, \apj, 186, 249

\bibitem[{{Peng} {et~al.}(2022){Peng}, {Bao}, {Lu}, \&
  {Zhang}}]{2022ApJ...926....7P}
{Peng}, Q.-Y., {Bao}, B.-W., {Lu}, F.-W., \& {Zhang}, L. 2022, \apj, 926, 7

\bibitem[{{Planck Collaboration} {et~al.}(2016){Planck Collaboration}, {Ade},
  {Aghanim}, {Argüeso}, {Arnaud}, {Ashdown}, {Aumont}, {Baccigalupi},
  {Banday}, \& {Barreiro}}]{PlanckCollab2016}
{Planck Collaboration}, {Ade}, P.~A.~R., {Aghanim}, N., {et~al.} 2016, \aap,
  594, A26

\bibitem[{{Popescu} {et~al.}(2017){Popescu}, {Yang}, {Tuffs}, {Natale},
  {Rushton}, \& {Aharonian}}]{2017MNRAS.470.2539P}
{Popescu}, C.~C., {Yang}, R., {Tuffs}, R.~J., {et~al.} 2017, \mnras, 470, 2539

\bibitem[{{Porter} {et~al.}(2017){Porter}, {J{\'o}hannesson}, \&
  {Moskalenko}}]{2017ApJ...846...67P}
{Porter}, T.~A., {J{\'o}hannesson}, G., \& {Moskalenko}, I.~V. 2017, \apj, 846,
  67

\bibitem[{Porth {et~al.}(2013)Porth, Komissarov, \& Keppens}]{2013Porth}
Porth, O., Komissarov, S.~S., \& Keppens, R. 2013, Monthly Notices of the Royal
  Astronomical Society, 438, 278–306

\bibitem[{{Rees} \& {Gunn}(1974{\natexlab{a}})}]{RG74}
{Rees}, M.~J. \& {Gunn}, J.~E. 1974{\natexlab{a}}, \mnras, 167, 1–12

\bibitem[{{Rees} \& {Gunn}(1974{\natexlab{b}})}]{1974MNRAS.167....1R}
{Rees}, M.~J. \& {Gunn}, J.~E. 1974{\natexlab{b}}, \mnras, 167, 1

\bibitem[{{Ritacco} {et~al.}(2018){Ritacco}, {Mac{\'\i}as-P{\'e}rez},
  {Ponthieu}, {Adam}, {Ade}, {Andr{\'e}}, {Aumont}, {Beelen}, {Beno{\^\i}t},
  {Bideaud}, {Billot}, {Bourrion}, {Bracco}, {Calvo}, {Catalano}, {Coiffard},
  {Comis}, {D'Addabbo}, {De Petris}, {D{\'e}sert}, {Doyle}, {Goupy}, {Kramer},
  {Lagache}, {Leclercq}, {Lestrade}, {Mauskopf}, {Mayet}, {Maury},
  {Monfardini}, {Pajot}, {Pascale}, {Perotto}, {Pisano}, {Rebolo-Iglesias},
  {Rev{\'e}ret}, {Rodriguez}, {Romero}, {Roussel}, {Ruppin}, {Schuster},
  {Sievers}, {Siringo}, {Thum}, {Triqueneaux}, {Tucker}, {Wiesemeyer}, \&
  {Zylka}}]{2018A&A...616A..35R}
{Ritacco}, A., {Mac{\'\i}as-P{\'e}rez}, J.~F., {Ponthieu}, N., {et~al.} 2018,
  \aap, 616, A35

\bibitem[{{Rudy} {et~al.}(2015){Rudy}, {Horns}, {DeLuca}, {Kolodziejczak},
  {Tennant}, {Yuan}, {Buehler}, {Arons}, {Blandford}, {Caraveo}, {Costa},
  {Funk}, {Hays}, {Lobanov}, {Max}, {Mayer}, {Mignani}, {O'Dell}, {Romani},
  {Tavani}, \& {Weisskopf}}]{2015ApJ...811...24R}
{Rudy}, A., {Horns}, D., {DeLuca}, A., {et~al.} 2015, \apj, 811, 24

\bibitem[{{Schlegel} {et~al.}(1998){Schlegel}, {Finkbeiner}, \&
  {Davis}}]{1998ApJ...500..525S}
{Schlegel}, D.~J., {Finkbeiner}, D.~P., \& {Davis}, M. 1998, \apj, 500, 525

\bibitem[{{Shaposhnikov} {et~al.}(2012){Shaposhnikov}, {Jahoda}, {Markwardt},
  {Swank}, \& {Strohmayer}}]{2012ApJ...757..159S}
{Shaposhnikov}, N., {Jahoda}, K., {Markwardt}, C., {Swank}, J., \&
  {Strohmayer}, T. 2012, \apj, 757, 159

\bibitem[{{Shibata} {et~al.}(2003){Shibata}, {Tomatsuri}, {Shimanuki}, {Saito},
  \& {Mori}}]{2003MNRAS.346..841S}
{Shibata}, S., {Tomatsuri}, H., {Shimanuki}, M., {Saito}, K., \& {Mori}, K.
  2003, \mnras, 346, 841

\bibitem[{{Smith}(2003)}]{2003MNRAS.346..885S}
{Smith}, N. 2003, \mnras, 346, 885

\bibitem[{{Strong} \& {Collmar}(2019)}]{2019MmSAI..90..297S}
{Strong}, A. \& {Collmar}, W. 2019, \memsai, 90, 297

\bibitem[{{Str{\"u}der} {et~al.}(2001){Str{\"u}der}, {Briel}, {Dennerl},
  {Hartmann}, {Kendziorra}, {Meidinger}, {Pfeffermann}, {Reppin}, {Aschenbach},
  {Bornemann}, {Br{\"a}uninger}, {Burkert}, {Elender}, {Freyberg}, {Haberl},
  {Hartner}, {Heuschmann}, {Hippmann}, {Kastelic}, {Kemmer}, {Kettenring},
  {Kink}, {Krause}, {M{\"u}ller}, {Oppitz}, {Pietsch}, {Popp}, {Predehl},
  {Read}, {Stephan}, {St{\"o}tter}, {Tr{\"u}mper}, {Holl}, {Kemmer}, {Soltau},
  {St{\"o}tter}, {Weber}, {Weichert}, {von Zanthier}, {Carathanassis}, {Lutz},
  {Richter}, {Solc}, {B{\"o}ttcher}, {Kuster}, {Staubert}, {Abbey}, {Holland},
  {Turner}, {Balasini}, {Bignami}, {La Palombara}, {Villa}, {Buttler},
  {Gianini}, {Lain{\'e}}, {Lumb}, \& {Dhez}}]{2001A&A...365L..18S}
{Str{\"u}der}, L., {Briel}, U., {Dennerl}, K., {et~al.} 2001, \aap, 365, L18

\bibitem[{{Tanaka} \& {Asano}(2017)}]{2017ApJ...841...78T}
{Tanaka}, S.~J. \& {Asano}, K. 2017, \apj, 841, 78

\bibitem[{{Tanaka} {et~al.}(2018){Tanaka}, {Toma}, \&
  {Tominaga}}]{2018MNRAS.478.4622T}
{Tanaka}, S.~J., {Toma}, K., \& {Tominaga}, N. 2018, \mnras, 478, 4622

\bibitem[{{Tavani} {et~al.}(2011){Tavani}, {Bulgarelli}, {Vittorini},
  {Pellizzoni}, {Striani}, {Caraveo}, {Weisskopf}, {Tennant}, {Pucella},
  {Trois}, {Costa}, {Evangelista}, {Pittori}, {Verrecchia}, {Del Monte},
  {Campana}, {Pilia}, {De Luca}, {Donnarumma}, {Horns}, {Ferrigno}, {Heinke},
  {Trifoglio}, {Gianotti}, {Vercellone}, {Argan}, {Barbiellini}, {Cattaneo},
  {Chen}, {Contessi}, {D'Ammando}, {DeParis}, {Di Cocco}, {Di Persio},
  {Feroci}, {Ferrari}, {Galli}, {Giuliani}, {Giusti}, {Labanti}, {Lapshov},
  {Lazzarotto}, {Lipari}, {Longo}, {Fuschino}, {Marisaldi}, {Mereghetti},
  {Morelli}, {Moretti}, {Morselli}, {Pacciani}, {Perotti}, {Piano}, {Picozza},
  {Prest}, {Rapisarda}, {Rappoldi}, {Rubini}, {Sabatini}, {Soffitta},
  {Vallazza}, {Zambra}, {Zanello}, {Lucarelli}, {Santolamazza}, {Giommi},
  {Salotti}, \& {Bignami}}]{2011Sci...331..736T}
{Tavani}, M., {Bulgarelli}, A., {Vittorini}, V., {et~al.} 2011, Science, 331,
  736

\bibitem[{{Temim} \& {Dwek}(2013)}]{2013ApJ...774....8T}
{Temim}, T. \& {Dwek}, E. 2013, \apj, 774, 8

\bibitem[{{Temim} {et~al.}(2006){Temim}, {Gehrz}, {Woodward}, {Roellig},
  {Smith}, {Rudnick}, {Polomski}, {Davidson}, {Yuen}, \& {Onaka}}]{Temim2006}
{Temim}, T., {Gehrz}, R.~D., {Woodward}, C.~E., {et~al.} 2006, \aj, 132,
  1610–1623

\bibitem[{{Tomsick} {et~al.}(2021){Tomsick}, {Boggs}, {Zoglauer}, {Wulf},
  {Mitchell}, {Phlips}, {Sleator}, {Brandt}, {Shih}, {Roberts}, {Jean}, {von
  Ballmoos}, {Martinez Oliveros}, {Smale}, {Kierans}, {Hartmann}, {Leising},
  {Ajello}, {Burns}, {Fryer}, {Saint-Hilaire}, {Malzac}, {Tavecchio},
  {Fioretti}, {Bulgarelli}, {Ghirlanda}, {Chang}, {Takahashi}, {Nakazawa},
  {Matsumoto}, {Melia}, {Siegert}, {Lowell}, {Lazar}, {Beechert}, \&
  {Gulick}}]{2021arXiv210910403T}
{Tomsick}, J.~A., {Boggs}, S.~E., {Zoglauer}, A., {et~al.} 2021, arXiv
  e-prints, arXiv:2109.10403

\bibitem[{{Toor} \& {Seward}(1974)}]{1974}
{Toor}, A. \& {Seward}, F.~D. 1974, \aj, 79, 995

\bibitem[{{Trimble}(1973)}]{1973PASP...85..579T}
{Trimble}, V. 1973, \pasp, 85, 579

\bibitem[{{Tsujimoto} {et~al.}(2011){Tsujimoto}, {Guainazzi}, {Plucinsky},
  {Beardmore}, {Ishida}, {Natalucci}, {Posson-Brown}, {Read}, {Saxton}, \&
  {Shaposhnikov}}]{2011A&A...525A..25T}
{Tsujimoto}, M., {Guainazzi}, M., {Plucinsky}, P.~P., {et~al.} 2011, \aap, 525,
  A25

\bibitem[{{{\v{C}}ade{\v{z}}} {et~al.}(2004){{\v{C}}ade{\v{z}}},
  {Carrami{\~n}ana}, \& {Vidrih}}]{2004ApJ...609..797C}
{{\v{C}}ade{\v{z}}}, A., {Carrami{\~n}ana}, A., \& {Vidrih}, S. 2004, \apj,
  609, 797

\bibitem[{{V\'eron-Cetty} \& {Woltjer}(1993)}]{1993A&A...270..370V}
{V\'eron-Cetty}, M.~P. \& {Woltjer}, L. 1993, \aap, 270, 370

\bibitem[{{Weekes} {et~al.}(1989){Weekes}, {Cawley}, {Fegan}, {Gibbs},
  {Hillas}, {Kowk}, {Lamb}, {Lewis}, {Macomb}, {Porter}, {Reynolds}, \&
  {Vacanti}}]{1989ApJ...342..379W}
{Weekes}, T.~C., {Cawley}, M.~F., {Fegan}, D.~J., {et~al.} 1989, \apj, 342, 379

\bibitem[{{Weiland} {et~al.}(2011){Weiland}, {Odegard}, {Hill}, {Wollack},
  {Hinshaw}, {Greason}, {Jarosik}, {Page}, {Bennett}, {Dunkley}, {Gold},
  {Halpern}, {Kogut}, {Komatsu}, {Larson}, {Limon}, {Meyer}, {Nolta}, {Smith},
  {Spergel}, {Tucker}, \& {Wright}}]{weinland2011}
{Weiland}, J.~L., {Odegard}, N., {Hill}, R.~S., {et~al.} 2011, \apjs, 192, 19

\bibitem[{{Weisskopf} {et~al.}(2012){Weisskopf}, {Elsner}, {Kolodziejczak},
  {O'Dell}, \& {Tennant}}]{2012ApJ...746...41W}
{Weisskopf}, M.~C., {Elsner}, R.~F., {Kolodziejczak}, J.~J., {O'Dell}, S.~L.,
  \& {Tennant}, A.~F. 2012, \apj, 746, 41

\bibitem[{{Weisskopf} {et~al.}(2010){Weisskopf}, {Guainazzi}, {Jahoda},
  {Shaposhnikov}, {O'Dell}, {Zavlin}, {Wilson-Hodge}, \&
  {Elsner}}]{2010ApJ...713..912W}
{Weisskopf}, M.~C., {Guainazzi}, M., {Jahoda}, K., {et~al.} 2010, \apj, 713,
  912

\bibitem[{{Weisskopf} {et~al.}(2016){Weisskopf}, {Ramsey}, {O'Dell}, {Tennant},
  {Elsner}, {Soffitta}, {Bellazzini}, {Costa}, {Kolodziejczak}, {Kaspi},
  {Muleri}, {Marshall}, {Matt}, \& {Romani}}]{2016SPIE.9905E..17W}
{Weisskopf}, M.~C., {Ramsey}, B., {O'Dell}, S., {et~al.} 2016, in Society of
  Photo-Optical Instrumentation Engineers (SPIE) Conference Series, Vol. 9905,
  Space Telescopes and Instrumentation 2016: Ultraviolet to Gamma Ray, ed.
  J.-W.~A. {den Herder}, T.~{Takahashi}, \& M.~{Bautz}, 990517

\bibitem[{{Weisskopf} {et~al.}(1978){Weisskopf}, {Silver}, {Kestenbaum},
  {Long}, \& {Novick}}]{1978ApJ...220L.117W}
{Weisskopf}, M.~C., {Silver}, E.~H., {Kestenbaum}, H.~L., {Long}, K.~S., \&
  {Novick}, R. 1978, \apjl, 220, L117

\bibitem[{{Weisskopf} {et~al.}(2013){Weisskopf}, {Tennant}, {Arons},
  {Blandford}, {Buehler}, {Caraveo}, {Cheung}, {Costa}, {de Luca}, {Ferrigno},
  {Fu}, {Funk}, {Habermehl}, {Horns}, {Linford}, {Lobanov}, {Max}, {Mignani},
  {O'Dell}, {Romani}, {Striani}, {Tavani}, {Taylor}, {Uchiyama}, \&
  {Yuan}}]{2013ApJ...765...56W}
{Weisskopf}, M.~C., {Tennant}, A.~F., {Arons}, J., {et~al.} 2013, \apj, 765, 56

\bibitem[{Wells(2019)}]{wells_veritas}
Wells, R.~M. 2019, PhD thesis, Iowa State University, Graduate Theses and
  Dissertations. 1706

\bibitem[{{Wilson-Hodge} {et~al.}(2011){Wilson-Hodge}, {Cherry}, {Case},
  {Baumgartner}, {Beklen}, {Narayana Bhat}, {Briggs}, {Camero-Arranz},
  {Chaplin}, {Connaughton}, {Finger}, {Gehrels}, {Greiner}, {Jahoda}, {Jenke},
  {Kippen}, {Kouveliotou}, {Krimm}, {Kuulkers}, {Lund}, {Meegan}, {Natalucci},
  {Paciesas}, {Preece}, {Rodi}, {Shaposhnikov}, {Skinner}, {Swartz}, {von
  Kienlin}, {Diehl}, \& {Zhang}}]{2011ApJ...727L..40W}
{Wilson-Hodge}, C.~A., {Cherry}, M.~L., {Case}, G.~L., {et~al.} 2011, \apjl,
  727, L40

\bibitem[{{Woosley} \& {Weaver}(1995)}]{1995ApJS..101..181W}
{Woosley}, S.~E. \& {Weaver}, T.~A. 1995, \apjs, 101, 181

\bibitem[{{Yeung} \& {Horns}(2019)}]{paul_paper}
{Yeung}, P. K.~H. \& {Horns}, D. 2019, \apj, 875, 123

\bibitem[{{Yeung} \& {Horns}(2020)}]{2020A&A...638A.147Y}
{Yeung}, P. K.~H. \& {Horns}, D. 2020, \aap, 638, A147

\end{thebibliography}
   
\begin{appendix}
\section{Individual VHE data sets}
\label{appendix:VHE}
The  observation of VHE ($E>100~\mathrm{GeV}$) gamma-rays
is the domain of ground based air shower detection. In this energy range, the Crab Nebula 
is the brightest steady source in the sky and therefore, the detection and observation
of the Crab Nebula is the litmus test for VHE gamma-ray detectors.
The fact that the very first TeV gamma-ray source clearly detected was the Crab Nebula 
\citep{1989ApJ...342..379W}
underlines the importance of this particular source for ground based gamma-ray 
astronomy. 
The routine monitoring  of this standard candle 
has lead to a large number of published data sets in this energy range with widely
different detection techniques. We have selected observations which cover both 
a wide range of observation times (from 1997 to 2020) and different 
techniques (imaging air Cherenkov telescopes as well as shower front sampling 
with extended air shower arrays). This is not a complete collection but it is meant
to provide a good cross section of the spectroscopic data from the past two decades.

\begin{figure}
    \centering
    \includegraphics[width=\linewidth]{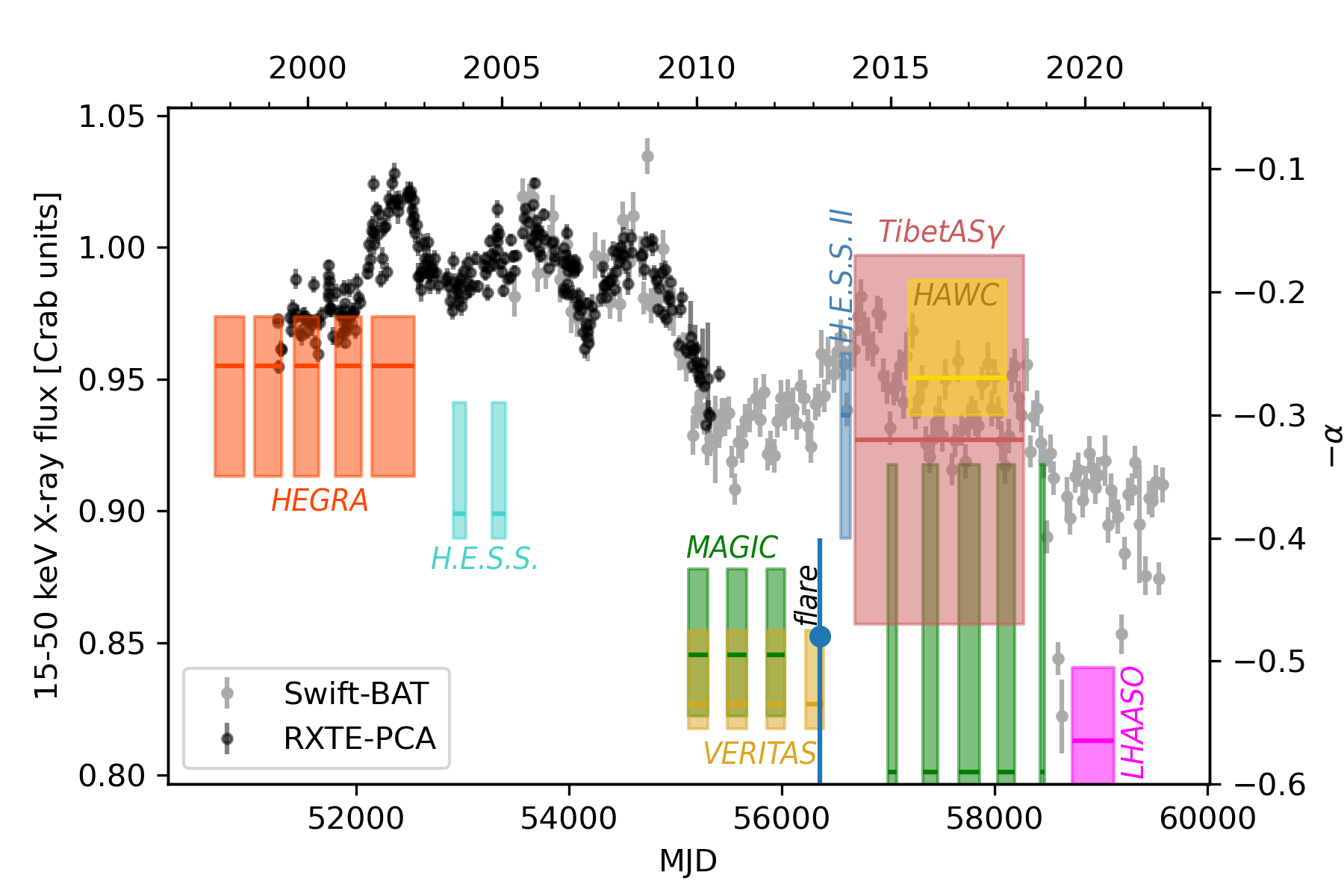}
    \caption{Observed hard X-ray light curve (\textit{RXTE}-PCA data from 
    \citep{2011ApJ...727L..40W}, \textit{Swift}-BAT \citep{2013ApJS..209...14K} available through
    the BAT transient monitor page)  from the Crab Nebula. The observed flux 
     varies on time-scales of several years.
    The observations at VHE (listed in Table~\ref{tab:data_sets}) 
    cover different parts
    of the X-ray flux states.}
    \label{fig:timeline}
\end{figure}

The individual data sets are listed in Table~\ref{tab:data_sets}. 
The 
observational times as well as ranges of best-fitting values of
$\alpha$ are included with the hard X-ray light curve from the nebula
in Fig.~\ref{fig:timeline}. The contemporaneous data sets are
consistent within the uncertainties of $\alpha$. 
When looking at the evolution of the hard X-ray flux and the 
changes of $\alpha$, there is at least a trend visible:
during higher flux states of hard X-rays, the VHE spectra 
tend to follow the shape given for a slower radial drop
of the magnetic field. During lower flux states, the
magnetic field seems to go down quicker with increasing distance.

\begin{table}[]
    \centering
\caption{Summary of VHE gamma-ray data sets used here.} 
\resizebox{\linewidth}{!}{
\begin{tabular}{l|ccccc|r}
Data set & Start & End &  $\chi^2(dof)$ & $\alpha$ & $\zeta$ &Ref.\\ 
\hline\hline
HEGRA (IACT)    & 1997  & 2002 & 18(15)      & $0.26^{+0.09}_{-0.04}$ & $1.197\pm0.005$ & (1) \\
H.E.S.S. (IACT) & 2003  & 2005 & 24(10)      & $0.38^{+0.02}_{-0.09}$ & $1.065\pm0.005$ & (2) \\
MAGIC  (IACT)   & 2009  & 2011 & 11(13)      & $0.495^{+0.05}_{-0.07}$ & $0.896\pm0.006$ & (3) \\
VERITAS (IACT)  &$2009$&$2013$ &9(11)      & $0.535^{+0.02}_{-0.06}$ & $1.038\pm0.002$ & (4) \\ 
HAWC    (WCD)   &6/2015&12/2017 &  7(8)     & $0.27^{+0.03}_{-0.08}$ & $1.093\pm0.003$ & (5) \\
MAGIC (VLZA)    & 2014  & 2018   &2(6)       & $0.59\pm0.25$          & $1.09\pm0.02$ & (6) \\
Tibet AS$\gamma$ (EAS, WCD) & 2014 & 2017 &4(9) & $0.32\pm0.15$           & $1.06\pm0.02$ & (7) \\
VERITAS (flare) & 03/2013 & 03/2013 & 13(13) & $0.42\pm0.17$ & $1.09\pm0.01$ & (8) \\
H.E.S.S. II (flare) & 03/2013 & 03/2013& 17(31) & $0.68\pm0.3$ &$0.96\pm0.01$  & (9) \\
H.E.S.S. II (IACT) & $11/2013$ & $1/2014$ &  93(37)  & $0.3^{+0.1}_{-0.05}$      & $1.100\pm0.007$ & (10)  \\
LHAASO (EAS, WCD) & 2019 & 2020 & 11(16) & $0.56\pm0.06$                  & $0.932\pm0.006$ & (11) \\
\hline
\end{tabular}
}
\tablefoot{Listed data sets include imaging air Cherenkov telescopes'
(IACTs) observations as well
as data taken with 
extended air shower front sampling (EAS) arrays with scintillators
or water Cherenkov detectors (WCD). 
By observing sources with IACTs at very
large zenith angles (VLZA) between 70$^\circ$ and 80$^\circ$, the
collection area at large energies beyond 10 TeV is substantially
increased.  The uncertainty on $\alpha$ is given for 
    a $90~\%$ confidence level and has been estimated from the discrete sampled values of $\alpha$. For 
    some data sets the $\chi^2$ curve is under-sampled while for others, the maximum range of $\alpha$ is not sufficient
    to cover the confidence interval.}
    \tablebib{
    (1) \citet{Aharonianetal2004}, 
    (2) \citet{2006A&A...457..899A},
    (3) \citet{MagicCrab}, 
    (4) \citet{wells_veritas}, 
    (5) \citet{hawc},
    (6) \citet{magic_2020}, 
    (7) \citet{PhysRevLett.123.051101}, 
    (8) \citet{2014ApJ...781L..11A}, (i) \citet{2014A&A...562L...4H},
    (9) \citet{hess_data}, \citet{holler_hess}, 
    (10) \citet{2021Sci...373..425L}.
}
    \label{tab:data_sets}
\end{table}
\FloatBarrier
\section{Pair-production opacity towards the Crab Nebula}
\label{appendix:tau}
The soft photon fields present in the Galactic disk include the cosmic microwave background (CMB) as well as 
dust and stellar emission. Energetic gamma-rays will be absorbed by pair-production processes when propagating in
the soft photon background \cite{1967PhRv..155.1404G}. Using a recent radiation model of the Galaxy \citep{2017MNRAS.470.2539P}, the optical depth
for the line of sight towards the Crab Nebula is calculated  to correct its apparent gamma-ray
brightness for absorption \citep{2021Sci...373..425L}. 
The CMB leads to a reduction by $\approx 24~\%$ at $E_\gamma=2$~PeV (assuming a distance of 2~kpc).
At lower energies, the CMB photons are not sufficiently energetic to create pairs such that at $E_\gamma=100$~TeV
pairs are mainly produced with thermal dust emission. 
  Given that  the photon density of the dust emission is substantially smaller at
the relevant wavelength of $\approx 100~\mu$m, the resulting absorption is less pronounced than the CMB-related absorption
at higher energies ($1-\exp(-\tau_{\gamma\gamma}(E_\gamma=100~\mathrm{TeV})) \approx 1.5~\%$. However, the result on the opacity used 
by \citet{2021Sci...373..425L} is not taking into account the anisotropy of the soft photon field. 

Here, we provide an estimate of the previously neglected effect
of anisotropy by using the following simplifications: (i) the relevant specific energy density $\lambda u_\lambda$ of 
the photon field is dominated by the dust component with a peak at $\lambda\approx 100~\mu$m, 
(ii) the angular distribution of the anisotropic background field follows the measured local intensity, (iii) 
both, the intensity as well angular distribution remain constant along the line of sight towards the Crab Nebula.

The resulting optical depth $\tau_{\gamma\gamma}(E_\gamma)$ for the Crab Nebula located at a distance
$d$ and in the direction along the normal vector $\mathbf{e}_C$ is given by:
\begin{equation}
    \tau(E_\gamma) = 
      \int\limits_{0}^d \mathrm{d}x~ \int\limits_{\varepsilon_{min}}^{\varepsilon_{max}} \mathrm{d}\varepsilon~ n(\varepsilon)\int \mathrm{d}\Omega~ f(\mathbf{e}_\Omega) \sigma_{\gamma\gamma}(s)(1-\mathbf{e}_\Omega\cdot \mathbf{e}_{C}), 
\end{equation}
where $\sigma_{\gamma\gamma}$ is the pair production cross section (see e.g. Eq.~1 of \citet{1967PhRv..155.1404G}),
$n(\varepsilon)$ is the total (sky integrated) differential number density of photons $\mathrm{d}n = n(\varepsilon)\mathrm{d}\varepsilon \mathrm{d}V$. The angular distribution  $f(\mathbf{e}_\Omega)$
was normalised  over the entire sphere, such that:
\begin{equation}
    \int \mathrm{d}\Omega\, f(\mathbf{e}_\Omega) = 1.
\end{equation}
The unit vector $\mathbf{e}_\Omega$ points towards a direction given in Galactic coordinates. 
This way, $\mu:=\mathbf{e}_{\Omega}\cdot \mathbf{e}_{C}$ corresponds to the cosine of the 
angular distance between the photon directions of the gamma ray from the Crab Nebula and the soft photon from the Galactic background emission.  
The centre of momentum energy $s$ depends on the photon energy $E_\gamma$, 
the soft photon's energy $\varepsilon$ as well as $\mu$:
\begin{equation}
    s = 2\varepsilon E_\gamma (1-\mu).
\end{equation}

Following the approximation (i),  the dust photon field was taken to be  a single function
\begin{equation}
    n(\varepsilon) = n_0 \left(\frac{\varepsilon}{\varepsilon_0}\right)^{3.5}
    \frac{1}{\exp(\varepsilon/\varepsilon_0)-1},
\end{equation}
with $n_0 = 1223.78~\mathrm{cm^{-3}\,eV^{-1}}$ and $\varepsilon_0=1.7\times 10^{-3}~\mathrm{eV}$. This is a reasonably close approximation of
the model proposed by \citet{2017MNRAS.470.2539P} around the peak at $\lambda\approx 100~\mu$m. 

\begin{figure}
    \centering
    \includegraphics[width=\linewidth]{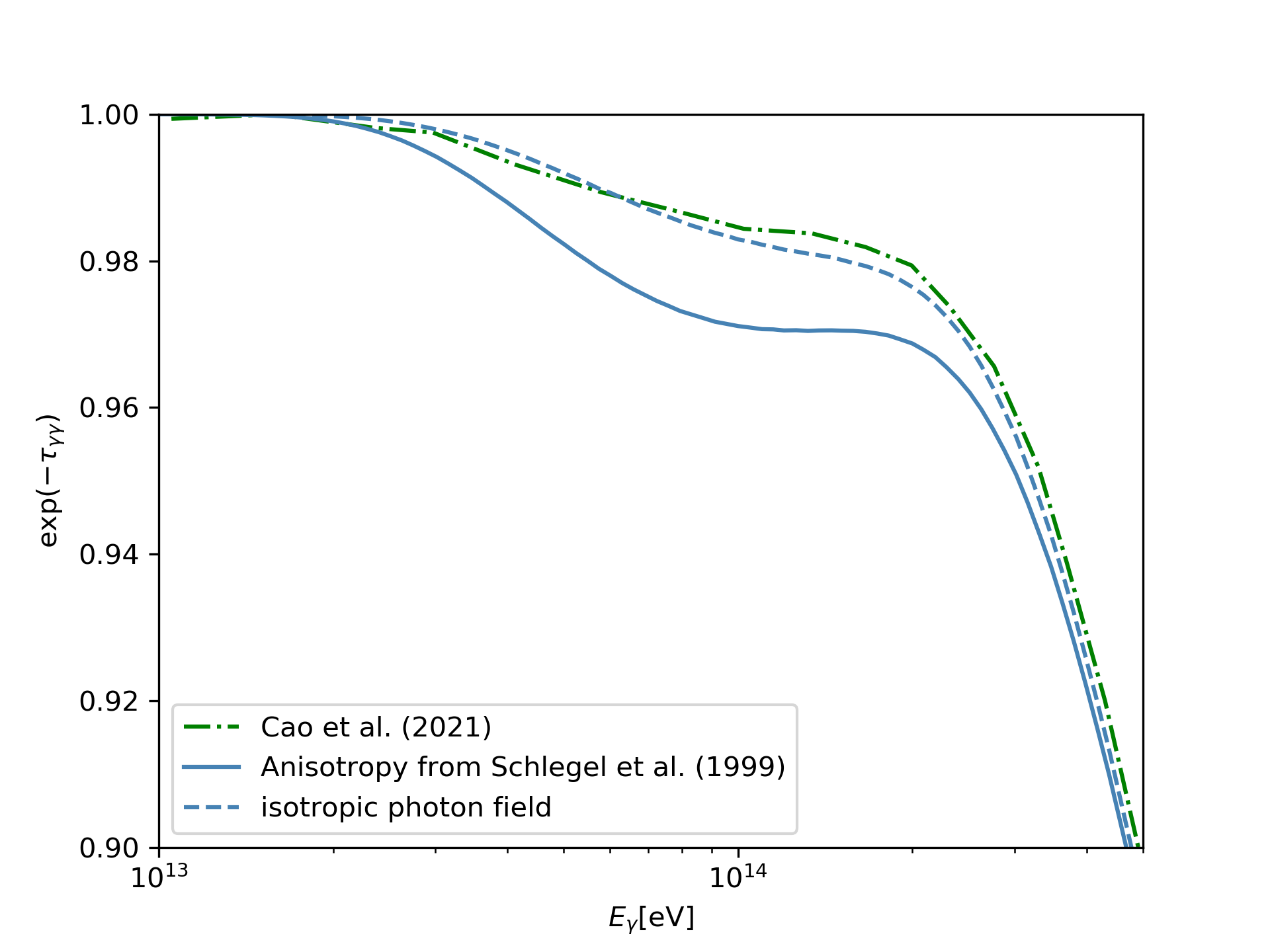}
    \caption{Energy-dependent attenuation of gamma rays
    for the line of sight towards the Crab Nebula: The solid and dashed lines are the 
    attenuation calculated here (see text for details) for the anisotropic
    radiation field (solid) and for the assumption of an isotropic radiation
    field with identical energy density (dashed line). The dash-dotted
    line is the attenuation as calculated in the framework of a detailed 
    Galactic dust emission model, however assuming an isotropic radiation field \citep{2021Sci...373..425L}}
    \label{fig:optical_depth}
\end{figure}

The angular distribution at $\lambda=100~\mu$m was taken from the all-sky map from DIRBE measurements \citep{1998ApJ...500..525S} with 
the zodiacal foreground and point sources removed. 
In order to highlight the effect of the anisotropy, we calculated the optical depth for both cases (anisotropic and isotropic photon field)
with the same energy density $\lambda u_\lambda$ and compare 
the results in 
Fig.~\ref{fig:optical_depth}. While the simplified model reproduces well the more detailed calculation, taking additional photon fields  as well as the spatial dependence 
of the photon density \citep{2017MNRAS.470.2539P} into account, the anisotropic case increases the attenuation as expected. At $E_\gamma=100~\mathrm{TeV}$, the flux is attenuated by
about 3~\% (as compared to about 1.7~\% for the isotropic case). It
is important to note, that this approach is particular to sources close to the Sun's position, such that the
approximations (ii) and (iii) allow the simplified treatment of the anisotropic case. The general anisotropic case has been treated in \citet{2017ApJ...846...67P} (and references
therein). 
\FloatBarrier
\section{Low $\alpha$ VHE solution}
\label{appendix:lowalpha}
The combination of VHE data sets shown in Sect.~\ref{section:VHE} favours a value of $\alpha=0.51\pm0.03$. However, the
VHE data sets show individually quite different fitting results as demonstrated by the values in Table~\ref{tab:data_sets}. 
While there is a large number of possible combinations conceivable, there seems to be at least one additional combination of data sets
that would favour a smaller value of $\alpha$ while still maintaining a good fit to the data. Excluding therefore the 
two data sets obtained with the H.E.S.S. telescopes which are not fit well by the model
(see Table~\ref{tab:data_sets}), we combined the spectra measured
with HEGRA, HAWC, and the Tibet AS$\gamma$ experiments. The resulting estimate of $\alpha=0.29^{+0.03}_{-0.06}$ is significantly 
smaller than the value found for the data sets shown in Sect.~\ref{section:VHE}. 

\begin{figure}
    \centering
    \includegraphics[width=\linewidth]{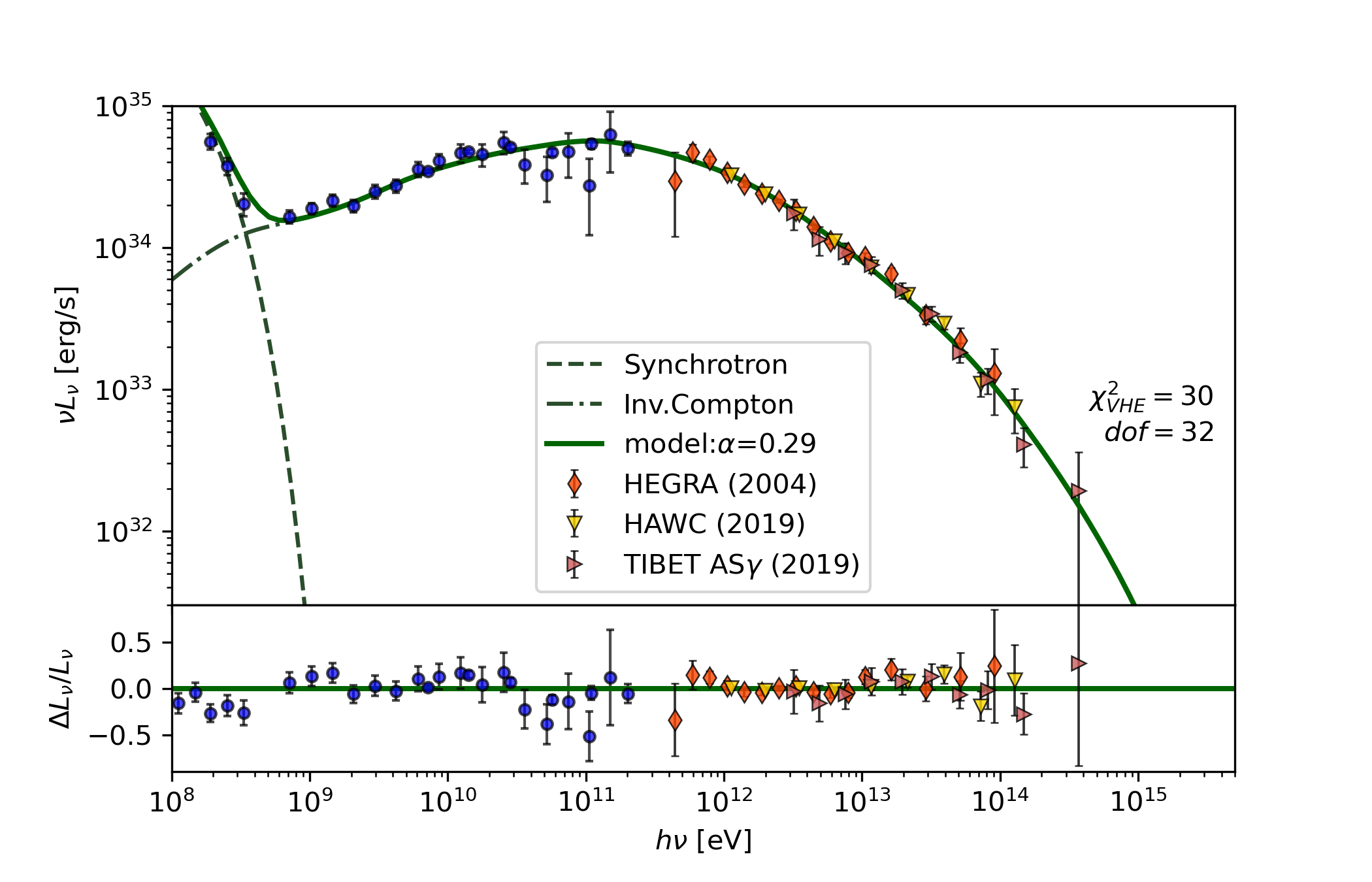}
    \caption{When combining the energy spectra obtained with HEGRA, HAWC, and Tibet~AS$\gamma$, the data-sets are consistent
    with each other (after scaling the energy) and can be described with the model reasonably well: $\chi^2_{VHE}(dof)=30(32)$.
    }
    \label{fig:sed_lowalpha}
\end{figure}

The resulting fit (shown in Fig.~\ref{fig:sed_lowalpha}) is still acceptable with $\chi^2_{VHE}(dof)=30(32)$.  The Tibet AS$\gamma$ 
data requires an up-scaling of the energy scale ($\zeta=1.07(2)$) to match the model while in the case of the 
$\alpha=0.51$ fit, $\zeta=0.92(2)$ (see Table~\ref{tab:lowalpha} for the results of this fit). 
The best-fitting parameters listed in Table~\ref{table:lowalpha_parameters} are systematically different to the 
parameters listed for the solution with $\alpha=0.51$ (see Table~\ref{table:parameters}). For  the solution shown here, 
the magnetic field at the shock is smaller and drops off slower. To compensate, the spatial particle distribution 
does not extend as far as in the case of $\alpha=0.51$ (see parameters $\Psi_3$, $\Psi_{14}$). This in turn reduces the size of the IC nebula (see
also Fig.~\ref{fig:sed_varalpha_size}). The resulting $\chi^2_{IC,ext}$ is therefore worse than in the solution with $\alpha=0.51$. Combined with 
the worse VHE fit, the probability to obtain this large value of $\chi^2(dof)=289(262)$ is only $p(>\chi^2,dof)=0.12$. Even though the goodness of
fit is noticeably worse than in the $\alpha=0.51$ case, it is still acceptable.

\begin{table}
\caption{Energy-scaling for the solution with $\check \alpha=0.29$.}
\begin{tabular}{ c|c|c|c|c }
 \textbf{Data set} & $\zeta$ &   
                     $\sigma_\zeta$   & 
                     $\chi^{2}_{before}(dof)$   & 
                     $\chi^{2}_{after}(dof)$ \\
($\check\alpha=0.29$) & & 68~\% c.l. & & \\
 \hline
 \hline
\textit{Fermi}-{LAT}  & 1.00  & fixed  & 26.6(25) &26.6(25)  \\
HEGRA               & 1.216 & 0.006  & 3050.7(16)   &18.1(15) \\  
HAWC                & 1.083 & 0.004  & 556.8(9)    &7.7(8) \\
Tibet AS$\gamma$    & 1.073 & 0.020  & 21.4(10)     &4.2(9) \\ 
\hline
\textbf{Combined} &&                 & 3655.5(60) & 56.6(57) \\
\hline
\end{tabular}
\label{tab:lowalpha}
\end{table}

\begin{table}[h]
\caption{Best-fitting parameters for the solution with $\alpha=0.29$.\label{table:lowalpha_parameters}} 
    \centering
    \begin{tabular}{|l|c|}
\hline          
\textbf{Parameter}  & \multicolumn{1}{c|}{\textbf{best-fitting values}} \\
                                  & (68~\% c.l.) \\
\hline
\multicolumn{2}{c}{\textit{Radio electrons}} \\
\hline
$\Psi_1 = s_r$          & $1.56\pm0.03$     \\
$\Psi_2 = \ln(N_{r,0})$     & $114.8\pm0.2$  \\
$\Psi_3 = \ln(\gamma_1)$ & $11.5\pm0.1$  \\ 
$\Psi_4 = \rho_r$ [$^{\prime\prime}$] & $83\pm3$    \\
\hline
\multicolumn{2}{c}{\textit{Wind electrons}} \\
\hline
$\Psi_5 = s_1$     & $2.9  \pm 0.1$  \\
$\Psi_6 = s_2$     & $3.3 \pm 0.01$  \\
$\Psi_7 = s_3$     & $3.62 \pm 0.04$  \\
$\Psi_8 = \ln(\gamma_{w0})$ & $12.7\pm0.2$  \\
$\Psi_9 = \ln(\gamma_{w1})$ & $15.8\pm0.8$  \\
$\Psi_{10} = \ln(\gamma_{w2})$ & $19.3\pm0.2$  \\
$\Psi_{11} = \ln(\gamma_{w3})$ & $22.5\pm0.04$  \\
$\Psi_{12} = \ln(N_{w,0})$         & $74.3\pm0.5$  \\
$\Psi_{13} = \beta$        &    $0.14\pm0.01$  \\
$\Psi_{14} = \rho_0$[$^{\prime\prime}$]        & $83\pm3$ \\
\hline 
\multicolumn{2}{c}{\textit{Dust parameters}}  \\
\hline
$\Psi_{15} = r_\mathrm{out}$ [pc] & $1.57\pm0.12$  \\
$\Psi_{16} = \log_{10} (M_1/M_\odot)$ & $-4.4\pm0.1$  \\
$\Psi_{17} = \log_{10} (M_2/M_\odot)$ & $-1.2\pm0.1$  \\
$\Psi_{18} = T_1$ [K] & $148\pm8$  \\
$\Psi_{19} = T_2$ [K] & $39\pm3$  \\
\hline
\multicolumn{2}{c}{\textit{Magnetic field parameters}}\\
\hline
$\Psi_{20} = B_0$ [$\mu$G]    & $167\pm5$   \\
$\Psi_{21} = \alpha$          & $0.29^{+0.03}_{-0.06}$  \\
\hline
\multicolumn{2}{c}{\textit{Goodness of fit}}\\
\hline
$\chi^2_{sync,SED}(dof)$           & $185$ (184) \\
$\chi^2_{sync,ext}(dof)$           & $17$ (15)   \\
$\chi^2_{IC,SED}(dof)$             & $27$ (23)   \\
$\chi^2_{IC,ext}(dof)$             & $30$ (8)    \\
$\chi^2_{VHE}(dof)$                & $30$ (32)   \\
\hline 
$\chi^2_{tot} (dof)$         &     $289$ (262) \\
\hline
\end{tabular}

\end{table}

\FloatBarrier
\section{UHE gamma-ray event statistic}
\label{appendix:photstat}
 In the main section, the expected number of events for the LHAASO observation \citep{2021Sci...373..425L} were used to compare with the 
 observed number of events detected at PeV energies. The underlying calculation is performed using the exposure time ($T_\mathrm{obs}=10^7~\mathrm{s}$)
 and the collection areas listed in the published data\footnote{\url{http://english.ihep.cas.cn/lhaaso/pdl/202110/t20211026_286779.html}}.
 The resulting event numbers $N_i$ in bin $i$ are calculated by integrating 
 the differential photon flux predicted by the model $\Phi_{model}(E)$ within the bin interval $E_{i,low}$ and $E_{i,high}$:
 \begin{equation}
     N_i = T_\mathrm{obs} A_{eff,i} \int\limits_{E_{i,low}}^{E_{i,high}}\mathrm{d}E\,\Phi_{model}(E/\zeta) e^{-\tau_{\gamma\gamma}(E/\zeta)}.
 \end{equation}
 In Tab.~\ref{tab:event_statistics}, the resulting event numbers are compared with the background subtracted excess events from the data. 
 For the two bins just below and above
 1 PeV, the expected number of events is $\lambda=0.7$ while the number of observed events (background free) is $k=2$. The probability to
 observe 2 or more events is calculated with the Poissonian distribution 
 \begin{equation}
     p(k\ge 2) = 1-p_{\lambda}(0)-p_\lambda(1) = 1 - \exp(-\lambda)(1 + \lambda) = 15.6~\%.
 \end{equation}
 \begin{table}[t]
     \caption{Event statistics for the LHAASO data set: comparison of model and data.}
     \centering
     \begin{tabular}{c|c|c|c}
    $\log_{10}(E_{i,low}/\mathrm{TeV})$      & 
    $\log_{10}(E_{i,high}/\mathrm{TeV})$ & $N_i$ & $N_{i,obs}$    \\
    \hline \hline
    1 & 1.2 & 3305 & 4011 \\
    1.2 & 1.4 & 1718 & 1979 \\
    1.4 & 1.6 & 601  & 649  \\
    1.6 & 1.8 & 255  & 278 \\
    1.8 & 2 & 145  & 151 \\
    2 & 2.2 & 53.2   & 53.6 \\
    2.2 & 2.4 & 19.1  & 22.5  \\
    2.4 & 2.6 & 6.3  & 5.8   \\
    2.6 & 2.8 & 1.9  & 2.9   \\
    2.8 & 3.0 & 0.5  & 1.0   \\
    3.0 & 3.2 & 0.2  & 1.0   \\
    3.2 & 4.0 & 0.02 & 0.0   \\
\hline \hline
     \end{tabular}
     \label{tab:event_statistics}
 \end{table}
\end{appendix}

\end{document}